\documentclass[12pt,preprint]{aastex}
\usepackage{graphicx}
\usepackage{color}
\usepackage{natbib}  
\usepackage{lscape}

\begin{document}

\title{Indicators of Intrinsic AGN Luminosity: a Multi-Wavelength Approach}

\author{Stephanie M. LaMassa$^{1}$, Tim M. Heckman$^{1}$, Andrew Ptak$^{2}$, Lucimara Martins$^{3}$, Vivienne Wild$^{4}$, Paule Sonnentrucker$^1$, Christy Tremonti$^{5}$}

\affil{$^1$The Johns Hopkins University, Baltimore, MD, USA
$^2$NASA Goddard Space Flight Center
$^3$ NAT - Universidade Cruziero do Sul, S\~{a}o Paulo, Brazil 
$^{4}$ Institut d'Astrophysique de Paris, 75014 Paris, France
$^{5}$ Department of Astronomy, University of Wisconsin-Madison, Madison, WI, USA}

\begin{abstract}
Active Galactic Nuclei (AGN) consist of an accretion disk around a supermassive black hole which is in turn surrounded by an obscuring torus of dust and gas. As the resulting geometry of this system affects the observable properties, quantifying isotropic indicators of intrinsic AGN luminosity is important in selecting unbiased samples of AGN. In this paper we consider five such proxies: the luminosities of the [OIII]$\lambda$5007 line, the [OIV]25.89$\mu$m line, the mid-infrared (MIR) continuum emission by the torus, and the radio and hard X-ray (E $>$ 10keV) continuum emission. We compare these different proxies using two complete samples of low-redshift type 2 AGN selected in a homogeneous way based on different indicators: an optically selected [OIII] sample and a mid-infrared selected 12$\mu$m sample. To assess the relative merits of these proxies, we have undertaken two analyses. First, we examine the correlations between all five different proxies, and find better agreement for the [OIV], MIR, and [OIII] luminosities than for the hard X-ray and radio luminosities. Next, we compare the ratios of the fluxes of the different proxies to their values in unobscured Type 1 AGN. The agreement is best for the ratio of the [OIV] and MIR fluxes, while the ratios of the hard X-ray to [OIII], [OIV], and MIR fluxes are systematically low by about an order-of-magnitude in the Type 2 AGN, indicating that hard X-ray selected samples do not represent the full Type 2 AGN population. In a similar spirit, we compare different optical and MIR diagnostics of the relative energetic contributions of AGN and star formation processes in our samples of Type 2 AGN. We find good agreement between the various diagnostic parameters, such as the equivalent width of the MIR polycyclic aromatic hydrocarbon features, the ratio of the MIR [OIV]/[NeII] emission-lines, the spectral index of the MIR continuum, and the commonly used optical emission-line ratios. Finally, we test whether the presence of cold gas associated with star-formation leads to an enhanced conversion efficiency of AGN ionizing radiation into [OIII] or [OIV] emission. We find that no compelling evidence exists for this scenario for the luminosities represented in this sample (L$_{bol}$ $\approx$ 10$^{9}$ - 8 $\times$ 10$^{11}$ L$_{\sun}$). 

\end{abstract}

\section{Introduction}

Active galactic nuclei (AGN) are powered by accretion onto a central supermassive blackhole. According to the unified model (e.g. Antonucci 1993), an optically-thick torus of dust and gas surrounds this central engine. Hence orientation of the system plays a central role in determining the observable features of AGN. In Type 1 AGN, the system is oriented face-on, leaving an unobstructed view of the central engine and the broad line region. In contrast, Type 2 AGN are oriented edge-on, blocking the accretion disk and the broad line region. These obscured AGN can be identified by their narrow optical and mid-infrared emission lines which originate in gas photoionized by accretion disk photons. This narrow line region (NLR) extends hundreds of parsecs away from the central source and is therefore not significantly affected by torus obscuration.

The luminosity of emission lines formed in the narrow line region can therefore be used as isotropic indicators of intrinsic AGN luminosity. The flux of the [OIII]$\lambda$5007 line is commonly used as such a diagnostic (e.g. Bassani et al. 1999, Heckman et al. 2005) as it is one of the most prominent lines and suffers little contamination from star formation processes in the host galaxy. This line can be attenuated by dust in the host galaxy, though this effect can be somewhat remedied by applying a reddening correction using the observed Balmer decrement (i.e. the observed ratio of the narrow H$\alpha$/H$\beta$ emission-lines compared to the intrinsic ratio) and the extinction curve for galactic dust \citep{oster}.

Isotropic indicators of AGN luminosity also exist in the infrared band and are much less affected by dust extinction than the optical [OIII] line. Recently, the luminosity of the [OIV] 25.89$\mu$m line has been shown to be a robust proxy of AGN power (e.g. Mel\'{e}ndez et al. 2008, Rigby et al. 2009, Diamond-Stanic et al. 2009): it is formed in the NLR, so it is not affected by torus obscuration, and with an ionization potential of 54.9 eV, starburst activity does not significantly contribute to this line. AGN also emit over 20\% of their bolometric flux in the mid-infrared (MIR), where photons produced by the continuum are absorbed by the torus and re-radiated (e.g. Spinoglio \& Malkan 1989). This MIR emission from the dusty torus can also be a proxy for the intrinsic AGN luminosity. Two potential issues with the MIR continuum are contamination by emission from dust heated by stars (e.g. Buchanan et al. 2006; Deo et al. 2009) and possible anisotropy in the torus emission (e.g. Pier \& Krolik 1992, Buchanan et al. 2006, Elitzur \& Shlosman 2006, Nenkova et al. 2008). 

Radio and hard X-ray (E $>$ 10 keV) flux can serve as proxies of the intrinsic AGN continuum. Radio emission has been shown to be similar between type 1 and type 2 AGN (e.g. Giuricin et al. 1990, Diamond-Stanic et al. 2009, Mel\'endez et al. 2010) and correlated with optical luminosity, in particularly the [OIII] flux \citep{Xu}, making high resolution radio observations that isolate emission from the nucleus another diagnostic of intrinsic AGN power. Hard X-rays can pierce through the obscuring torus, provided that the object is not heavily Compton thick (N$_H < 10^{25}$ cm$^{-2}$), and has therefore been used as a method to select AGN samples (e.g. Winter et al. 2008, Treister et al. 2009).

Connections have been observed between star formation activity in the host galaxy and the central AGN (e.g. Kauffmann et al. 2003). Starburst activity can be parametrized by various IR features, such as the equivalent width (EW) of polycyclic aromatic hydrocarbons (PAHs). IR and optical data, such as the ratio of fine structure lines and the shape of the spectral slope, can also reveal the relative amount of AGN to starburst activity. Samples of Seyfert 2 galaxies (the predominant local class of type 2 AGN) are useful in examining the relationships among these star formation vs. AGN-to-starburst indicators as the obscuration of the central engine allows detailed study of the host galaxy.

To address these issues, we will use two complete and homogeneous Sy2 samples selected based on isotropic indicators of AGN luminosity (one [OIII]-selected sample and one MIR-selected sample). We will compare the various diagnostics of intrinsic AGN luminosity and probe for biases resulting from sample selection criteria, starburst contamination, errors introduced from extinction correction, and scatter due to the various physical mechanisms producing these emission features.  Such biases are likely minimized in the diagnostic ratios with the smallest dispersion. Where available, we compare these ratios with the Sy1 values to probe for differences due to the inclination of the system, thus testing to which extent these indicators of intrinsic AGN luminosity are truly ``isotropic.'' We will also test the agreement among mid-infrared and optical star formation indicators. Finally, we will examine the possibility that the fraction of the AGN ionizing luminosity that is converted into [OIII] and [OIV] emission is systematically higher in systems in which there is a copious supply of dense gas associated with starburst activity.

\section{The Data}

\subsection{Sample Selection}
The selection of the SDSS [OIII] sample is discussed in detail in LaMassa et al. (2009). In brief, Type 2 AGN were drawn from a parent sample of approximately 480,000 galaxies in SDSS Data Release 4 = DR4 (Adelman-McCarthy et al. 2006). The Type 2 AGN with an observed [OIII] flux greater than 4 $\times$ 10$^{-14}$ erg s$^{-1}$ cm$^{-2}$ were selected, providing a complete sample of 20 Sy2s (hereafter the ``[OIII] sample,'' listed in Table \ref{o3_sample}).

The mid-IR sample comprises the Seyfert 2 galaxies from the original {\it IRAS} 12$\mu$m survey \citep{12m}. This represents a complete sample of Sy2s down to a flux-density limit of 0.3 Jy at 12 $\mu$m, drawn from the {\it IRAS} Point Source Catalog (Version 2), with latitude $|b| > 25^o$ to avoid contamination from the Milky Way. We have dropped NGC 1097 from this original sample as it has since been classified as a Type 1 Seyfert (Storchi-Bergmann et al. 1993), leaving 31 mid-IR selected Sy2s (hereafter the ``12$\mu$m sample,'' listed in Table \ref{12m_sample}).

\subsection{Optical Data}
The optical data for the [OIII]-sample were drawn from SDSS DR4, whereas the optical data for the 12$\mu$m sample were collected from the literature or from SDSS Data Release 7 (DR7) where available. The reddening corrected [OIII] flux (F$_{[OIII],corr}$) was calculated using the observed H$\alpha$/H$\beta$ ratio and an intrinsic ratio of 3.1 with R=3.1 extinction curve for galactic dust \citep{oster}. Tables \ref{o3_opt} and \ref{12m_opt} list the optical emission line fluxes and ratios utilized for this study, as well as the relevant literature sources for the 12$\mu$m sample. The black hole masses (M$_{BH}$) were derived for the [OIII] sample by the SDSS velocity dispersion ($\sigma$) and the M-$\sigma$ relation (M$_{BH}$ = 10$^{8.13}$($\sigma$/200 km s$^{-1}$)$^{4.02}$ M$_{\sun}$, Tremaine et al. 2002). We used literature values for M$_{BH}$ for the 12$\mu$m sources, with most of the masses derived using the M-$\sigma$ relation cited above. For F04385-0828, F05189-2524 and TOLOLO 1238-364, the full width half max (FWHM) of the [OIII] line was used as a proxy for the velocity dispersion (Wang \& Ahang 2007, Greene \& Ho, 2005), and photometry of the host galaxy was used to estimate M$_{BH}$ for F08572+3915 (see Veilleux et al. 2009 for details).

\subsection{Infrared Data}
The infrared data presented here were obtained from the Infrared Spectrograph ({\it IRS}, Houck et al. 2004) on board the {\it Spitzer Space Telescope}. Low-resolution spectra were obtained using the Short-Low (SL, 3.6''$\times$57'' aperture size) and the Long-Low (LL, 10.5''$\times$168'' aperture size) modules and high-resolution spectra were provided by the Short-High (SH, 4.7''$\times$11.3'' aperture size) and the Long-High (LH, 11.1''$\times$22.3'') modules.

The Sy2s in the [OIII]-sample were observed in IRS staring mode in both high and low resolution under Program ID 30773. For the 12-$\mu$m sample, high resolution data existed for all 31 Sy2s but low resolution data were only available for 30 galaxies (IRAS 000198-7926 lacked low resolution data). The high resolution data were obtained in IRS staring mode for 30 of the Sy2s (NGC 5194 had only IRS spectral mapping mode high-resolution data in the archive). Several galaxies had multiple IRS observations: we analyzed these observations independently and compared our results between the two observations.  For the low-resolution data, IRS staring mode was used when available with the remainder observed in IRS spectral mapping mode. The Spectral Modeling Analysis and Reduction Tool (SMART, Higdon et al. 2004) was used to reduce the staring mode observations for the 12$\mu$m-sample to be consistent with previous IRS analysis of the 12 $\mu$m sample (e.g. Tommasin et al. 2008, Wu et al. 2009, Buchanan et al. 2006), Spitzer IRS Custom Extraction (SPICE) was used to analyze the staring mode observations for the [OIII]-sample\footnote{Though the IRS staring data for the 12$\mu$m sample was reduced using SMART and the [OIII]-sample with SPICE, the effect of the different reduction software is expected to be negligible on the derived parameters used in this analysis.}, and the Cube Builder for IRS Spectra Maps (CUBISM, Smith et al. 2007a) was utilized to analyze the spectral mapping observations. Table \ref{Spitzer_info} lists the Program ID(s) for each galaxy, the IRS mode used, and spectral extraction area for low-resolution spectral mapping mode data (discussed below).

\subsubsection{High-Resolution IRS Staring Spectra}
We used the basic calibrated data (BCD) pipeline products as the starting point for our analysis. Rogue pixels were removed using the IDL routine IRSCLEAN\_MASK and the rogue pixel mask matching the campaign number of the observation. Dedicated off-source background observations were taken for all sources in the [OIII]-sample and for most of the Sy2s in the 12$\mu$m sample. Multiple background observations, if present, were coadded within each nod and subsequently subtracted from the source image. The background-subtracted source images were then coadded between the two nods. The galaxies in the 12$\mu$m sample that had dedicated background observations and thus were background subtracted are marked with a ``b'' in Table \ref{Spitzer_info}. For sources in the 12$\mu$m sample without dedicated off-source observations, no background subtraction was performed.

Spectra were extracted from these combined observations, using the full aperture extraction mode. The edges of each order were then inspected, removing any data points that fell outside of the calibrated range for that order (IRS Data Handbook, Version 3.1, Table 5.1). The orders were then combined using a 2.5-$\sigma$ clipping mean, resulting in a final cleaned spectrum.

\subsubsection{Low-Resolution IRS Staring Spectra}
The low-resolution data were processed in a similar fashion as the high-resolution data, i.e. we started with the BCD products and removed the rogue pixels with IRSCLEAN\_MASK. However, for these observations a background data set was built for each nod and order by coadding the off-source order and nod position. The background-subtracted nods (following the same procedure as above) were combined for each order and the spectra then extracted using tapered column extraction. The orders were combined using a 2.5-$\sigma$ clipping average. This procedure was executed separately for the SL and LL module. Fourteen of our galaxies had low resolution IRS staring mode data; the rest were acquired in spectral mapping mode. We note that IRAS 00198-7926 did not have archival low resolution spectral data.

\subsubsection{Spectral Mapping Spectra}
The IRS spectral mapping observations were analyzed with CUBISM \citep{CUBISM}, which uses the BCD data to create 3-D spectral cubes (one spectral dimension and 2 spatial dimensions). For the low resolution data, background observations were built from the other order of the on-source module (e.g. SL 2 was used as the background for SL 1, etc.). After the rogue pixels were removed, using the default ``autogen bad pixels'' option in CUBISM, a spectral cube was built. Spectra were then extracted using matched apertures among the detectors and centered on the nucleus.  The aperture extraction size for these low-resolution spectral mapping observations are listed in Table \ref{Spitzer_info}. The low resolution spectral mapping data for NGC 1068 was saturated near the nucleus and consequently not included in this analysis.

For the IRS spectral mapping high resolution observation of NGC 5194, no background subtraction was performed. The spectrum was extracted over the full cube, corresponding to a size of 31.5''$\times$45'' in the LH module and 13.8''$\times$27.6'' in the SH module.

\subsection{Radio and Hard X-ray Data}
The radio and hard X-ray data were drawn from the literature; VLA radio data at 8.4 GHz were only available for the 12$\mu$m sample (Thean et al. 2000). In several cases, multiple radio components were analyzed; we included only the flux for the component that was nearest the published center of the galaxy. Twenty-six of the 31 12$\mu$m sources had radio data, with 3 additional sources having upper limits. The hard X-ray fluxes originated from the 22-month Swift-BAT Sky Survey \citep{Tueller} and from BeppoSax \citep{Dadina}. Only 11 out of the 31 12$\mu$m sources and one of the 20 [OIII] sources (IC 0486) have X-ray detections in the 14-195 keV range. We adopted an upper limit of 3.1$\times 10^{-11}$ erg cm$^{-2}$ s$^{-1}$, the flux limit of BAT, for the remainder of the sample when an upper limit was not quoted in either Tueller et al. (2009) or Dadina et al. (2007).

\section{Measurements}

\subsection{IR Emission Line Fluxes}
The high resolution spectra were utilized to measure the emission line fluxes: a Gaussian profile was fit to the emission line feature, with the local continuum, centered on the line's rest-frame wavelength, fit by a zero- or first-order polynomial. The errors were estimated by calculating the root-mean-square (RMS) around this local continuum and measuring the flux values with the continuum shifted by $\pm$ the RMS. In the cases where an emission line was not present, a 3-$\sigma$ upper limit was estimated from the RMS around the best-fit local continuum (where the RMS is assumed to be the 1-$\sigma$ error). In the cases with multiple observations per galaxy, we measured the emission line fluxes independently and averaged the resulting values; these flux measurements agreed within several percent between most of the individual observations, with at most a factor of $\sim$1.5 discrepancy, which was only present in one of the sources.\footnote{We note that NGC 1143/4 has two high resolution archival observations, one centered at RA=43.8004, Dec=-0.1839, and the other at RA=43.7985 and Dec=-0.1807, a distance of $\sim$13.4''. We present the line fluxes from the first region as this corresponds to NGC 1144 which is classified as a Sy2 in SIMBAD. The optical data are for NGC 1144 as well.} Tables \ref{oiii_ir} and \ref{12m_ir} list the emission line flux values for the [OIII] and 12$\mu$m samples, respectively. Comparing our line flux values with Tommasin et al. (2008, 2010), we find that our [OIV] flux values largely agree within a factor of 1.5 (with the exception of NGC 1667 and NGC 7582 where their values are a greater than a factor of 2 higher than ours). However, their [NeII] flux values are generally systematically higher by a factor of $\sim$2.5 - $\sim$4.5, though we do obtain consistent values for NGC 424, NGC 5135 and NGC 5506. Despite these differences in the measured [NeII] line strength, we obtain similar results to Tommasin et al (2010), namely that as the relative contribution of the AGN to the ionization field increases (parameterized by [OIV]/[NeII]), the starburst strength (parameterized by the PAH equivalent width) decreases.

\subsection{IR Continuum Flux and PAHs}
The MIR continuum flux values (F$_{MIR}$) and PAH equivalent widths (EWs) were measured using the low resolution spectra. For the galaxies that had multiple observations, we utilized the observations that had consistent flux values in the overlap region between the SL and LL modules: Program ID 30572 for For NGC 1386, NGC 4388, NGC 5506 and NGC 7130; Program ID 0086 for NGC 5135; and Program IDs 00086 and 30572 for NGC 5347 (for this source, the analysis was done separately for each observation and the results averaged together). The MIR continuum flux was measured at 13.5 $\mu$m (rest-frame), averaged over a 3$\mu$m window; these flux values are listed in Table \ref{oiii_ir} for the [OIII]-selected sample and Table \ref{12m_ir} for the 12$\mu$m sample. This window was chosen as it is free from strong emission line and PAH features.\footnote{Though this range does include the [NeII] 12.81$\mu$m line, in most cases this comprises less than 1\% of the MIR flux, with the exception of NGC 7582 where the [NeII] line is $\sim$1.5\% of the MIR flux.} In LaMassa et al. (2009), we included the flux centered at 30$\mu$m as part of the MIR flux diagnostic. However, emission at this higher wavelength can be strongly affected by star formation processes in the host galaxy (e.g. Deo et al. 2009, Baum et al. 2010), so here we use F$_{13.5\mu m}$ as F$_{MIR}$.

We used PAHFIT \citep{PAHFIT} to measure the PAH EWs, a program which uses a model consisting of several components: a starlight component represented by blackbody emission at T = 5000 K, a thermal dust continuum constrained to default temperature bins (35, 40, 50, 65, 135, 200 and 300 K), IR emission lines, PAH (dust) features and extinction (we used a foreground extinction screen).  As PAHFIT requires a single stitched spectrum, the SL spectrum was scaled to match the LL spectrum, with typical adjustments under 20\% (though several galaxies were adjusted by $\sim$40\% and NGC 7582 by greater than a factor of 6, indicating the presence of extended IR emission in this object). Here we utilize the EW of the PAH features at 11.3$\mu$m  and 17$\mu$m, which consist of the features within the wavelength range 11.2-11.4$\mu$m and 16.4-17.9$\mu$m, respectively (Tables \ref{oiii_ir} and \ref{12m_ir}). However, we note that the current version of PAHFIT has a bug which assumes the PAH EW feature to be Gaussian rather than Drude, which could underestimate the PAH EW by a factor of 1.4. We report the EWs as reported from PAHFIT, with the caveat that these may be lower limits.

We compared our results with the 11.2$\mu$m feature from Wu et al. (2009) and the 11.3$\mu$m and 17$\mu$m features measured by Gallimore et al. (2010), where in the latter, we added their published 11.2$\mu$m and 11.3$\mu$m EW values. Wu et al. employed a spline fit between 10.8$\mu$m and 11.8$\mu$m to measure the EW. With this method, the results are widely influenced by the choice of anchor points for fitting the pseudo-continuum and can result in an underestimate of the EW compared to a method that utilizes spectral decomposition, such as PAHFIT \citep{PAHFIT}. Of the 28 sources we have in common with Wu et al. (2009), 12 of them had consistent 11.3$\mu$m EW values (within a factor of 2), 6 had lower values than we obtained (which would be expected from the disagreements between the spline vs. decomposition methods mentioned above) and 10 had higher values, where for 6 of these, PAHFIT had obtained an EW value of zero, yet the spline method yielded a measurement. Comparing our results with Gallimore et al. (2010) gave better results, though a discrepancy did still exist: of the 23 sources in common, we obtained consistent EW values (within a factor of 2) for 11 sources at 11.3$\mu$m and for 12 sources at 17$\mu$m. Though Gallimore used PAHFIT to measure these features, they modified the code to include more fine-structure lines, fit silicate emission features, and use the cold dust model from Ossenkopf et al. (1992); they also generated their own software to build spectral data cubes whereas we employed CUBISM. Such differences could account for the inconsistencies in our PAH EW measurements. Though the derived EWs are different from those reported by Wu et al. and Gallimore et al. for at least half the sources we have in common, our main conclusions based on PAH EWs agree qualitatively with Wu et al. and Gallimore et al.: PAH features are associated with other star formation activity indicators (Gallimore et al. 2010, Wu et al. 2009) and the EWs are inversely correlated to the strength of the ionization field (Wu et al. 2009 where they use the \textit{IRAS} colors to parameterize AGN strength).

In the discussion that follows, we divide the 12$\mu$m-sample into two classes, those with weak PAH emission (``PAH-weak'' sources) and those with strong PAH emission (``PAH-strong'' sources, galaxies with EW $>$ 1 $\mu$m in either the 11.3 $\mu$m or 17$\mu$m band, with PAH EWs detected in both bands); the strong PAH emission is likely due to starburst activity in the host galaxy (see $\S$5.2).

\section{Diagnostics of Intrinsic AGN Luminosity}

Our goal is to evaluate the relative efficacy of the five different proxies for the AGN intrinsic luminosity under consideration in this paper. We expect that these different proxies will not agree perfectly, due to the different physical mechanisms that produce and affect the emission features as well as biases resulting from sample selection, starburst contamination, statistical errors and in some cases, uncertain application of extinction corrections.  To address this, we will undertake two kinds of comparions.

First, we will use our two Sy2 samples to inter-compare these proxies in a pair-wise fashion and measure the amount of scatter in the corresponding flux ratios. Which proxies agree best with one another? Second, we will compare these pairs of flux ratios to the corresponding values for unobscured Type 1 AGN to test which proxies are more ``isotropic,'' i.e. suffer the least AGN-viewing-angle dependence.

Figures \ref{hist_o4_o3} - \ref{xray_radio} show the histograms of a subset of ratios for the five proxies. In each plot, the solid black line represents both samples combined, the red dashed line and green dotted-dashed line delineate the 12$\mu$m sample (``PAH-weak'' and  ``PAH-strong'' sources respectively) and the cyan filled histogram reflects the [OIII]-sample. Adjacent to these histograms are the luminosity vs. luminosity plots, showing the correlation between these indicators: the cyan asterisks represent the [OIII] sample, the red diamonds (green triangles) depict the ``PAH-weak'' (``PAH-strong'') 12$\mu$m sources, and the dashed black line represents the best fit from multiple linear regression analysis (i.e. the REGRESS routine in IDL), where in the figure captions, $\rho$ is the linear regression coefficient and P$_{uncorr}$ is the probability that the two quantities are uncorrelated. Though the distance dependence in luminosity vs. luminosity plots enhances the correlation compared to flux vs. flux plots, we employed this method as the 12$\mu$m sample lies at a systematically lower redshift, and thus have higher flux values, than the [OIII] sample. One of our main goals is to examine the dispersion in the flux ratios, where this distance dependence cancels out. In $\S$4.3, we test if these ratios are affected by luminosity.

Where available, the values for Sy1s are included in these plots. The results are summarized in Tables \ref{diag_results1} and \ref{diag_results2} which lists the mean and sigma of each ratio for the combined sample and the sub-samples separately. In the histograms and luminosity plots, the upper limits are plotted but not included in the analysis of the mean and sigma (except for the ratios involving the hard X-ray flux). Since only 12 of the 51 AGN were detected in hard X-rays, we have employed survival analysis to quantify the correlations among the proxies and to calculate the mean of the ratios. This approach takes the upper-limits into account (ASURV Rev 1.2, Isobe and Feigelson 1990; LaValley, Isobe and Feigelson 1992; for univariate problems using the Kaplan-Meier estimator, Feigelson and Nelson 1985; for bivariate problems, Isobe et al. 1986). 

\subsection{Inter-Comparison of Proxies}

The isotropic luminosity diagnostics that agree best, and therefore may be least subject to the uncertainties and errors discussed above, are F$_{[OIII],obs}$, F$_{[OIV]}$ and F$_{MIR}$. A wider spread is present between the radio and hard X-ray fluxes compared with the optical and MIR values.

In all cases, a wider dispersion is present between all the flux ratios in the 12$\mu$m sample as compared to the [OIII] sample. Below, we examine whether such a scatter could be due to aperture effects, extinction corrections applied to the [OIII] flux, starburst contamination to the the MIR flux or if it represents a real difference between AGN selected on the basis of [OIII] flux versus MIR flux.

Since the 12$\mu$m Sy2s are typically more nearby than the [OIII]-selected galaxies, aperture effects can potentially play a significant role when comparing flux values by either missing NLR flux or obtaining too much host galaxy contamination. However, we find no evidence in our data for such an effect (see Appendix). Another possible explanation for this wider dispersion is that the optical data for the 12$\mu$m sample are drawn from the literature, which can introduce scatter into the optical diagnostics as such data are not taken and reduced in a uniform matter. The most striking example of this is the comparison of the [OIV] flux with the observed and extinction corrected [OIII]-flux: de-reddening the [OIII]-flux widens the dispersion in the 12$\mu$m-sample (see Figure \ref{hist_o4_o3}). This can be an artifact of using literature H$\alpha$ and H$\beta$ values and to a lesser extent can be due to large amounts of dust in the host galaxies of the 12$\mu$m sample as evidenced by the wide range of Balmer decrements. Goulding and Alexander (2009) note that galaxies that would not be identified optically as AGN (i.e. have a low ``D'' value, see $\S$5.1) tend to have similar Balmer decrements  yet higher F$_{[OIV]}$/F$_{[OIII],obs}$ ratios than optically identified AGN. This result suggests that applying a reddening correction using the Balmer decrement still under-represents the intrinsic [OIII] flux. However, our 12$\mu$m sources with lower ``D'' values ($<$1.2) do not show systematically higher F$_{[OIV]}$/F$_{[OIII],obs}$ ratios (see Figure \ref{balmer}), indicating that ``extra'' extinction that Goulding and Alexander observe in their sources is not present in ours. As expected, the ratio of F$_{[OIV]}$/F$_{[OIII],obs}$ increases with H$\alpha$/H$\beta$, denoting that both quantities trace host galaxy extinction, though with wide scatter. Comparison with the locus of points for the [OIII] sample shows that the Balmer decrement is systematically higher for similar 12$\mu$m F$_{[OIV]}$/F$_{[OIII],obs}$ values, suggesting that the the 12$\mu$m Balmer decrements are over-estimating the amount of dust present rather than under-estimating. This result indicates that the literature Balmer decrements, which are not analyzed in a systematic and homogenous way, are introducing uncertainties that bias the results and do not better recover the truly intrinsic AGN luminosity. 

However, this can not be the only cause of the greater scatter in the 12$\mu$m sample, since the ratio of MIR/[OIV] fluxes shows more scatter in the 12 $\mu$m sample (Figure \ref{mir2_oiv}), though these data were analyzed homogeneously. Could the presence of Sy2s in the 12$\mu$m sample that have significant contributions from starburst activity create the wider dispersion in these diagnostics? To address this issue, we isolated the ``PAH-strong'' sources, which have greater amounts of star formation activity (discussed in detail below). The distributions between the ``PAH-strong'' and ``PAH-weak'' sub-samples are similar, suggesting that starburst processes are not responsible for the wider dispersion. We also focused on sources with a limited ``D'' value, which as noted above indicates the relative contribution of AGN to starburst activity. Repeating the calculation of mean and standard deviations on the flux ratios for the 12$\mu$m sources with 1.2$\leq$ D $\leq$ 1.7 did not result in a significant decrease (a factor of 2 or more) in the dispersion with the exception of log (F$_{MIR}$/F$_{[OIII],obs}$) ($\sigma$=0.42 dex), log (F$_{[OIII],obs}$/F$_{8.4GHz}$) ($\sigma$=0.60 dex) and log (F$_{[OIII],corr}$/F$_{8.4GHz}$) ($\sigma$=0.51 dex). For these first two ratios, this is largely due to the removal of the 3 outliers (F04385-0828, F08572+3914 and Arp 220) with systematically higher (lower) F$_{MIR}$/F$_{[OIII],obs}$ (F$_{[OIII],obs}$/F$_{8.4GHz}$) values from the full sample. The dispersions for the other ratios were still systematically higher than the [OIII]-sample.  We conclude that there is a real difference between the AGN selected on the basis of [OIII] emission-lines and MIR continuum.

We also compared our results with two other samples of Seyfert 2 galaxies: one a complete sample down to a flux limit of (1-3) $\times 10^{-11}$ erg cm$^{-2}$ s$^{-1}$ at 14 - 195 keV drawn from the 3 and 9 month Swift-BAT survey (Mel\'endez et al. 2008 and references therein) and the other drawn from the revised Shapley-Ames catalog (Shapley \& Ames 1932; Sandage \& Tammann 1987), consisting of galaxies with $B_T \leq$ 13 \citep{DS09}. Here we include those radio quiet Seyfert types 1.8 - 2 that have measured [OIII] and [OIV] fluxes, giving 12 and 56 Sy2s, respectively. The log of the ratios of [OIV] to [OIII],obs for both samples are  higher than our combined sample, 0.60 $\pm$ 0.74 dex (Mel\'{e}ndez et al. 2008) and 0.57 $\pm$ 0.67 dex (Diamond-Stanic et al. 2009) vs. 0.08 $\pm$ 0.41 dex, but the differences are not statistically significant. A wider dispersion is present in these comparison samples than the [OIII]-selected sample (as was the case for the 12$\mu$m sample). This may indicate that selection based on [OIII] leads to better agreement between between the [OIII] and [OIV] flux rather than selection based on other methods. This effect could be due to the [OIII]-bright sources having less extinction in the NLR than Sys selected in other ways.

As the samples in Weaver et al. (2010) and Winter et al. (2010) samples were selected based on their hard (14-195 keV) X-ray flux from the Swift-BAT 9 month (Winter et al. (2010) and Weaver et al.(2010)) and 22 month (Weaver et al. (2010)) catalog, we can compare our Sy2 X-ray flux ratios. Using the values in Winter et al. (2010), we find log (F$_{14-195 keV}$/F$_{[OIII],obs}$)  = 2.76$\pm$0.59 dex and log (F$_{14-195 keV}$/F$_{[OIII],corr}$)\footnote{We note that for the cases where the Winter et al. 2010 Balmer decrement was less than the assumed intrinsic value (3.1), we did not apply a redenning correction, but rather used the observed value, both here and in $\S$4.2.} = 2.34$\pm$ 0.69 dex for Sy2 galaxies. The log (F$_{14-195 keV}$/F$_{[OIV]}$) ratio from Weaver et al. (2010) for Sy2s is 2.38$\pm$0.45 dex. All three values are systematically higher than what we obtain for our samples of Sy2 galaxies by roughly an order of magnitude (see Table \ref{diag_results2}). This is depicted graphically in Figure \ref{bat_hist}. Employing survival analysis, we compared these ratios between the BAT-selected Sy2s and the [OIII] and 12$\mu$m samples separately and find that they differ significantly (i.e. p $<$0.05, corresponding to the 2$\sigma$ level, that they are drawn from the same parent population), with the caveat that with only one [OIII] Sy2 detected by BAT, the comparison  between BAT and [OIII] selected Sy2s may be less robust. These differences suggest that the samples selected in hard X-rays do not fairly sample the population of Type 2 AGN selected in the MIR and possibly the optical, however comparisons with BAT-selected Sy1s reveal mixed results (see $\S$4.2).

\subsection{Comparison with Sy1s}

In order to determine if the proxies we are considering are affected by the orientation of the AGN, and evaluate the extent to which they may be considered truly ``isotropic,'' we compared our results for Sy2 with the corresponding values for Sy1, using data taken from the literature. The Sy1 MIR fluxes were calculated from the 14.7$\mu$m flux densitites reported in Deo et al. (2009), where they analyzed a heterogeneous sample of Sy1 and Sy2 galaxies available in the \textit{Spitzer} archive, ranging in redshift 0.002 $< z <$ 0.82. The radio flux values were derived from the high-resolution 8.4-GHz flux density values from Thean et al.(2000), which presented analysis of the extended 12$\mu$m sample.\footnote{The extended 12$\mu$m sample probes to a lower flux density limit than the original 12$\mu$m sample: 0.22 Jy vs. 0.30 Jy, giving a total of 118 detected Sys over the 59 detected in the original sample \citep{Rush}.} The hard X-ray data (14-195 keV) are drawn from Mel\'endez et al. (2008, sample selection described above), the 22-month Swift-BAT Catalog \citep{Tueller} and Rigby et al. (2009, same parent sample as Diamond-Stanic et al. 2009, with X-ray data derived from the 22-month Swift-BAT Catlog, BeppoSAX (Dadina 2007) and Integral (Krivonos et al. 2007)). The comparison Sy1 [OIII] and [OIV] flux values are derived from Mel\'endez et al. (2008) and Tommasin et al. (2008, 2010), which presents high resolution \textit{Spitzer} spectroscopy of the extended 12$\mu$m sample.  As only Winter et al. (2010) quote Balmer decrements, we only have comparison Sy1 F$_{[OIII],corr}$ values for the samples selected from the BAT catalog (i.e. F$_{14-195 keV}$ and F$_{[OIV]}$ from Weaver et al. 2010). We utilize both the Kolmogorov-Smirnov test (``K-S'' test) and Kuiper's test on the detected data points (excluding the three [OIV] and three radio upper limits in the 12$\mu$m data) to determine to which extent the flux ratios are significantly inconsistent between the Sy1 and Sy2 galaxies: a lower Kuiper ``D'' value indicates that these two populations are drawn from the same parent population, suggesting that such fluxes are independent of viewing angle. The Kuiper test is similar to the more often used K-S test but with the following modification: the ``D'' statistic of the K-S test represents the maximum deviation between the cumulative distribution functions (CDFs) of the two samples, whereas the ``D'' statistic in Kuiper's test is the sum of the maximum and minimum deviations between the CDFs of the two samples, so that this statistic is as sensitive to the tails as to the median of the distribution. The results of the K-S test and Kuiper test agree in that they do not lead us to reject the null hypothesis that the two samples are drawn from the same parent population, with the exception of the F$_{[OIV]}$/F$_{[OIII],obs}$ ratio, where the tests lead to conflicting results. We note that two-sample tests work better for larger data sets, so the probabilities quoted in Table \ref{Kuiper} should be interpreted as approximate. 

The comparisons of the Sy2 and Sy1 samples are shown in Figures \ref{hist_o4_o3} through \ref{xray_radio}. In each, the dotted-dashed and (in the cases of more than one comparison sample) dashed line(s) on these plots indicate the mean values for the Sy1 diagnostic ratios and the correlations from linear regression we calculated from the literature values.

A mild disagreement between the average Sy1 and Sy2 F$_{[OIV]}$/F$_{[OIII],obs}$ ratio (up to a factor of $\sim$2) is evident. Here the comparison Sy1 data come from the Diamond-Stanic et al. (2009) and the Mel\'endez et al. (2008) samples.  We obtain mixed results as to the significance of this difference based on the statistical test used: according to the K-S test, the F$_{[OIV]}$/F$_{[OIII],obs}$ ratio for the Sy1 and Sy2 populations are statistically significantly different (D=0.301, p=0.042), but not according to the Kuiper test (D=0.310, p=0.223). We find similar disparate results when we run these tests on the F$_{[OIV]}$/F$_{[OIII],obs}$ ratio for the detected points between the Sy1s and Sy2s in the Mel\'endez et al. (2008) and Diamond-Stanic et al (2009) samples, namely that the K-S test implies different parent populations (p=0.005 and p=0.004, respectively) but not the Kuiper test (p=0.097 and p=0.126, respectively). Mel\'endez et al. (2008) and Diamond-Stanic et al. (2009) (as well has Haas et al. 2005, who compared seven quasars with seven Fanaroff-Riley II (FRII) radio galaxies) have reported significant differences between the observed [OIII] and [OIV] flux between Sy1s (quasars for Haas et al. 2005) and Sy2s (FRIIs for Haas et al. 2005), with the type 2 sources having higher F$_{[OIV]}$/F$_{[OIII],obs}$ ratios. These authors have attributed the diminumtion of [OIII] in type 2 AGN to extinction in the NLR. Baum et al. (2010) suggests that such [OIII] obscuration results from the AGN torus: using the 12 $\mu$m sample, they find a correlation between the F$_{[OIV]}$/ F$_{[OIII],obs}$ ratio and the Sil 10$\mu$m feature, which probes torus obscuration.\footnote{They define the Sil strength by the natural logarithm of the observed flux of the feature divided by the interopolated continuum flux at 10$\mu$m.} In type 1 Sy1s, this silicate feature is in emission, whereas Sy2s exhibit Sil absorption, making the Sil strength a probe of system orientation. The ratio of F$_{[OIV]}$ to F$_{[OIII],obs}$ increases with Sil absorption (parameterized by negative values of the Sil strength) which could suggest that the torus is extincting part of the [OIII] emission. Our results may confirm these previous studies as we find that Sy2s tend to have lower observed [OIII] emission as compared to Sy1s and this may be due to NLR extinction. We note, however, that such extinction affects the [OIII] line only up to a factor of 2 on average between our Sy2 and comparison Sy1 samples, albeit with a wide dispersion, and this difference between the two populations may not be significant.

The average log (F$_{MIR}$/F$_{[OIII],obs}$) ratio is consistent between Sy1s (2.56 $\pm$ 0.50 dex) and Sy2s (2.62 $\pm$ 0.62 dex), which could seemingly contradict the results cited above where NLR extinction causes attenuation of the [OIII] flux in Sy2s but not in Sy1s. The clumpy torus model of Nenkova et al. (2008) and smooth torus model of Pier \& Krolik (1992) predicts a slight anistropy in emission at 12$\mu$m depending on viewing angle: as the viewing angle increases from 0$^{\circ}$ (Sy1) to 90$^{\circ}$ (Sy2), the torus flux decreases by a factor of $\sim$2.  The effects of depressed MIR emission in Sy2s and enhanced MIR emission in Sy1s, assuming [OIII] is more extincted in the former than the latter, would therefore result in F$_{MIR}$/F$_{[OIII],obs}$ ratios that are more consistent than F$_{[OIV]}$/F$_{[OIII],obs}$, which is indeed what we observe. However, the average differences between F$_{MIR}$/F$_{[OIII],obs}$ and F$_{[OIV]}$/F$_{[OIII],obs}$ are within the scatter of these ratio values, and we are unable to rule this out as the main driver for the disagreement, rather than invoking anisotropies in torus emission. 

Interestingly, the relationship between L$_{MIR}$ and L$_{[OIV]}$ is nearly identical for Sy1 and Sy2 galaxies (Figure \ref{mir2_oiv}). Though the MIR flux is not corrected for starburst contamination (see Appendix), and the Sy2s in the 12$\mu$m sample are thought to harbor more star formation activity than Sy1s (e.g. Buchanan et al. 2006), we see no evidence that star formation activity is contributing significantly to the MIR emission. As the F$_{MIR}$/F$_{[OIV]}$ diagnostic ratio shows the smallest dispersion for the combined sample,  a similar relationship to Sy1 galaxies and no evidence for luminosity bias (see Sectin $\S$4.3), the MIR and [OIV] flux may be the most robust proxies for the intrinsic AGN luminosity in Type 2 AGN. However, the KS test and Kuiper's test indicates a lower probability that the Sy1 and Sy2 samples agree than the F$_{MIR}$/F$_{[OIII],obs}$ ratio, though this could be driven by the lower N$_{eff}$ value for F$_{MIR}$/F$_{[OIII],obs}$ rather than a robust statistical agreement. 

The different slopes between Sy1s and Sy2s in the luminosity plots of the radio data against other intrinsic AGN flux proxies (Figures \ref{o3_radio}, \ref{oiv_radio} and \ref{mir2_radio}) suggest disagreements between these samples. However, the F$_{[OIV]}$/F$_{8.4GHz}$ and F$_{MIR}$/F$_{8.4GHz}$ flux ratios are consistent between Sy1 and Sy2 galaxies, indicating that the disparate slopes are perhaps influenced by scatter due to the wide range of radio loudness in AGN. Results of the KS test and Kuiper's test (Table \ref{Kuiper}) also indicate that the differences in the radio flux ratios between Sy1s and Sy2s are not statistically significant. Diamond-Stanic et al. (2009) compared the 6 cm radio data between Sy1s and Sy2s and found that for the Sy2s with a measured X-ray column density, these two samples show no statistically significantly differences, though they find a higher probability that they are drawn from the same distribution ($\sim$68 - 78\%) than we do ($\sim$55\%). Mel\'endez et al. (2010) also found that the  8.4 GHz and [OIV] fluxes between Sy1 and Sy2 galaxies are not significantly different, though sources dominated by star formation (i.e. less than 50\% of the [NeII] line attributable to AGN ionization) had statistically different F$_{[OIV]}$/F$_{8.4GHz}$ values than AGN dominated sources, indicating that radio emission may not accurately trace intrinsic AGN power. This latter result may agree qualitatively with our Figure \ref{oiv_radio}, where the ``PAH-strong'' sources lie at or below the best-fit line between L$_{[OIV]}$ and L$_{8.4GHz}$.

The hard X-ray proxy performs much more poorly (Figures 7 - \ref{xray_radio}), based on both the wider dispersion in the diagnostic flux ratios and the larger disagreement between the Sy1 and Sy2 flux ratios. The mean hard X-ray emission (normalized by other isotropic indicators) in Sy2s tends to be about an order of magnitude weaker than in Sy1s, though this is driven largely by the 12$\mu$m sample as only one source was detected by BAT in the [OIII] sample. This disagreement agrees with the results of Rigby et al. (2008) and Weaver et al. (2010), where the X-ray flux was normalized by the [OIV] emission. Indeed, using survival analysis, we find that F$_{14-195 keV}$/F$_{[OIV]}$ disagrees significantly between BAT-selected Sy1s and both the [OIII] and 12$\mu$m sub-samples. Such a large disagreement is not found between the Sy1s and Sy2s in the Winter et al. (2010) sample (see Table \ref{diag_results2}), which  is driven by the lower [OIII] flux observed in their Sy2s as compared to their Sy1s. The hard X-ray to [OIII] flux ratios, both observed and reddening corrected, do not differ significantly between the BAT-selected Sy1s and the [OIII]-selected Sy2s, but do for the 12$\mu$m sample.\footnote{F$_{14-195 keV}$/F$_{[OIII],obs}$ has mixed results for the 12$\mu$m sample: according to two of the tests (Logrank and Peto \& Prentice Generalized Wilcoxon Test), the difference is significant though with other tests, the null hypothesis that the two samples are drawn from the same parent distribution can only be discarded at the $\sim$12\% confidence level.} According to the Logrank and Peto \& Prentice Generalized Wilcoxon tests, the hard X-ray flux normalized by the MIR flux differs signficantly for both Sy2 subsamples and the BAT-selected Sy1s. Consistent with the results from $\S$4.1, hard X-ray selected AGN do not represent the population of those selected in the MIR, and there may be some evidence that they do not fully sample the optically selected sources. Compton scattering may be responsible for weakening the observed hard X-ray emission in Sy2s, as suggested by Weaver et al. (2010), which indicates that the 14-195 keV emission is not truly isotropic.

\subsection{Luminosity Dependence}

As we have seen above, there is significant scatter in the flux ratios of the different proxies for AGN intrinsic luminosity. Here we examine the possibility that some of this scatter is caused by systematic differences that correlate with the accretion rate of the black hole (in units of the Eddington limit). 

To make this test, for any pair of luminosity proxies we parameterized $L_{AGN}/L_{Edd}$ by the square root of the product of the luminosities of the two proxies divided by the mass of the central black hole (M$_{BH}$, listed in Tables \ref{o3_sample} and \ref{12m_sample}). Linear regression fits were performed, with the correlation coefficients and probability of uncorrelation listed in Table \ref{lum_dep}. 

We find three statistically significant anti-correlations (Figure \ref{ledd}):  F$_{[OIV]}$/F$_{[OIII],obs}$,  F$_{MIR}$/F$_{[OIII],obs}$ and F$_{MIR}$/F$_{8.4GHz}$. The anti-correlations for the ratios involving F$_{[OIII],obs}$ are largely driven by those galaxies with a high Balmer decrement. When we exclude the 6 sources with H$\alpha$/H$\beta \geq$ 9, which may be those systems with the most NLR etinction, these anti-correlations are no longer statistically significant. This may indicate that the bolometric correction to the observed [OIII] luminosity might have a weak dependence on the Eddington ratio. However, the observed [OIII] luminosity, which partly parameterizes the Eddington ratio, does not as accurately trace intrinsic AGN flux for these more dust-obscured sources. If the Eddington ratio is defined as just $L_{OIV}/M_{BH}$ and L$_{MIR}/M_{BH}$ in these relationships, the anti-correlations are no longer significant. Hence, the weak trends in Figure \ref{ledd} a) and b) is likely driven more by NLR extinction bias on the [OIII] flux rather than the accretion rate of the black hole. This latter mechanisms could be affecting the F$_{MIR}$/F$_{8.4GH}$ ratio, albeit with wide scatter.

\section{Starburst Activity in the Host Galaxy}

\subsection{Comparison of Different Diagnostics}

Given the strong connection between Type 2 AGN and star-formation (e.g. Kauffmann et al. 2003) we expect that the signature of both processes will be present in optical and MIR spectra of AGN. By analogy to the previous section (where we compared different proxies for the intrinsic AGN luminosity) we now undertake a comparison of different diagnostics of the relative energetic significance of a starburst vs. the AGN.
 
One such diagnostic involves the use of the MIR polycyclic aromatic hydrocarbon (PAH) features. These have been shown to be correlated with star formation activity (e.g. Smith et al. 2007) and possibly anti-correlated with the presence of an AGN (O'Dowd et al. 2009; Voit 1992). More specifically, we used the equivalent width (EW) of the PAH features to assess the relative amount of starburst activity in the host galaxy (e.g. Genzel et al. 1998). Another empirical diagnostic of the relative contribution of the starburst in the MIR can be parametrized by the MIR spectral index: $\alpha_{20-30\mu m}$\footnote{$\alpha_{\lambda_1 - \lambda_2}$ = log($f_{\lambda_1}$/$f_{\lambda_2}$)/log($\lambda_1$/$\lambda_2$)}. Larger values of $\alpha_{20-30\mu m}$ indicate the presence of cold dust from starburst activity (Deo et al. 2009 and references therein). The ratio of the [OIV] to [NeII] 12.81$\mu$m MIR emission-lines probes the hardness of ionizing spectrum and hence the relative importance of the AGN and starburst. A larger ratio ($\sim$1) implies the dominance of AGN activity whereas a lower ratio ($\sim$0.02) is pure starburst activity \citep{Genzel}. The analogous diagnostic from optical spectra is the distance a galaxy spectrum lies from the locus of star forming galaxies in the Baldwin, Phillips \& Televich BPT (1981, BPT) diagram (D = $\sqrt{([NII]/H\alpha + 0.45)^2 + ([OIII]/H\beta + 0.5)^2}$, Kauffmann et al. 2003). A larger ``D'' parameter indicates pure AGN activity while a smaller value implies a mixture of starburst and AGN processes in the host galaxy.

\begin{figure}
\epsscale{1.0}
\plotone{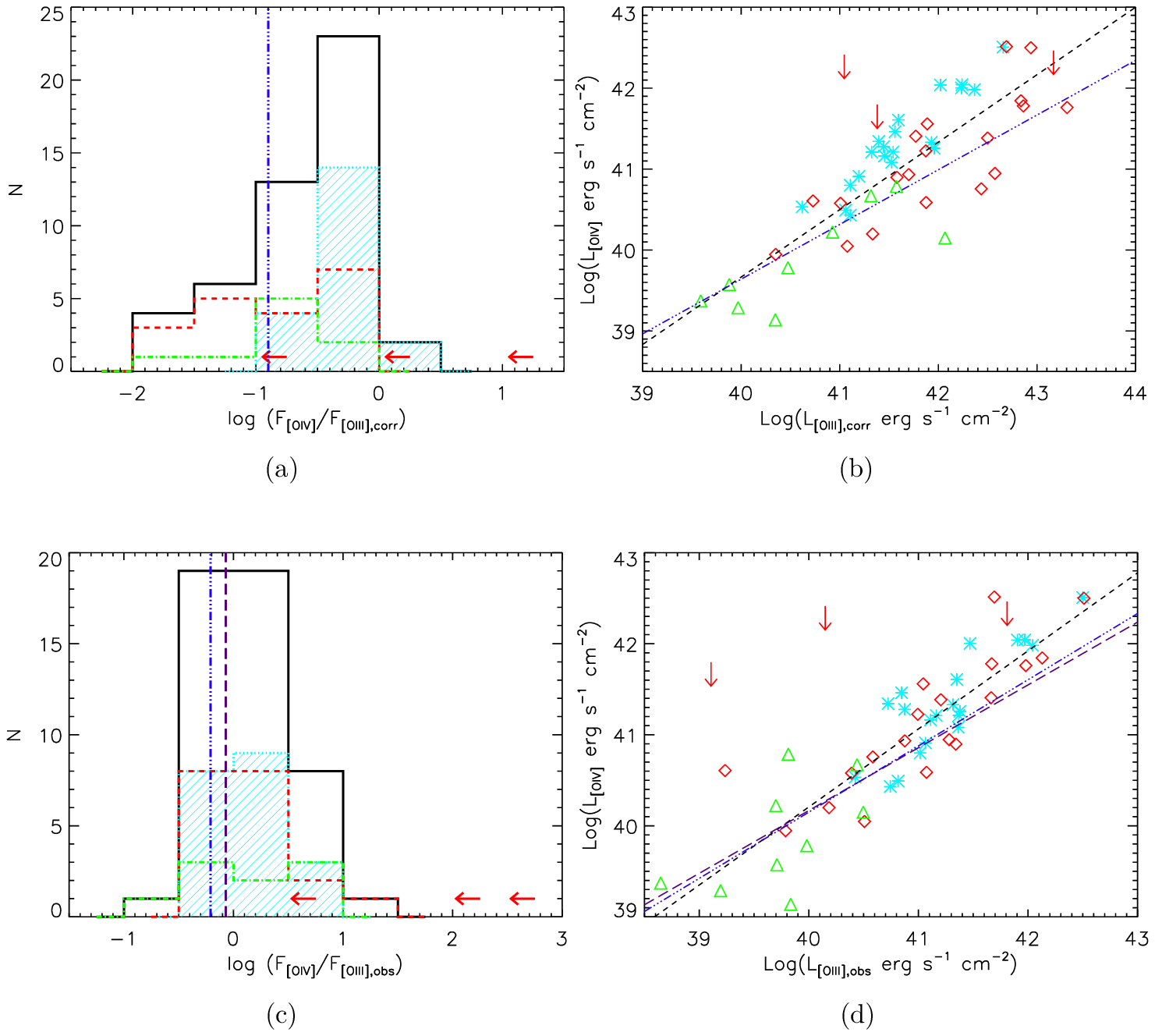}
\caption{\label{hist_o4_o3} a) Distribution of log (F$_{[OIV]}$/F$_{[OIII],corr}$). Black solid line shows the combined sample, filled cyan histogram depicts the [OIII]-sample, and red (green) dashed histogram are the weak PAH (strong PAH) 12$\mu$m sources.  b) Log (L$_{[OIV]}$) vs log (L$_{[OIII],corr}$); black dashed line represents best-fit from linear regression with $\rho$=0.837 giving P$_{uncorr}$=1.28$\times 10^{-13}$ (slope = 0.83 $\pm$ 0.08 dex with intercept at 6.32 dex). Cyan asterisks represent the [OIII] sample and red diamonds (green triangles) depict the weak PAH (strong PAH) 12$\mu$m Sy2s. The blue dotted-dashed line in a) and b) reflect the values for Sy1s, with F$_{[OIII],corr}$ from Winter et al. (2010) and F$_{[OIV]}$ from Weaver et al. (2010). c) Distribution of log (F$_{[OIV]}$/F$_{[OIII],obs}$). d) Log(L$_{[OIV]}$) vs log(L$_{[OIII],obs}$); black dashed line represents best-fit from linear regression with $\rho$=0.884 giving P$_{uncorr}$=8.52$\times 10^{-17}$ (slope = 0.86 $\pm$ 0.07 dex with intercept at 5.92 dex). The purple dashed line and blue dot-dashed line represent the Sy1 values from Diamond-Stanic et al. (2009) and Mel\'endez et al. (2008) samples, respectively.}
\end{figure}

\begin{figure}
\epsscale{1.0}
\plotone{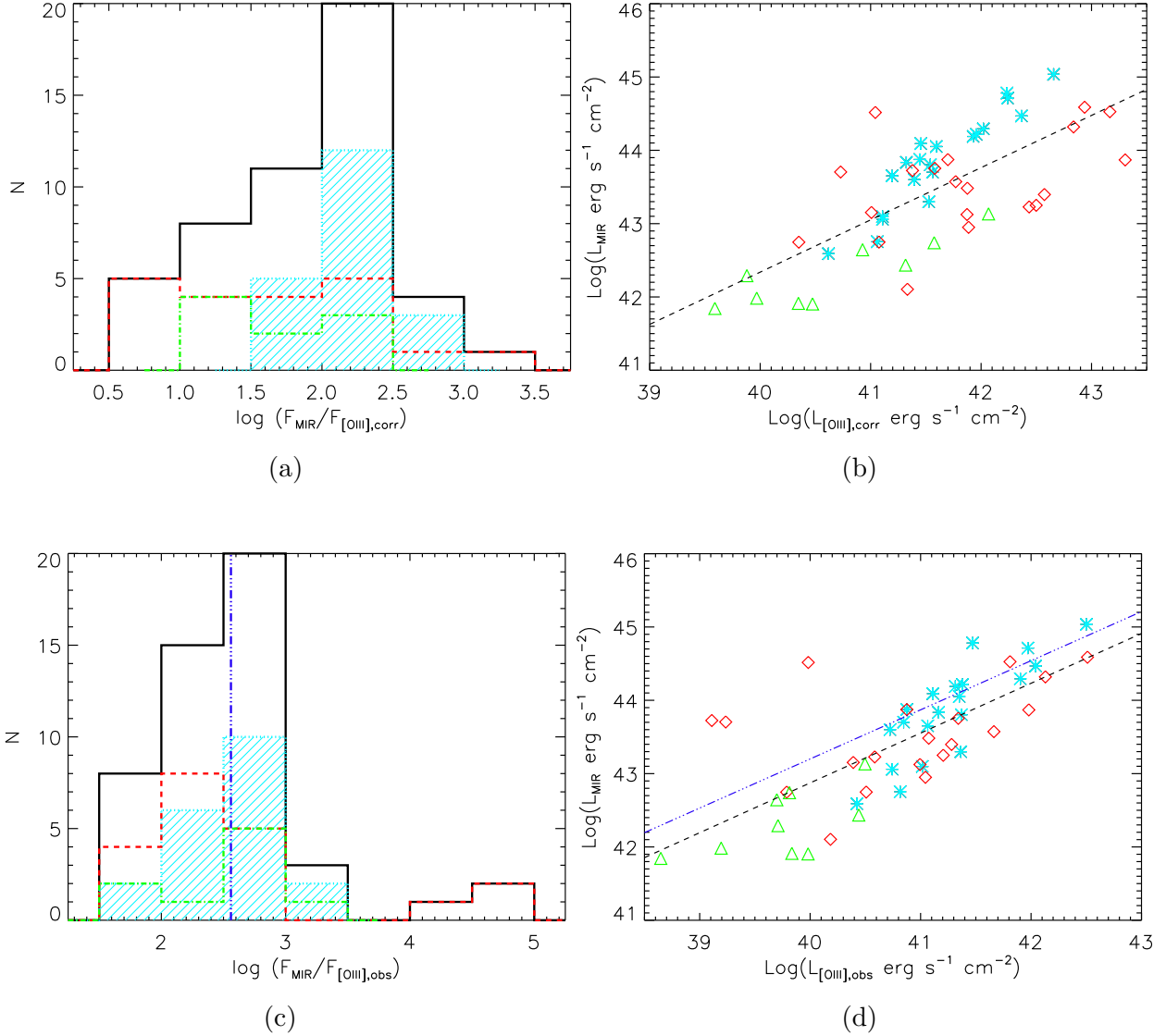}
\caption{\label{mir_o3} a) Distribution of log (F$_{MIR}$/F$_{[OIII],corr}$). b)  Log (L$_{MIR}$) vs log (L$_{[OIII],corr}$); $\rho$=0.711 giving P$_{uncorr}$=1.04$\times 10^{-8}$ (slope = 0.71 $\pm$ 0.10 dex with intercept at 13.8 dex). c) Distribution of log (F$_{MIR}$/F$_{[OIII],obs}$). d) Log (L$_{MIR}$) vs log (L$_{[OIII],obs}$); $\rho$=0.731 giving P$_{uncorr}$=2.46$\times 10^{-9}$ (slope = 0.68 $\pm$ 0.09 dex with intercept at 15.7 dex).  The blue dashed-dotted line represents the Sy1 values, with F$_{MIR}$ from Deo et al. (2009) and F$_{[OIII],obs}$  from Mel\'endez et al. (2008). Color and linestyle coding same as Figure \ref{hist_o4_o3}.}
\end{figure}

Figures \ref{alpha_ew} and \ref{oiv_neii_ew} illustrate the relationship between these AGN and star formation activity indicators for the Sy2s in our combined sample. The color coding is the same as in Figures \ref{hist_o4_o3} - \ref{xray_radio}. According to linear regression analysis, $\alpha_{20-30\mu m}$ and the PAH EWs are correlated and [OIV]/[NeII] and the PAH EWs are anti-correlated at greater than the 3.5$\sigma$ level: PAH EW 11.3 $\mu$m vs. $\alpha_{20-30\mu m}$ has $\rho$=0.609,  P$_{uncorr}$=1.47$\times 10^{-5}$; PAH EW 17 $\mu$m vs. $\alpha_{20-30\mu m}$ has $\rho$=0.600, P$_{uncorr}$=5.26$\times 10^{-6}$; PAH EW 11.3$mu$m vs. [OIV]/[NeII] has $\rho$=-0.677, P$_{uncorr}$=2.26$\times$10$^{-6}$;  PAH EW 17$\mu$m vs. [OIV]/[NeII] has $\rho$=-0.515, P$_{uncorr}$=4.89$\times10^{-4}$. We also note that the Sy2s with strong PAH emission mostly lie at systematically higher PAH EW values than the relation found between $\alpha_{20-30\mu m}$ and the PAH EWs. The majority of the Sy2s have D values between $\sim$1.2-2.0, with five of the strong PAH sources at systematically lower D values, $\sim$0.5-1.1 (see Figure \ref{d_figs}\footnote{The two ``PAH-weak'' sources with low D values are Arp 220 and F08572+3915, which have [OIV] upper limits and no measureable PAH EW at 11.3$\mu$m.}). The Sy2s with lower D values have higher PAH EW values and lower F$_{[OIV]}$/F$_{[NeII]}$ ratios, though they exhibit a weaker trend with the IR spectral index.

These results indicate that the various MIR and optical indicators of starburst activity agree both qualitatively and quantitatively. 

\subsection{The Spatial Scale of the MIR Emission and the Role of the Host Galaxy}

The results above refer to measurements of the central region enclosed by the IRS aperture. However, the 12$\mu$m sample was drawn from the \textit{IRAS} Point Source Catalog, where the aperture size (0.75 $\times$ 4.5' at 12$\mu$m) is much larger and will encompass contributions from the host galaxy. To quantify the extendedness of the MIR emission in the 12$\mu$m sample, we calculate the ratio of the \textit{IRAS} flux (at 12$\mu$m, Spinoglio \& Malkan 1989) to the \textit{IRS} flux. A ratio of $\sim$1 indicates the MIR emission is dominated by the galactic center, whereas higher ratios imply a greater amount of extended emission. In Figure \ref{iras}, we plot this ratio against the PAH EW at 11.3 $\mu$m and F$_{[OIV]}$/F$_{[NeII]}$. As expected, the relative amount of extended MIR emission decreases as the relative energetic importance of the AGN increases. 

Comparison of the ratio by which the SL module was rescaled to match the LL module (see $\S$ 4.2) reveals the presence of extended emission on smaller spatial scales. Inspection of this extendedness factor does not show any significant differences between the ``PAH-weak'' and ``PAH-strong'' sources (with the exception of NGC 7582).

\subsection{Are [OIII] and [OIV] Biased by Star Formation?}

In this section, we investigate whether the relative fraction of the AGN bolometric luminosity that emerges in [OIII] and [OIV] line emission is preferentially higher in galaxies with more star formation. This might be expected if the gas clouds in the NLR that are photoionized by the AGN and produce these lines are directly related to the gas clouds responsible for star-formation. If true, this would imply that AGN selected using [OIII] or [OIV] would be biased towards galaxies with higher star formation rates.

To test this, we have plotted the ratio of both F$_{[OIII],obs}$ and F$_{[OIV]}$ to F$_{MIR}$ versus the star formation indicators analyzed above (PAH EWs, IR spectral index and the ``D'' parameter). We find no strong trends between the star formation indicators and [OIV] and [OIII] emission, as illustrated in Figures \ref{sb_bias} and \ref{sb_bias2}. 

We conclude that there is no convincing evidence that host galaxies with a large star formation rate have preferentially higher relative luminosities of [OIII] and [OIV] at the luminosities represented in this sample, where the bolometric luminosity (L$_{bol}$) ranges from L$_{bol}$ $\approx$ 10$^{9}$ - 8 $\times$ 10$^{11}$ L$_{sun}$, which is $\sim$3$\times 10^{-5}$ to 0.5 of the Eddington luminosity (L$_{edd}$)\footnote{We estimated L$_{bol}$ by assuming the observed mid-infrared flux constitutes 20\% of the bolometric flux \citep{12m}.}.  Thus, these proxies of intrinsic AGN power are not biased by star formation activity at these Eddington ratios.

\section{Conclusions}
We have taken an empirical approach in analyzing the agreement among the various indicators of isotropic AGN luminosity for two complete and homogeneously selected samples of Sy2s, one selected based on observed [OIII] flux and the other on MIR flux.  The diagnostic ratios with the smallest spread are likely those where such biases from sample selection, starburst contamination, statistical errors, and scatter due to the various physical mechanisms that produce these emission features, are minimized. Such indicators, as well as those that agree the most with Sy1 relations, may be the most robust tracers of AGN activity. Our results on these indicators are summarized below.

- \textit{Sample Selection} The optically selected sample, picked on the basis of high [OIII] flux, shows tighter correlations among these diagnostics than the MIR selected sample. We investigated whether the inclusion of active star forming galaxies in the 12$\mu$m sample introduced the spread in these ratios by dividing the sample into galaxies that have large amounts of starburst activity (``PAH-strong'') and those that do not (``PAH-weak''). The distribution of the diagnostic ratios for the two sub-samples are similar. Isolating the 12$\mu$m sources with a limited D range (1.2$\leq$D$\leq$1.7) also results in large scatter that is systematically higher than that observed in the [OIII] sample for all but three of the flux ratios. A similarly wide spread between F$_{[OIV]}$/F$_{[OIII],obs}$ is present in other samples (i.e. Mel\'endez et al. 2008 and Diamond-Stanic et al. 2009), indicating that sample selection based on [OIII] is primarily responsible for the tighter correlations we observe. This may be due to less extinction in the NLR region which would be expected in sources that have high observed [OIII] flux.

- \textit{Extinction Correction} Applying an extinction correction to the [OIII] flux tightens the correlations with the other luminosity indicators for the [OIII]-selected sample, yet broadens the dispersion for the 12$\mu$m sample. It is not clear whether this difference is primarily due to the different sources of the emission-line data (homogenous SDSS data for the [OIII] sample and heterogeneous data for the 12$\mu$m sample), or whether it simply points out the limitations of extinction corrections in very dusty AGN. Comparison of the optical vs. MIR properties of the most dusty AGN in the SDSS suggest that the former effect is important (Wild et al. 2010).

- \textit{Agreement Among Sy2s} The observed [OIII], [OIV] and MIR luminosities agree the best in the combined Sy2 sample. The widest spread among the various proxies is seen in the radio emission. The X-ray data are dominated by upper limits, but also show a significantly larger dispersion than the optical and IR isotropic flux indicators. 

- \textit{Comparison with Sy1s} The mean ratio of the observed [OIII] flux to the [OIV] flux is lower in Sy2 than in Sy1s by a factor of 1.5-2, while the mean ratio of the observed [OIII] flux to MIR is consistent between Sy2s and Sy1s. The former result, which represents a statistically significant difference between Sy1s and Sy2s according to the KS test but not Kuiper's test, agrees with previous findings (e.g. Haas et al. 2005, Mel\'endez et al. 2008, Diamond-Stanic et al. 2009) and has been interpreted as extinction affecting [OIII] in the NLR, or even torus obscuration attenuating the [OIII] emission (Baum et al. 2010). However, the latter result can not be simply explained by a larger amount of dust extinction of [OIII] in the Sy2s, but it could be due to a slight anisotropy in the MIR emission as predicted by  Pier \& Krolik (1992) and Nenkova et al. (2008) where Sy1s could have up to a factor of two higher MIR flux as compared to Sy2s. The wide scatter in these ratios can also be responsible for the discrepancy between the F$_{[OIV]}$/F$_{[OIII],obs}$ and F$_{MIR}$/F$_{[OIII],obs}$ values, which may be the main driver for the mild disagreement rather than torus emission anisotropy. The F$_{[OIV]}$/F$_{8.4GHz}$ and F$_{MIR}$/F$_{8.4GHz}$ mean values are consistent between Sy1 and Sy2 galaxies (in agreement with Diamond-Stanic et al. 2009 and Mel\'endez et al. 2010 for the [OIV] and radio comparison), though the slopes of the luminosity plots show disagreements and there is wide scatter which is expected due to the wide range of radio loudness observed in AGNs. 

- \textit{Hard X-ray Selected Samples} We find that the hard X-ray flux (relative to the [OIII], [OIV], and MIR fluxes) is suppressed by about an order-of-magnitude in MIR selected Sy2s compared to both Sy1s (consistent with Rigby et al. 2008) and to hard X-ray selected Type 2 AGN (Winter et al. 2010 and Weaver et al. 2010). The comparison with the [OIII] sample is mixed, with statistically significant differences between the Sy2s and Sy1s when the X-ray flux is normalized by the [OIV] and MIR flux, but not when normalized by the [OIII] flux. However, hard X-ray selected Sy2s differ significantly from [OIII]-selected Sy2s (though with only one [OIII] Sy2 detected by BAT, this analysis may be less robust than the 12$\mu$m comparison). These results indicate that hard X-ray emission (E $>$ 10 keV) is anistropic and hard X-ray selected samples are biased against the more heavily obscured type 2 AGN that are present in MIR and possibly [OIII] selected samples. As Weaver et al. (2010) note for sources detected in hard X-rays, Compton scattering could be responsible for the hard X-ray attenuation observed in Type 2 AGN as compared to Type 1. In more obscured sources, Compton scattering may be pushing them below the flux sensitivity of BAT.

F$_{MIR}$ and F$_{[OIV]}$ agree the best, both in comparison with the other indicators in the combined Sy2 sample (having the least scatter) and in having a nearly identical flux ratio in Sy2s and Sy1s as well as not suffering from a luminosity bias. Among the indicators we have considered, they are the best proxies for truly isotropic AGN emission.

We tested the level of agreement of various optical and infrared indicators of star formation activity compared with proxies of AGN activity. Similar to previous works, we find statistically significant relations among the various indicators of the relative energetic significance of star formation and an AGN. These include the MIR spectral slope (parametrized by $\alpha_{20-30\mu m}$),  the PAH EWs at 11.3$\mu$m and 17$\mu$m, the ratio of [OIV]/[NeII] fluxes, and the location of the galaxy in the commonly used diagnostic diagram based on optical emission-lines. We note that the last two diagnostics are a measure of the relative contribution of AGN vs. starburst activity to the incident ionizing radiation. As the incident radiation field becomes more dominated by the AGN activity, the PAHs become weaker relative to the MIR continuum. In part this is expected because of the``dilution'' of the MIR continuum by AGN-heated dust, but the AGN may also be directly affecting the PAH emission (e.g. O'Dowd et al. 2009). 

We also found that the Sy2s that were clearly starburst/AGN composites based on the above indicators were systematically those cases in which large-scale MIR emission from the host galaxy greatly exceeded that from the AGN. We quantified this by comparing the ratio of the 12$\mu$m flux from the large \textit{IRAS} aperture with the 15.5$\mu$m flux from the small \textit{IRS} aperture. A smaller aperture is therefore necessary to isolate AGN emission in systems with high rates of star formation on large scales and/or low AGN luminosities.

The ratios of the [OIII]/MIR and [OIV]/MIR fluxes show little if any evidence for a correlation with any of the above measures of the relative amount of star formation. This lack of a relationship suggests that the [OIII] and [OIV] lines are not biased to be a preferentially higher fraction of the AGN bolometric luminosity in host galaxies with more star formation activity (more dense gas).

\acknowledgments{This work is based in part on observations made with the Spitzer Space Telescope, which is operated by the Jet Propulsion Laboratory, California Institute of Technology under a contract with NASA. The authors thank the anonymous referee whose insightful comments improved the quality of this manuscript.}

\clearpage


\begin{figure}
\epsscale{1.3}
\plottwo{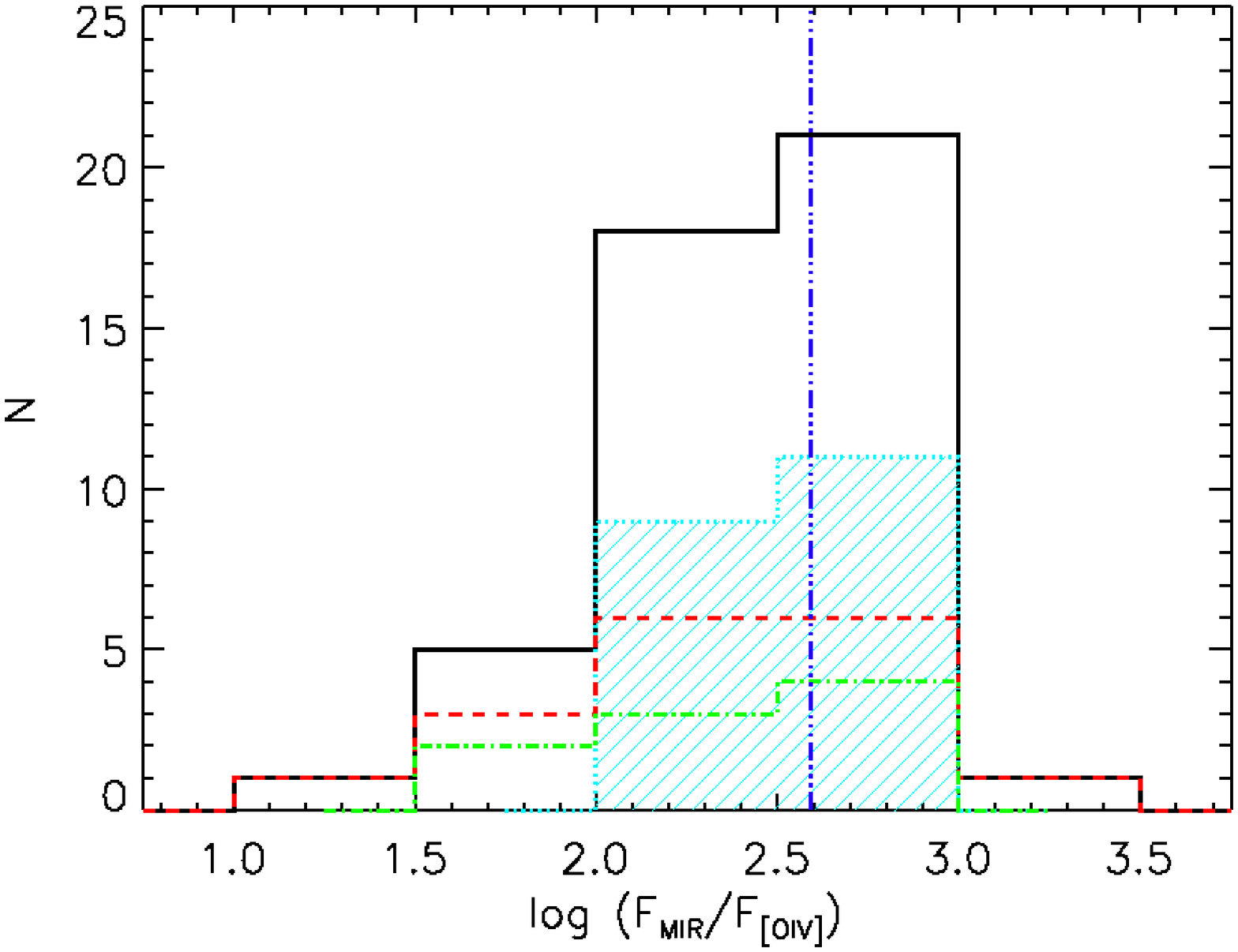}{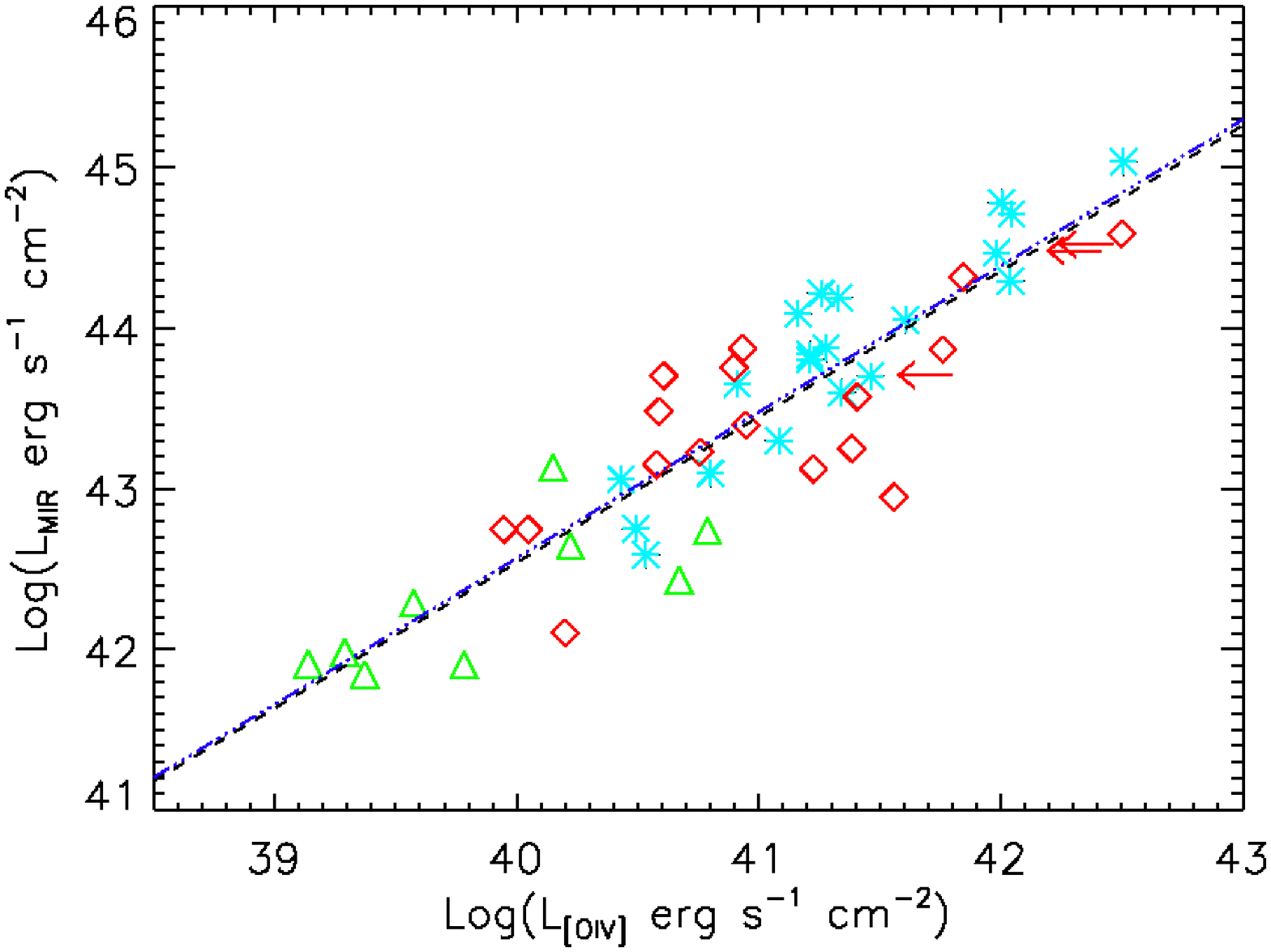}
\caption{\label{mir2_oiv}Left: Distribution of log (F$_{MIR}$/F$_{[OIV]}$). Right: Log (L$_{MIR}$) vs log (L$_{[OIV]}$); $\rho$=0.894 giving P$_{uncorr}$=5.76$\times 10^{-17}$ (slope = 0.91 $\pm$0.07 dex with intercept at 6.24 dex). The blue dashed-dotted line represents the Sy1 values, with F$_{MIR}$ from Deo et al. (2009) and F$_{[OIV]}$ from Tommasin et al. (2010) and Mel\'endez et al. (2008). The relationship between L$_{MIR}$ and L$_{[OIV]}$ is nearly identical for Sy1s and Sy2s. Color and linestyle coding same as Figure \ref{hist_o4_o3}.}
\end{figure}

\clearpage

\begin{figure}
\epsscale{1.0}
\plotone{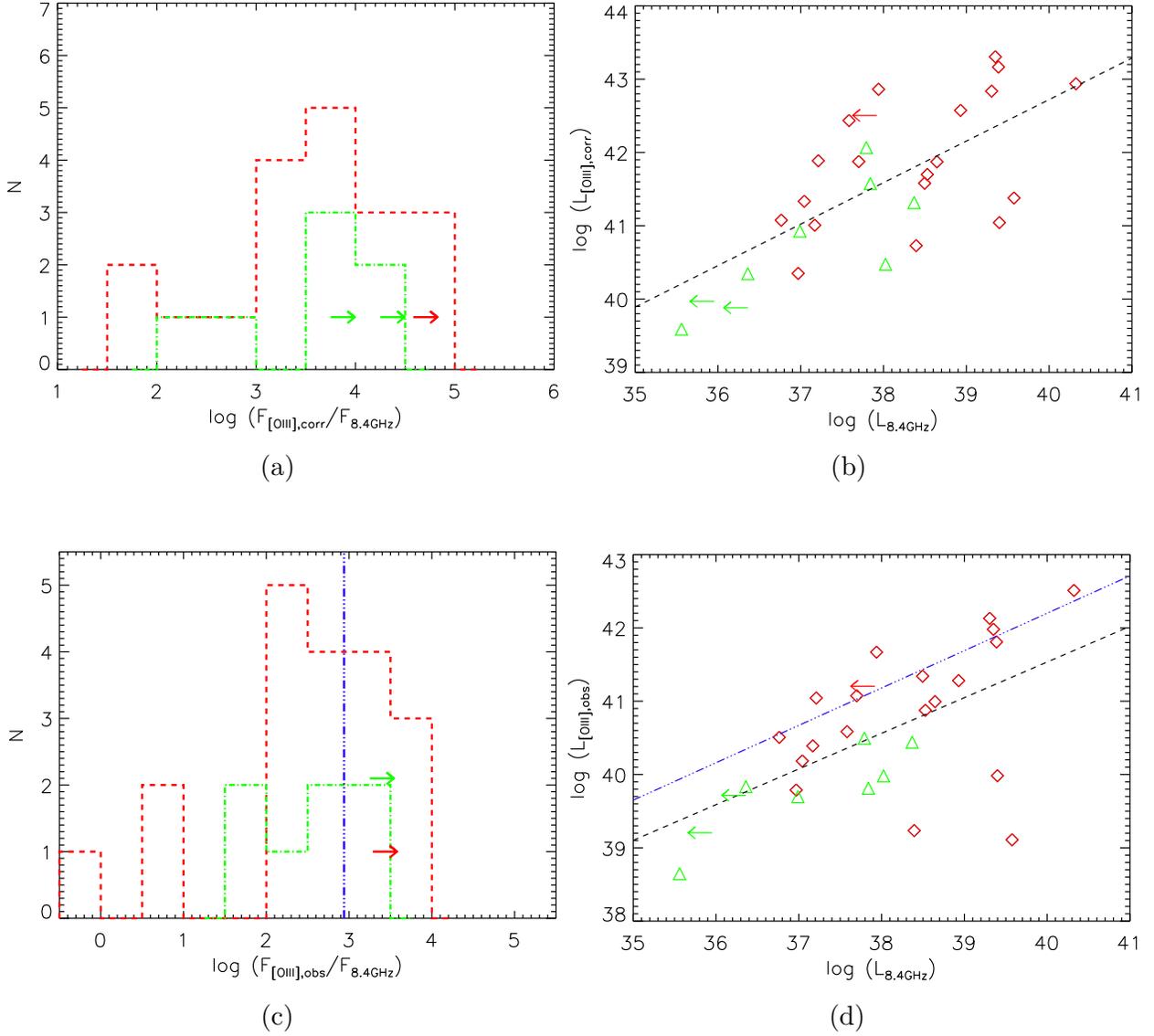}
\caption{\label{o3_radio} a) Distribution of log (F$_{[OIII],corr}$/F$_{8.4GHz}$). b) Log (L$_{[OIII],corr}$) vs log (L$_{8.4GHz}$); $\rho$=0.667 giving P$_{uncorr}$=2.00$\times 10^{-4}$ (slope = 0.57 $\pm$ 0.13 dex with intercept at 20.1 dex). c) Distribution of log (F$_{[OIII],obs}$/F$_{8.4GHz}$). d) Log (L$_{[OIII],obs}$) vs log (L$_{8.4GHz}$); $\rho$=0.57 giving P$_{uncorr}$=0.003 (slope = 0.49 $\pm$ 0.14 dex with intercept at 22.1 dex). The blue dashed-dotted line represents the Sy1 values, with the radio data from Thean et al. (2000) and F$_{[OIII],obs}$ from Diamond-Stanic et al. (2009) and  Mel\'endez et al. (2008). Color and linestyle coding same as Figure \ref{hist_o4_o3}.}
\end{figure}

\begin{figure}
\epsscale{1.3}
\plottwo{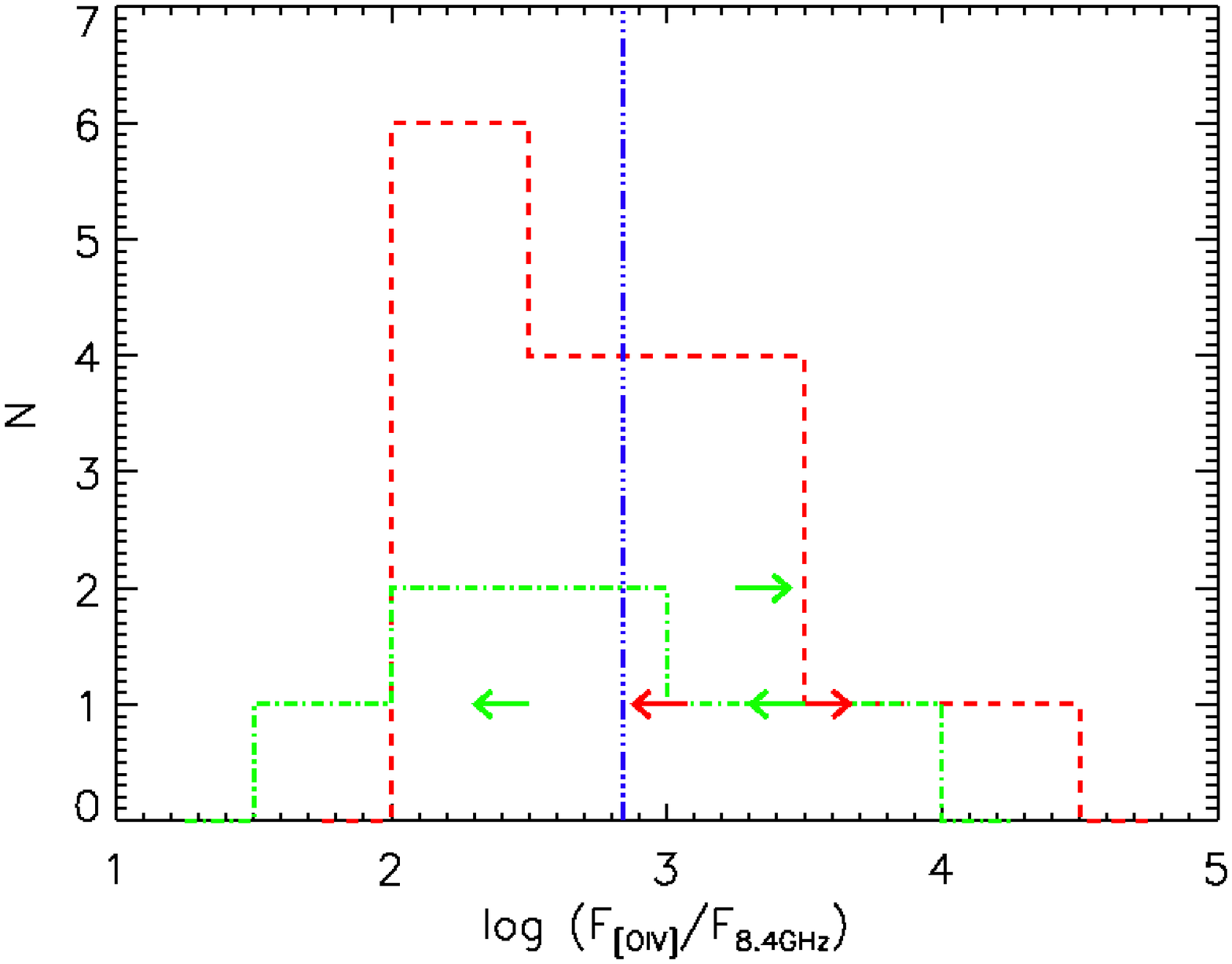}{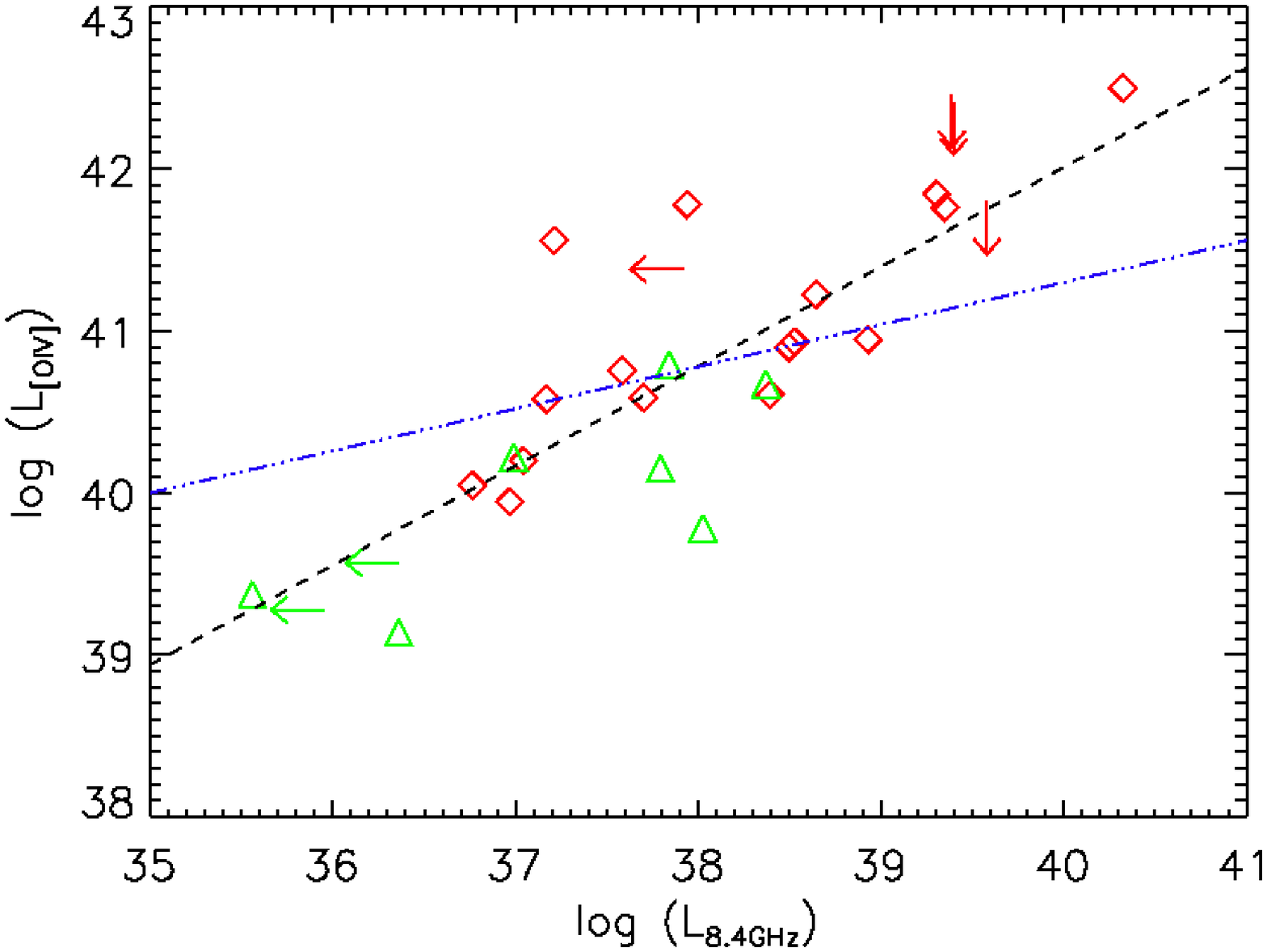}
\caption{\label{oiv_radio}Left: Distribution of log (F$_{[OIV]}$/F$_{8.4GHz}$). Right: Log (L$_{[OIV]}$) vs log (L$_{8.4GHz}$); $\rho$=0.804 giving P$_{uncorr}$=3.81$\times 10^{-6}$ (slope = 0.61 $\pm$ 0.10 dex with intercept at 17.5 dex). The blue dashed-dotted line represents the Sy1 values, with the radio data from Thean et al. (2000) and F$_{[OIV]}$ from Diamond-Stanic et al. (2009) and  Mel\'endez et al. (2008). Color and linestyle coding same as Figure \ref{hist_o4_o3}.}
\end{figure}

\begin{figure}
\epsscale{1.3}
\plottwo{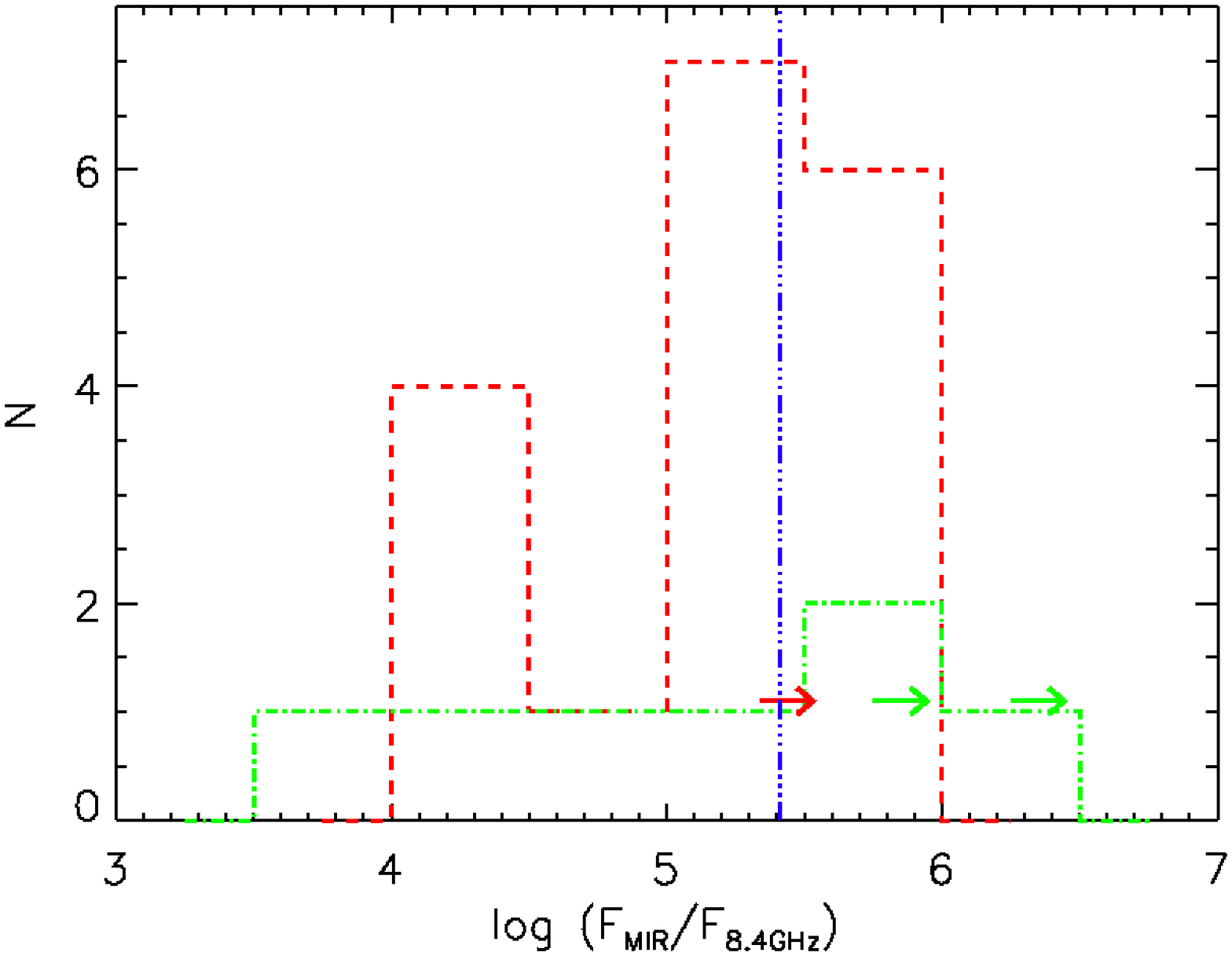}{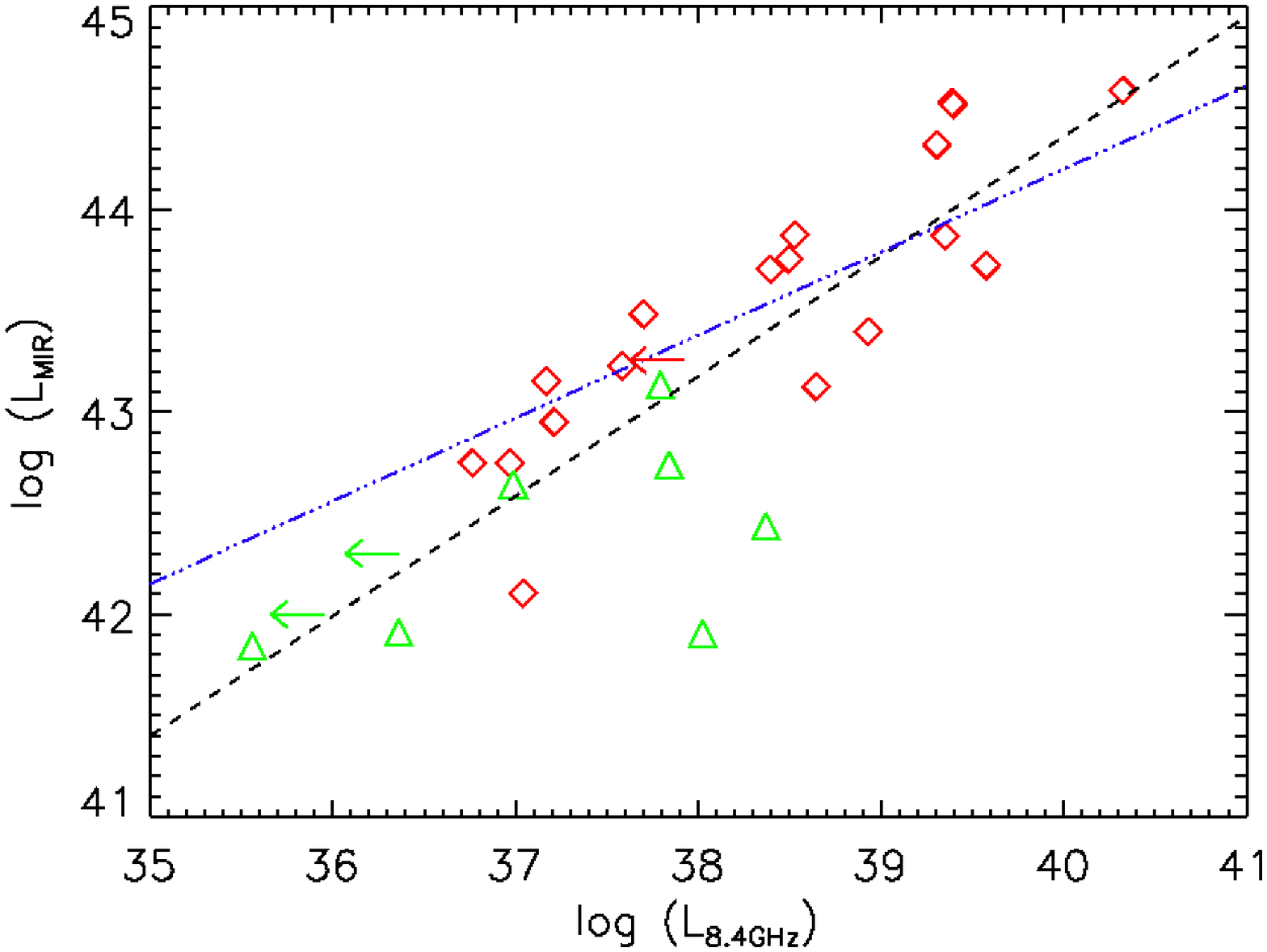}
\caption{\label{mir2_radio}Left: Distribution of log (F$_{MIR}$/F$_{8.4GHz}$). Right: Log (L$_{MIR}$) vs log (L$_{8.4GHz}$); $\rho$=0.826 giving P$_{uncorr}$=3.65$\times 10^{-7}$ (slope = 0.59 $\pm$ 0.08 dex with intercept at 20.7 dex). The blue dashed-dotted line represents the Sy1 values, with the radio data from Thean et al. (2000) and F$_{MIR}$ from Deo et al (2009). Color and linestyle coding same as Figure \ref{hist_o4_o3}.}
\end{figure}

\begin{figure}
\epsscale{1.0}
\plotone{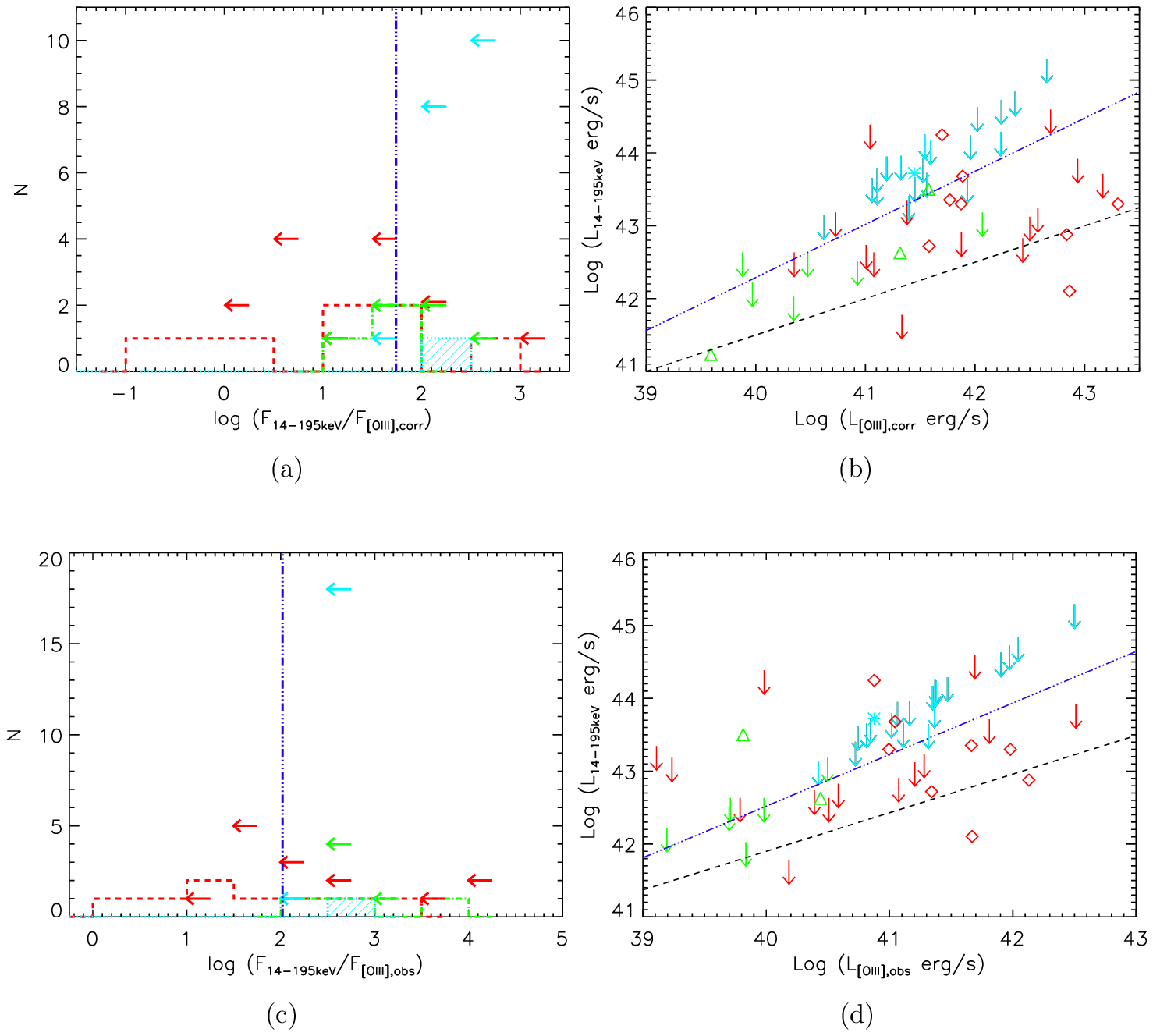}
\caption{\label{xray_o3} a)  Distribution of log (F$_{14-195keV}$/F$_{[OIII],corr}$). b)  Log (L$_{14-195keV}$) vs log (L$_{[OIII],corr}$) with dashed line from survival analysis; P$_{uncorr}$=0.0075 (slope = 0.50 $\pm$ 0.18 dex with intercept at 21.5 dex). c) Distribution of log (F$_{14-195keV}$/F$_{[OIII],obs}$). d)  Log (L$_{14-195keV}$) vs log (L$_{[OIII],obs}$) with dashed line from survival analysis; P$_{uncorr}$=0.0028 (slope = 0.53 $\pm$ 0.18 dex with intercept at 20.7 dex). In both luminosity vs. luminosity plots, the blue dotted-dashed line represents the Sy1 values from the Winter et al. sample (2010). Color and linestyle coding same as Figure \ref{hist_o4_o3}.}
\end{figure}

\begin{figure}
\epsscale{1.3}
\plottwo{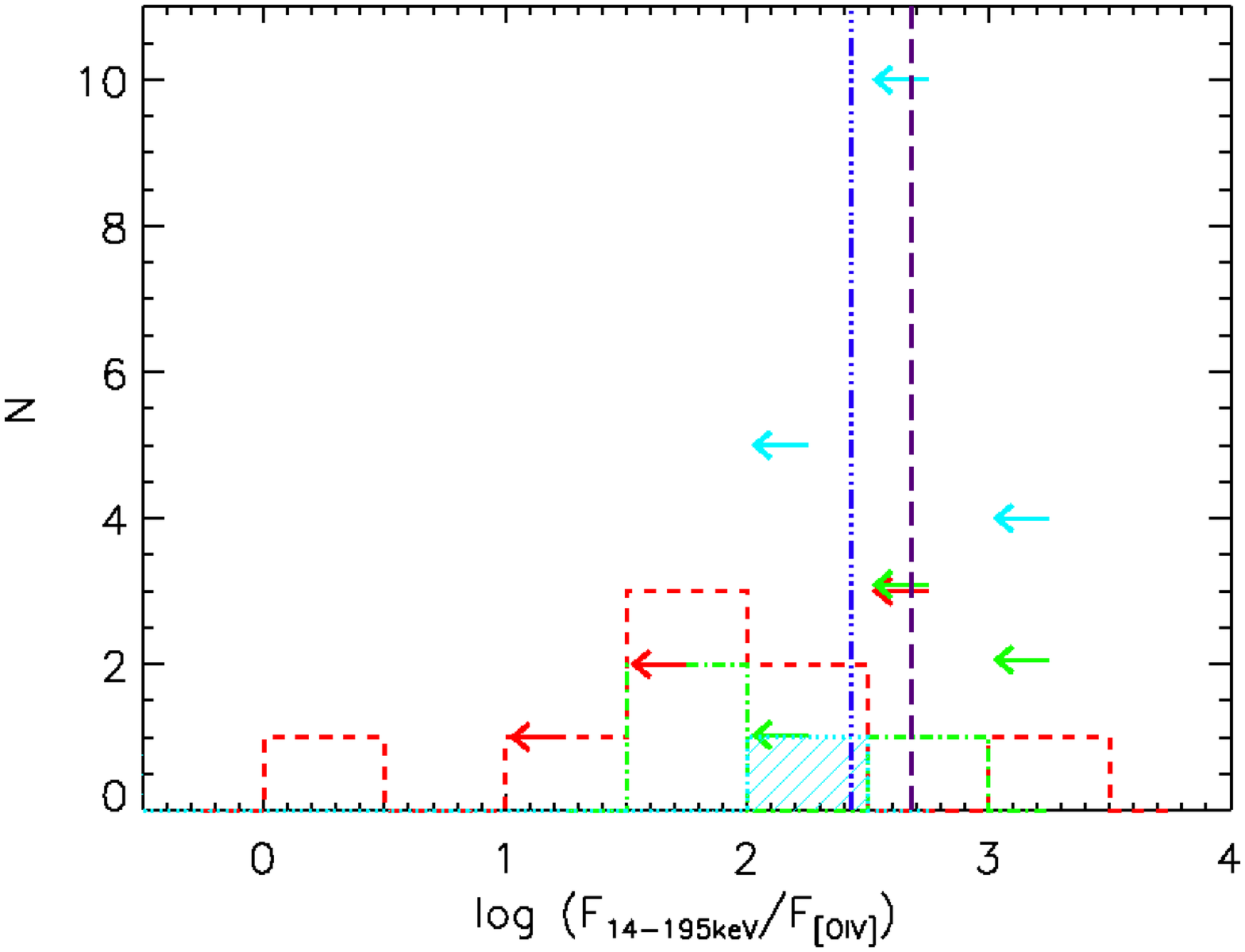}{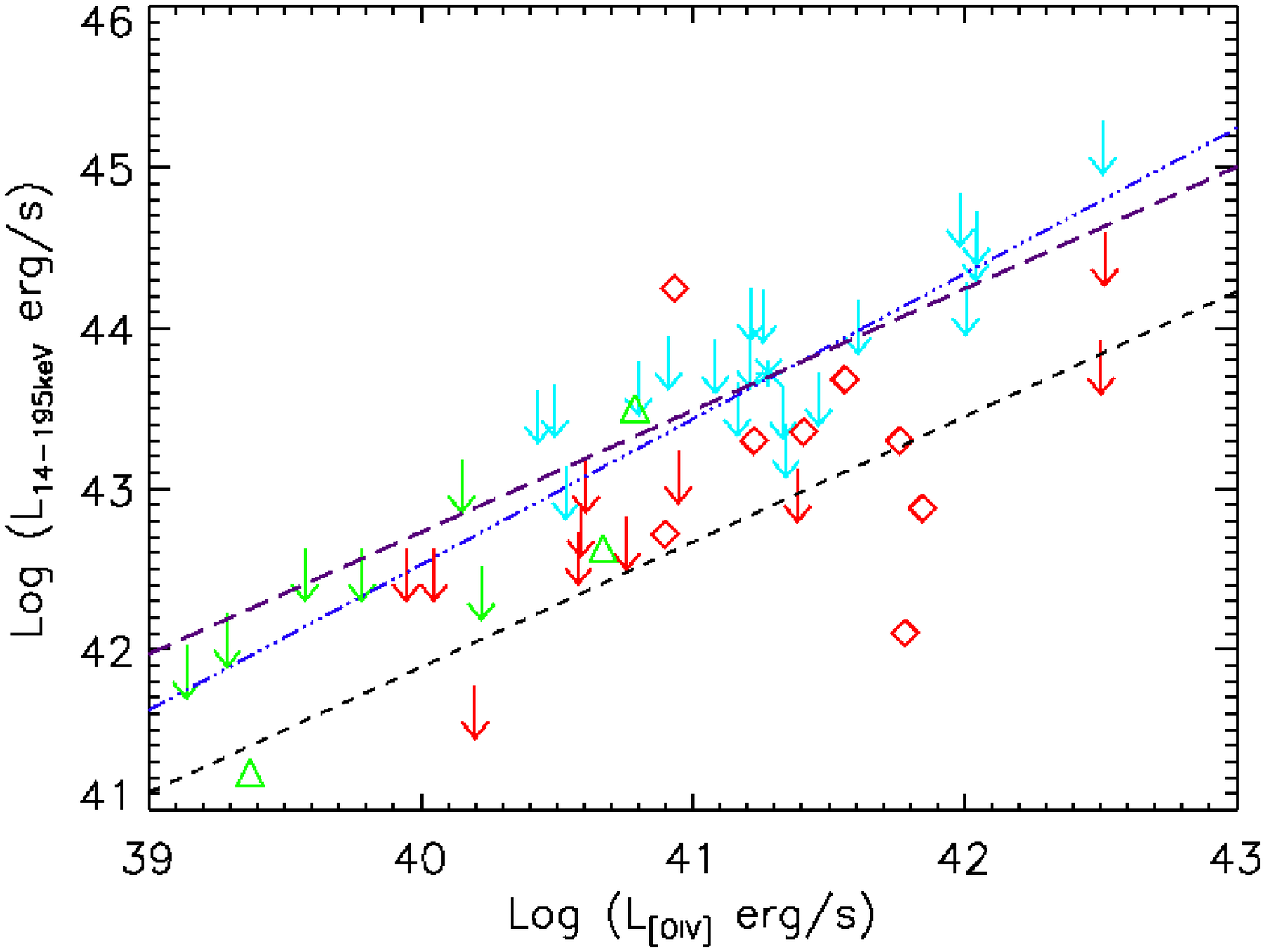}
\caption{\label{xray_o4}Left: Distribution of log (F$_{14-195keV}$/F$_{[OIV]}$). Right: Log (L$_{14-195keV}$) vs log (L$_{[OIV]}$) with dashed line from survival analysis; correlation probability $\sim$99.97\% (slope = 0.78 $\pm$ 0.19 dex with intercept at 10.7 dex). The blue dotted-dashed and purple dashed lines represent the Sy1 values from Weaver et al. (2010) and Rigby et al. (2009), respectively. Color and linestyle coding same as Figure \ref{hist_o4_o3}.}
\end{figure}

\begin{figure}
\epsscale{1.3}
\plottwo{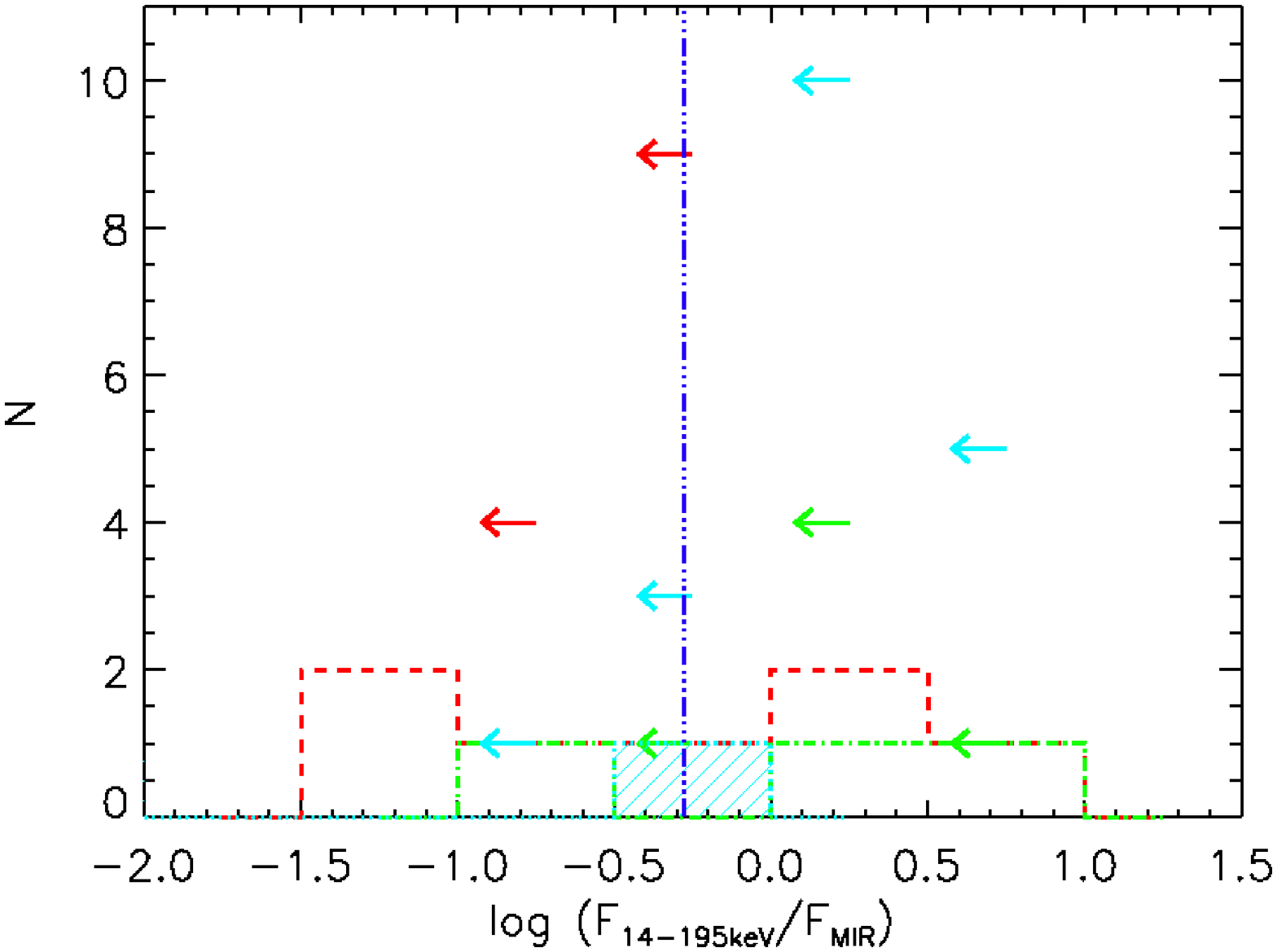}{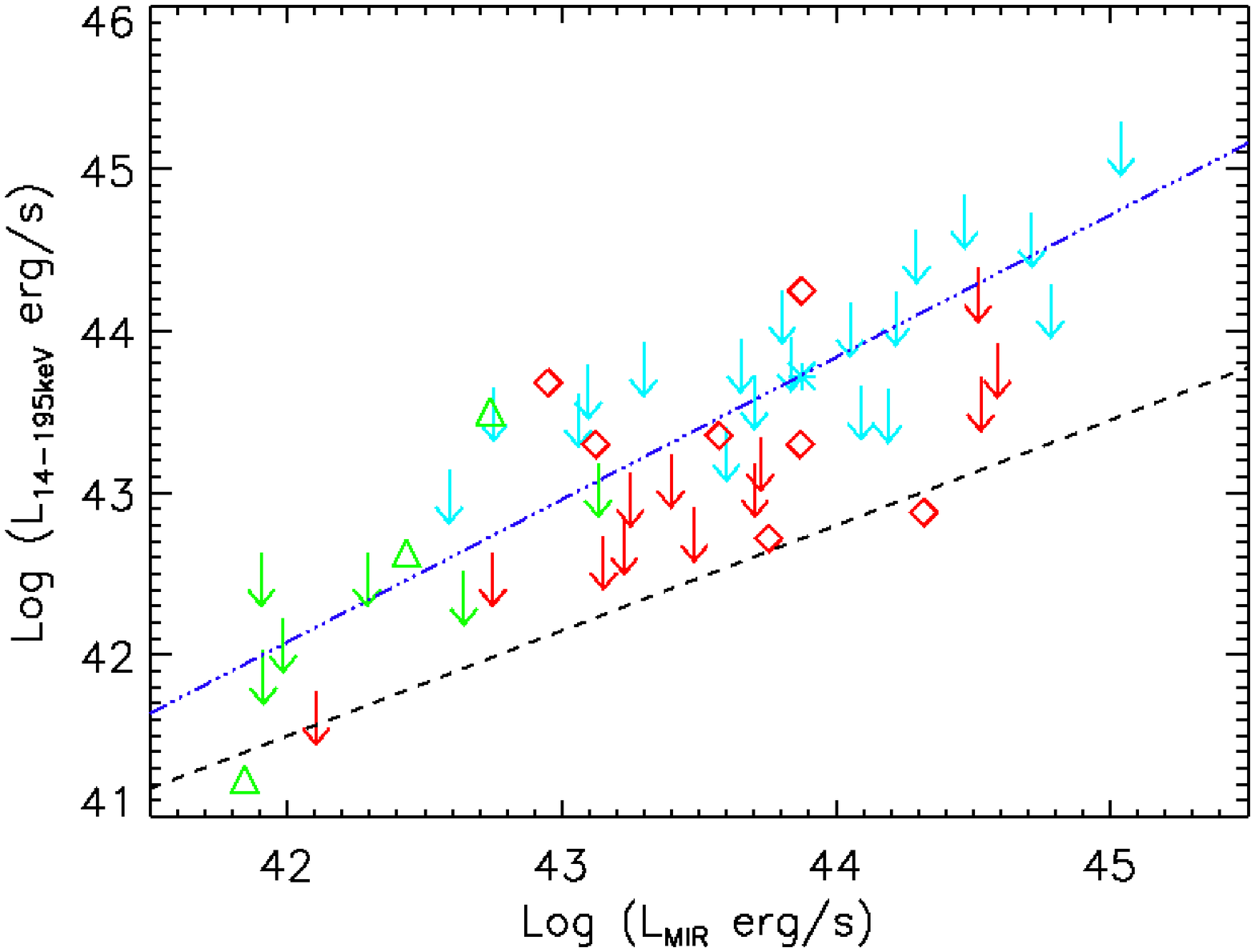}
\caption{\label{xray_mir2}Left: Distribution of log (F$_{14-195keV}$/F$_{MIR}$). Right: Log (L$_{14-195keV}$) vs log (L$_{MIR}$) with dashed line from survival analysis; P$_uncorr$=0.0064 (slope = 0.57 $\pm$ 0.19 dex with intercept at 17.8 dex). Blue dotted-dashed line represents the Sy1 values, with F$_{14-195keV}$ from Mel\'endez et al. (2008) and Tueller et al. (2009) and F$_{MIR}$ from Deo et al. (2009). Color and linestyle coding same as Figure \ref{hist_o4_o3}.}
\end{figure}

\begin{figure}
\epsscale{1.3}
\plottwo{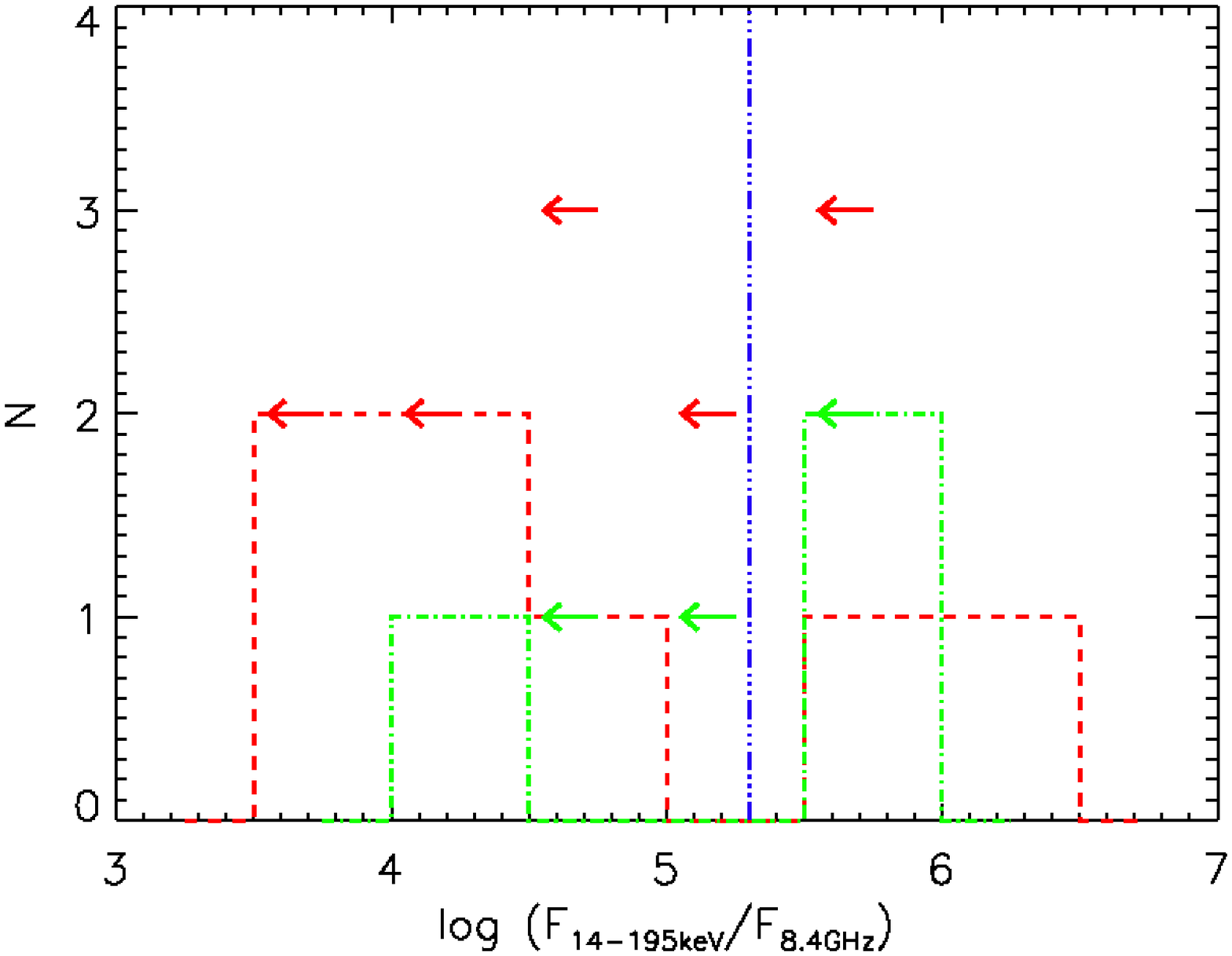}{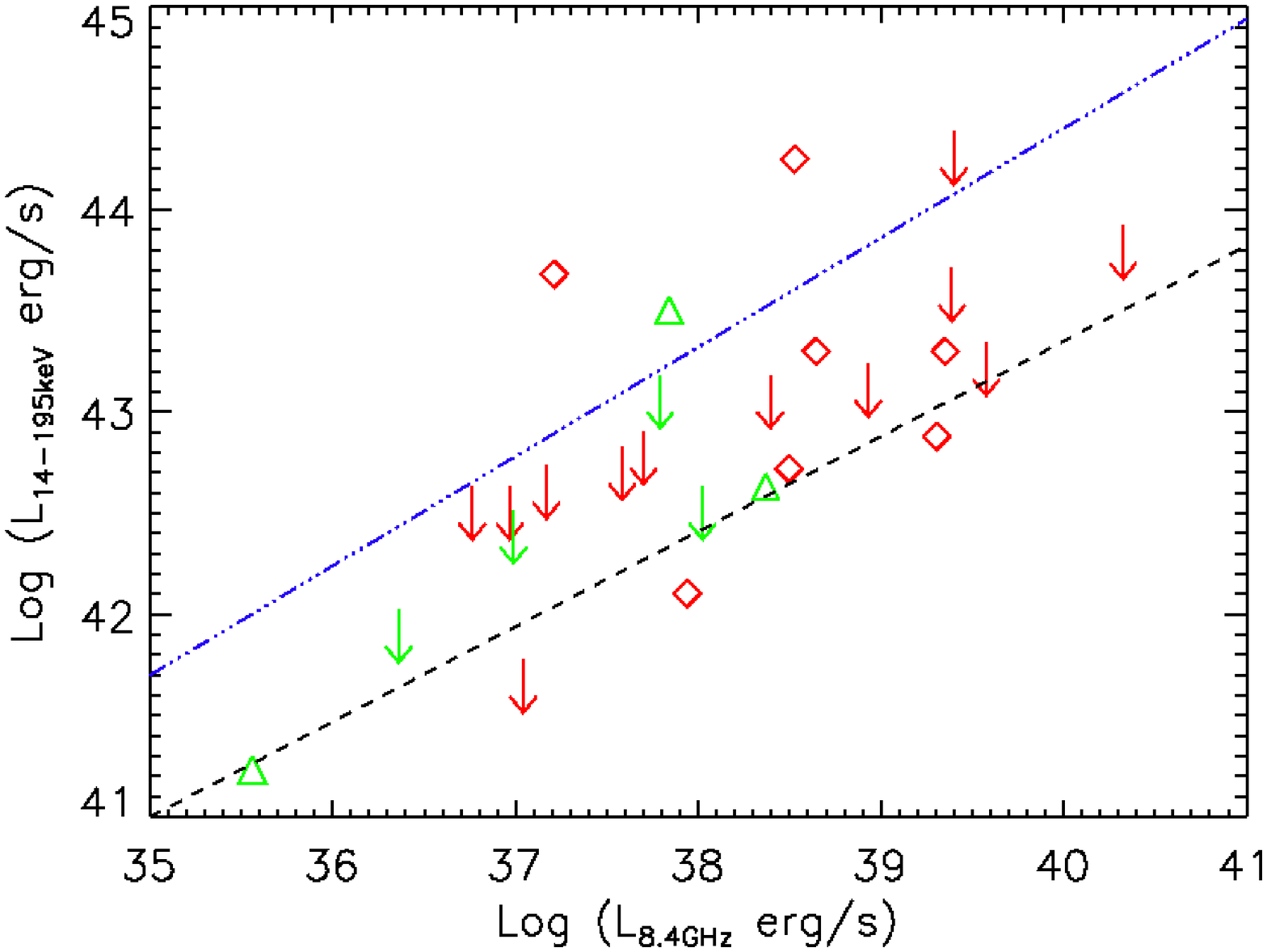}
\caption{\label{xray_radio}Left: Distribution of log (F$_{14-195keV}$/F$_{8.4GHz}$). Right: Log (L$_{14-195keV}$) vs log (L$_{8.4GHz}$) with dashed line from survival analysis; correlation probability $\sim$99.4\% (slope = 0.47 $\pm$ 0.16 dex with intercept at 24.6 dex). The blue dotted-dashed line represent the Sy1 values, with F$_{14-195keV}$ from Mel\'endez et al. (2008) and Tueller et al. (2009) and F$_{8.4GHz}$ from Thean et al. (2000). Color and linestyle coding same as Figure \ref{hist_o4_o3}.}
\end{figure}


\begin{figure}
\epsscale{0.8}
\plotone{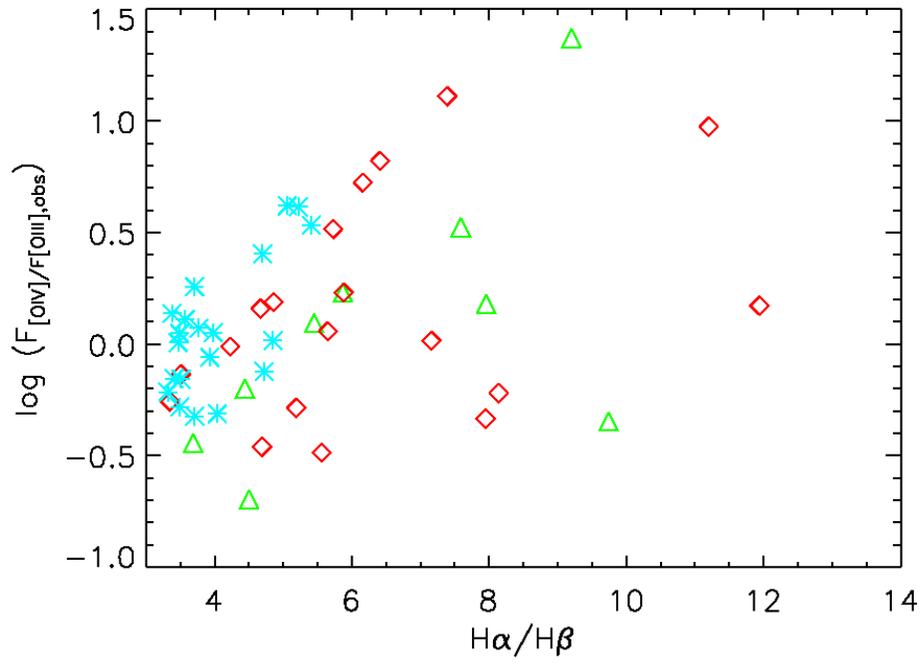}
\caption{\label{balmer} Log (F$_{[OIV]}$/F$_{[OIII],obs}$ vs H$\alpha$/H$\beta$. The green triangles are the 12$\mu$m sources with D $<$1.2, the red diamonds are the 12$\mu$m sources with D $\geq$ 1.2 and the cyan asterisks represent the [OIII] sample.}
\end{figure}


\begin{figure}
\epsscale{0.8}
\plotone{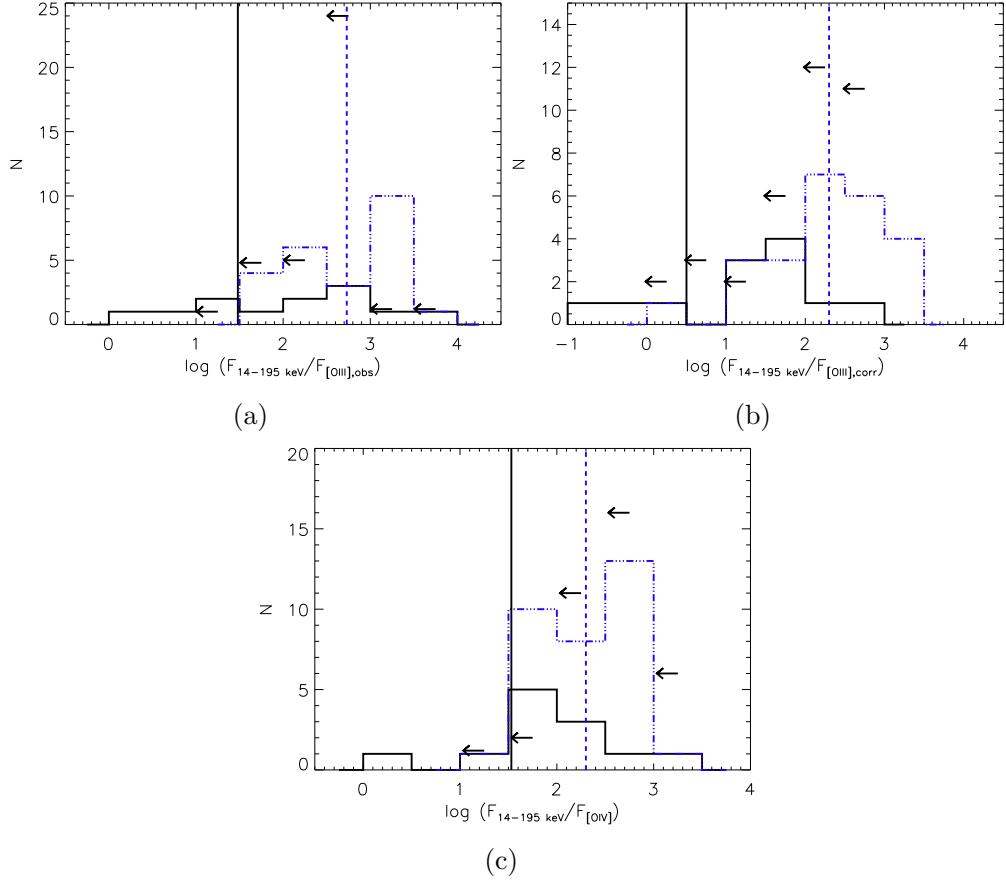}
\caption{\label{bat_hist} In all 3 plots, the solid black line (arrows) represents the combined [OIII] and 12$\mu$m sample with detected (upper limits) 14-195 keV emission and the dotted-dashed blue line represents the BAT selected Sy2s from Winter et al. (2010, panels a and b) and Weaver et al. (2010, panel c). The vertical black line reflects the mean values for our combined [OIII] and 12$\mu$m sample (from Survival Analysis) and the vertical dashed blue line delineates the mean for the Sy2 samples from Winter et al. (2010, panels a and b) and Weaver et al. (2010, panel c). In all cases, the BAT selected Sy2s have systematically higher hard X-ray emission when normalized by other intrinsic AGN proxies as compared to the optical and IR selected Sy2s, with the mean values differing by almost an order of magnitude or more.}
\end{figure}


\begin{figure}
\epsscale{0.8}
\plotone{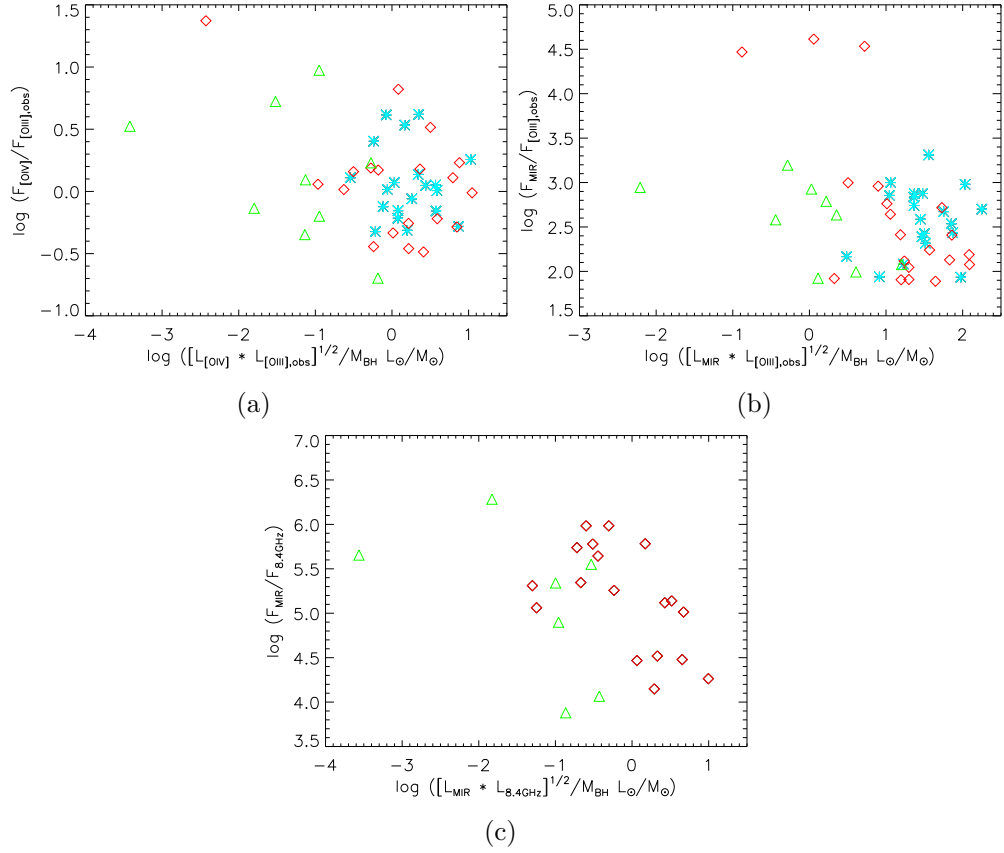}
\caption{\label{ledd} Flux ratios vs. $L_{AGN}/L_{Edd}$. The cyan asterisks represent the [OIII] sample, the red diamonds the ``PAH-weak'' 12$\mu$m sub-sample and the green triangles the ``PAH-strong'' 12 $\mu$m sub-sample.}
\end{figure}


\begin{figure}
\epsscale{1.3}
\plottwo{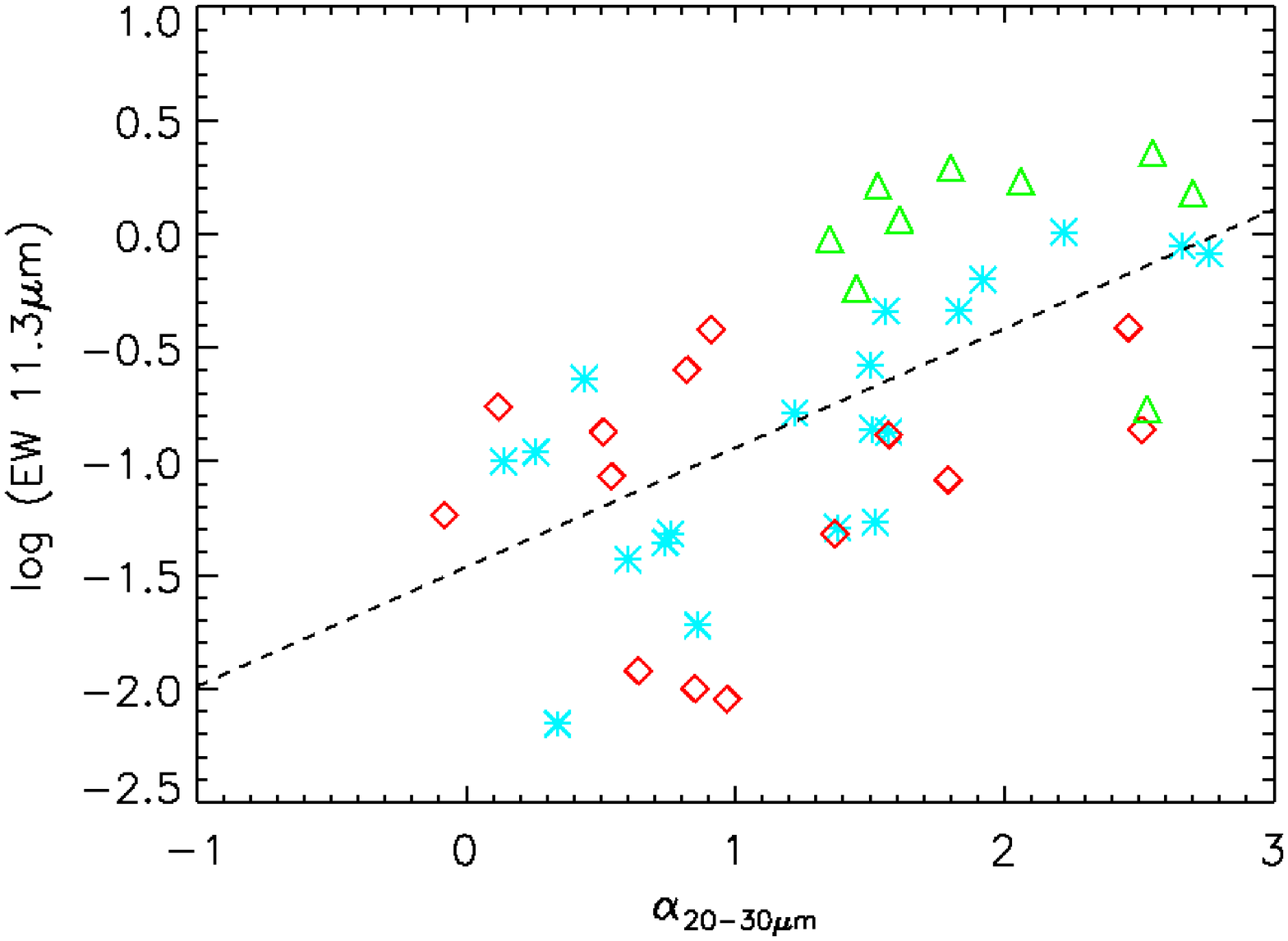}{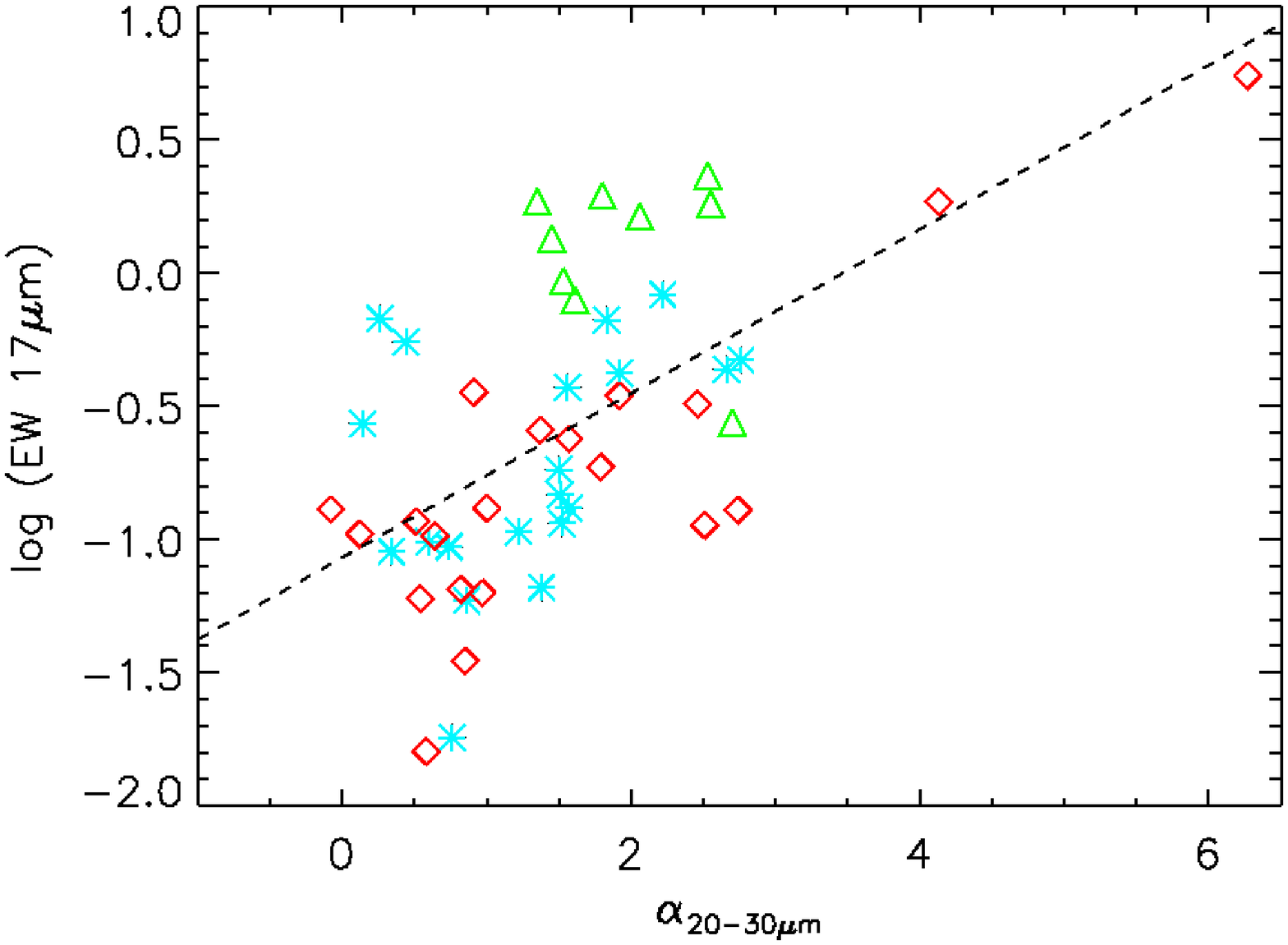}
\caption{\label{alpha_ew}Left:  Log (PAH EW 11.3$\mu$m) vs $\alpha_{20-30\mu m}$. Right:  Log (PAH EW 17$\mu$m) vs $\alpha_{20-30\mu m}$. The dashed line is the fit from linear regression, with $\rho$=0.609 and $\rho$=0.600, giving probabilities of uncorrelation of P$_{uncorr}$=1.47$\times 10^{-5}$ and P$_{uncorr}$=5.26$\times 10^{-6}$ and slopes of 0.52 $\pm$ 0.11 and 0.31 $\pm$ 0.06 with intercepts at -1.47 and -1.07, respectively. Color coding same as Figure \ref{ledd}}
\end{figure}

\begin{figure}
\epsscale{1.3}
\plottwo{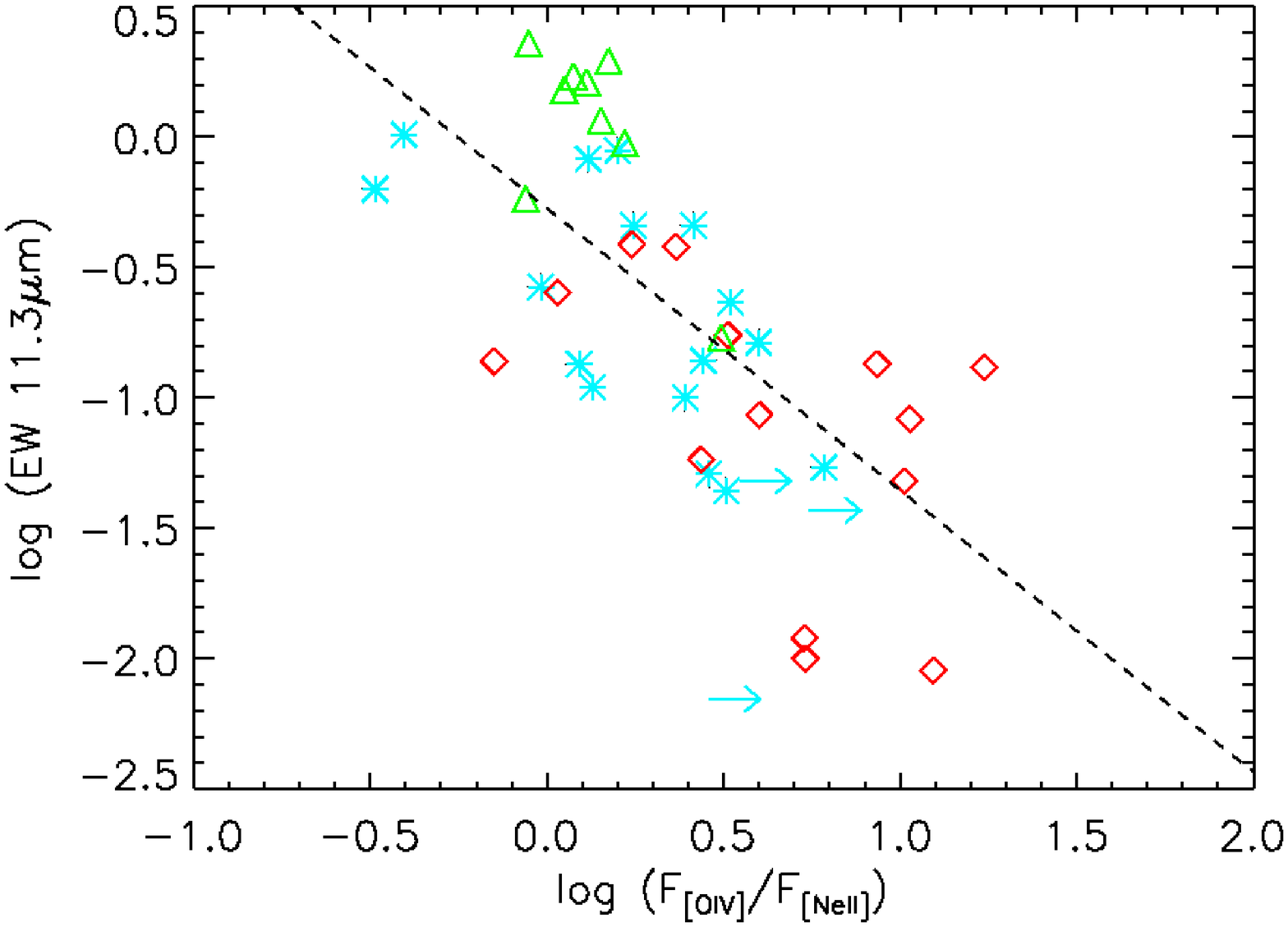}{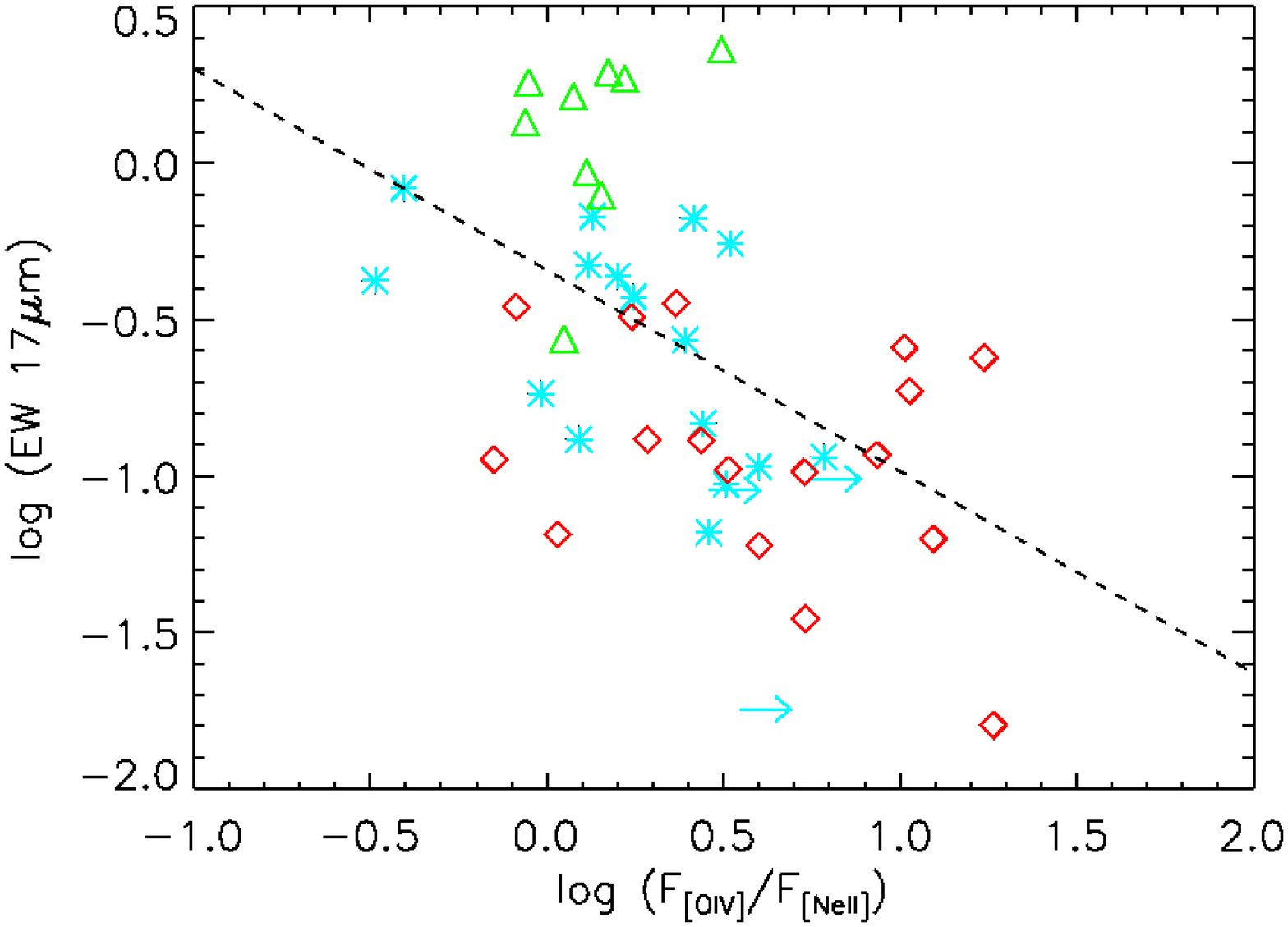}
\caption{\label{oiv_neii_ew}Left: Log (PAH EW 11.3$\mu$m) vs log (F$_{[OIV]}$/F$_{[NeII]}$). Right: Log (PAH EW 17$\mu$m) vs log (F$_{[OIV]}$/F$_{[NeII]}$). The dashed line is the fit from linear regression, with $\rho$=-0.677 and $\rho$=-0.515, giving probabilities of uncorrelation of P$_{uncorr}$=2.26$\times$10$^{-6}$ and P$_{uncorr}$=4.89$\times$10$^{-4}$ and slopes of -1.08 $\pm$ 0.19 and -0.64 $\pm$ 0.17 with intercepts at -0.27 and -0.34, respectively.  The color coding is the same as in Figure \ref{ledd}.}
\end{figure}

\begin{figure}
\epsscale{1.0}
\plotone{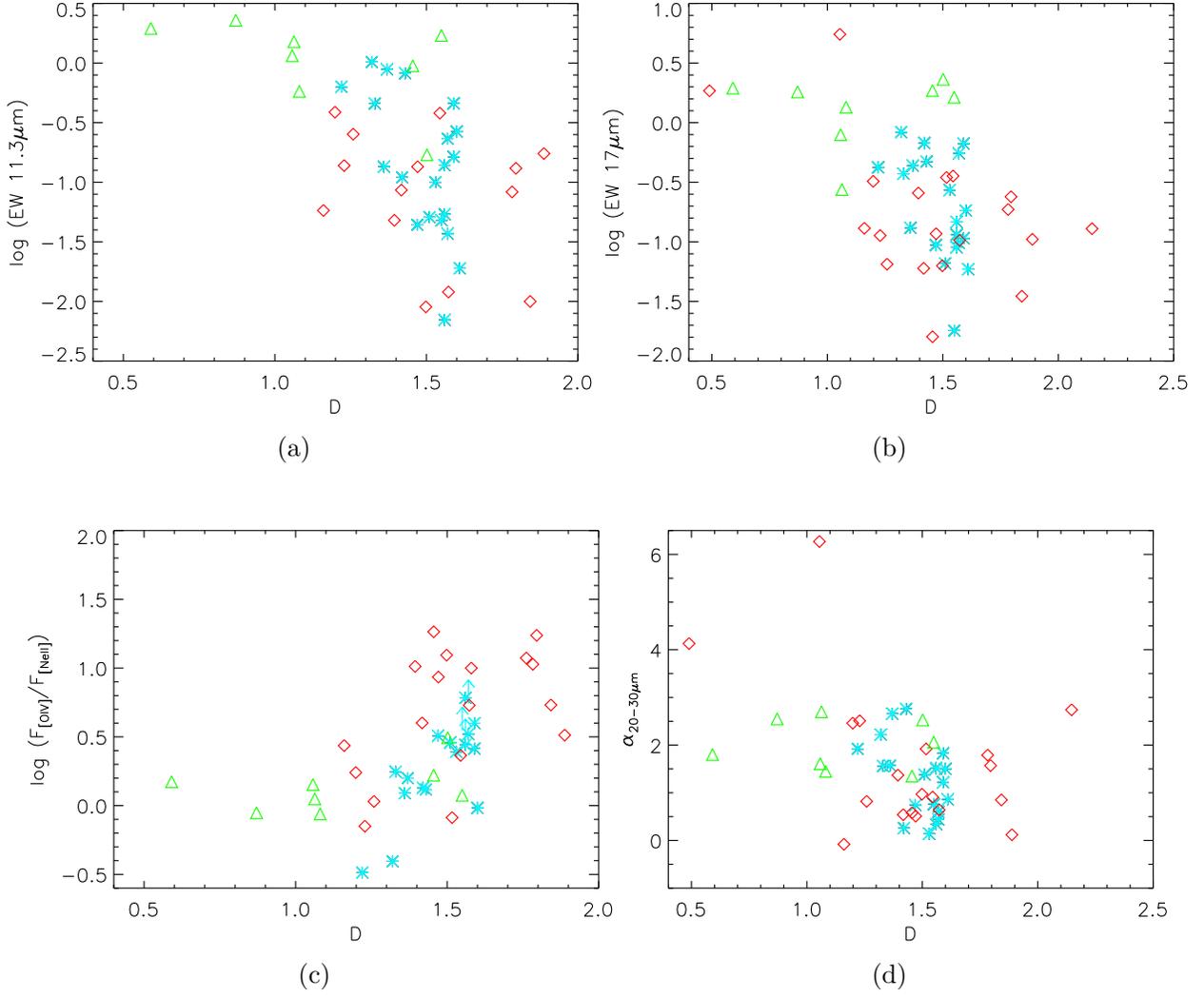}
\caption{\label{d_figs} a) Log (PAH EW 11.3$\mu$m) vs D. b) Log (PAH EW 17$\mu$m) vs D. c) Log (F$_{[OIV]}$/F$_{[NeII]}$) vs D. d) $\alpha_{20-30\mu m}$ vs D. The color coding is the same as in Figure \ref{ledd}.}
\end{figure}

\begin{figure}
\epsscale{1.3}
\plottwo{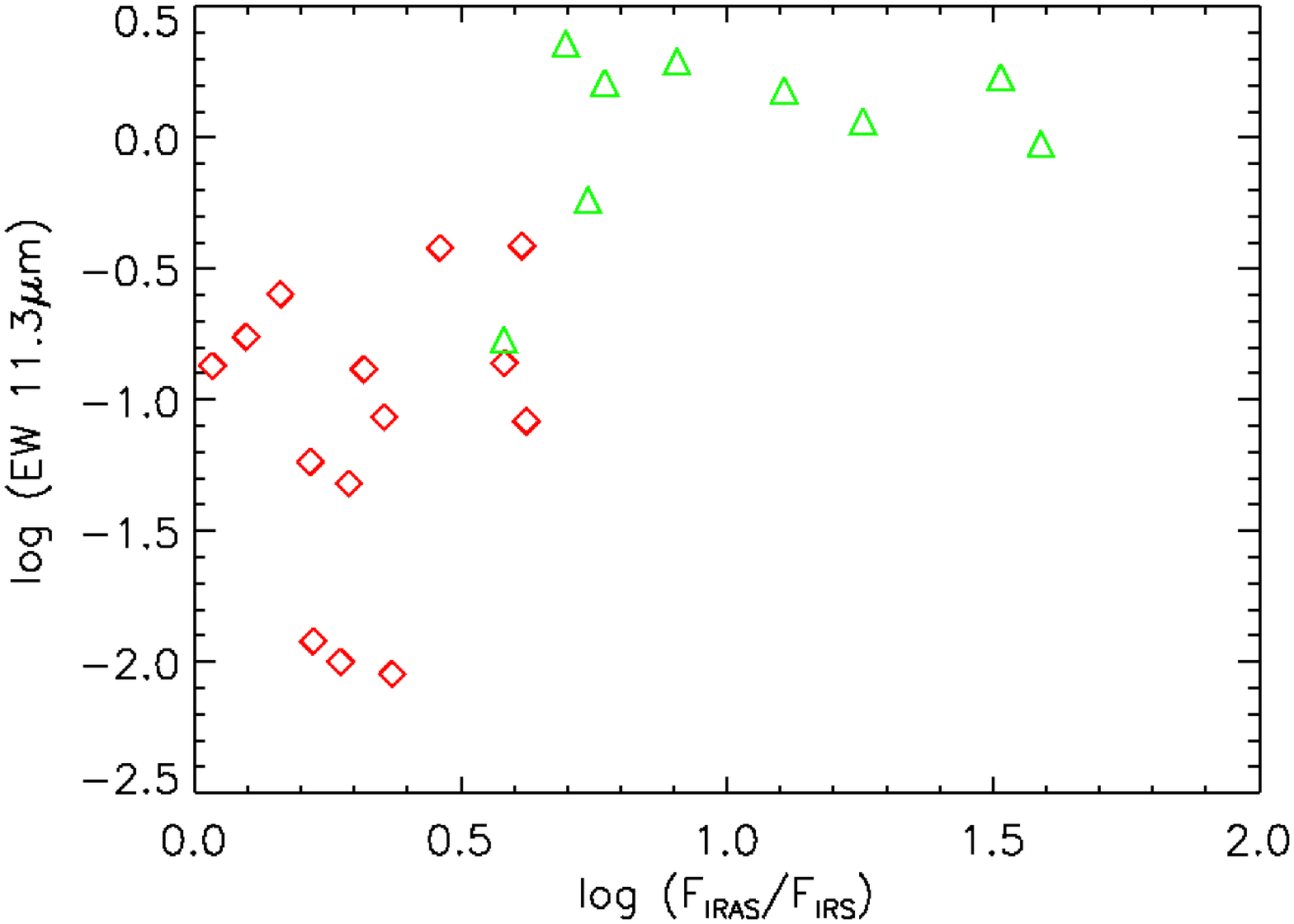}{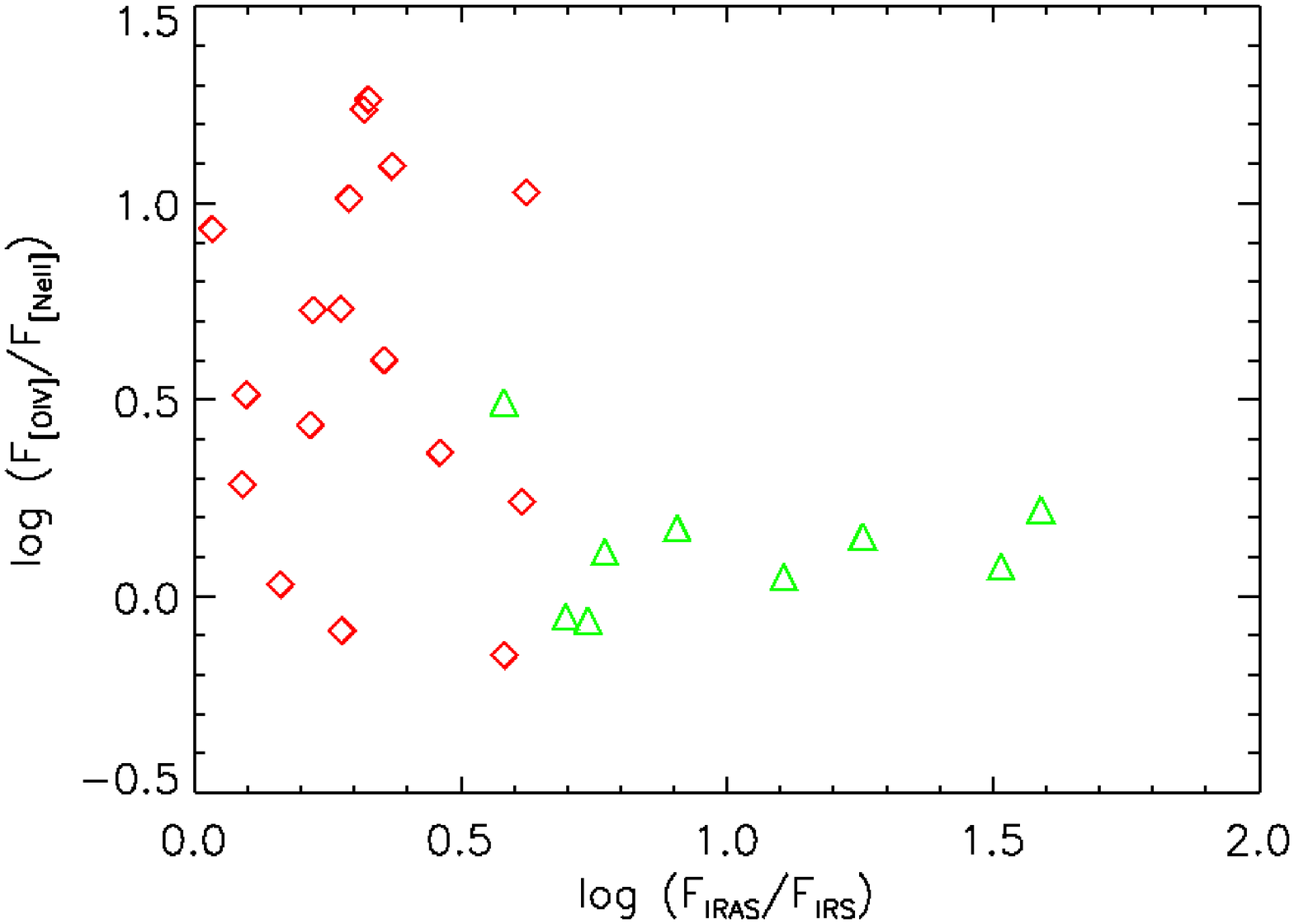}
\caption{\label{iras}Left: Log (PAH EW 11.3$\mu$m) vs log (F$_{IRAS}$/F$_{IRS}$). Right: Log (F$_{[OIV]}$/F$_{[NeII]}$) vs log (F$_{IRAS}$/F$_{IRS}$). The color coding is the same as in Figure \ref{ledd}.}
\end{figure}


\begin{figure}
\epsscale{1.0}
\plotone{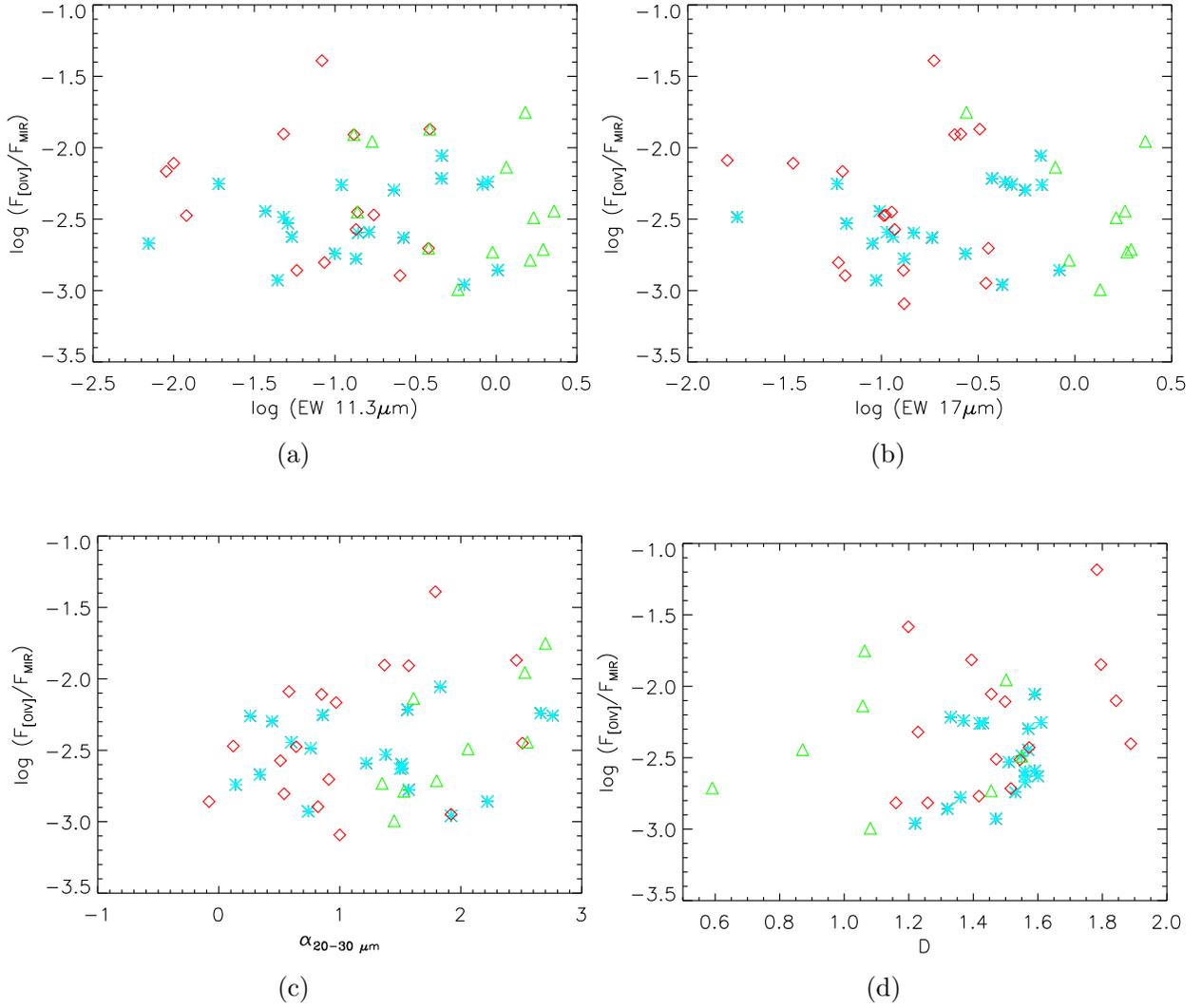}
\caption{\label{sb_bias} a) Log (F$_{[OIV]}$/F$_{MIR}$) vs log (PAH EW 11.3$\mu$m). b) Log (F$_{[OIV]}$/F$_{MIR}$) vs log (PAH EW 17$\mu$m). c) Log (F$_{[OIV]}$/F$_{MIR}$) vs $\alpha_{20-30\mu m}$. d) Log (F$_{[OIV]}$/F$_{MIR}$) vs. D. No strong trends are apparent. The color coding is the same as in Figure \ref{ledd}.}
\end{figure}

\begin{figure}
\epsscale{1.0}
\plotone{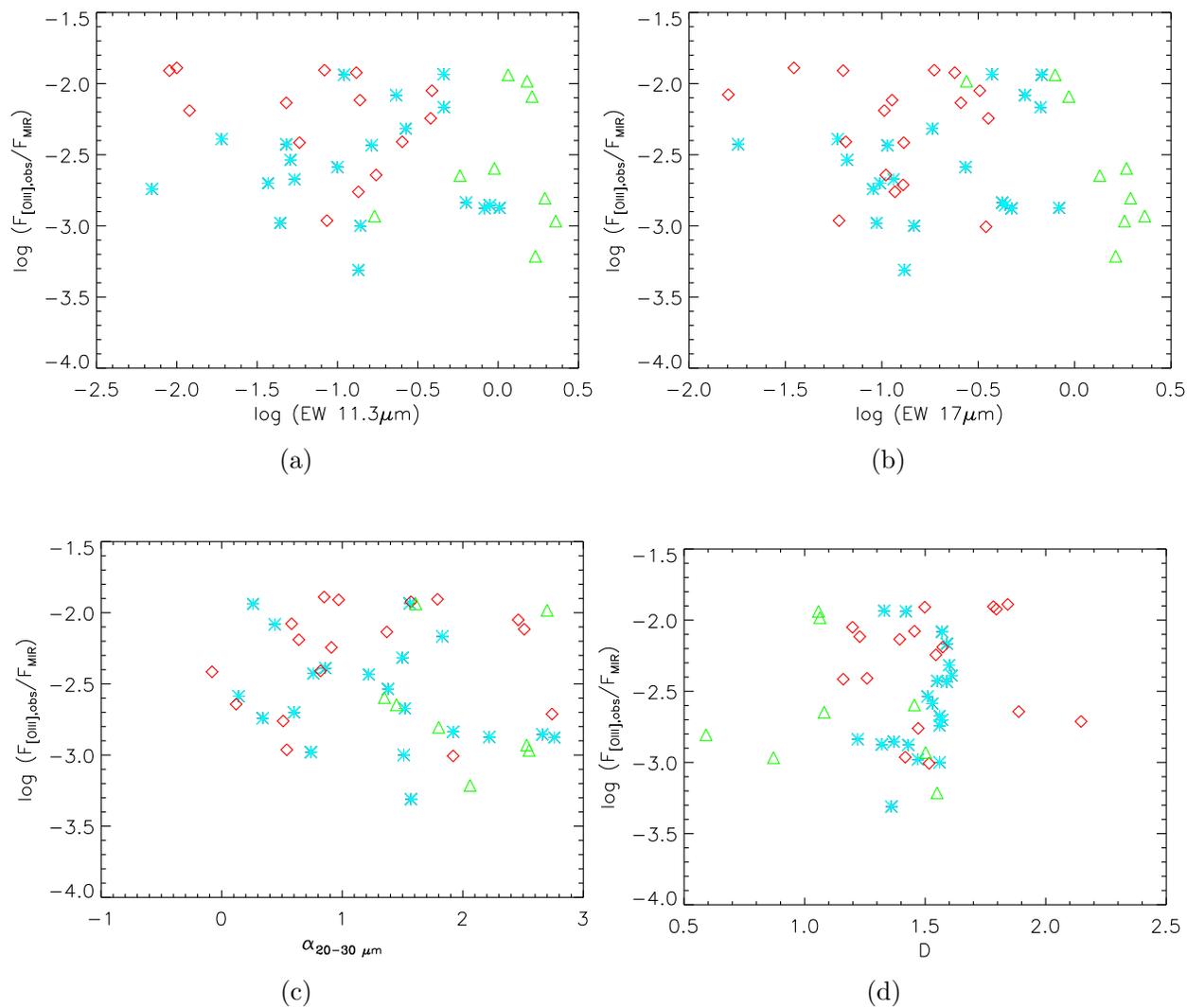}
\caption{\label{sb_bias2} a) Log (F$_{[OIII,obs]}$/F$_{MIR}$) vs log (PAH EW 11.3$\mu$m). b) Log (F$_{[OIII,obs]}$/F$_{MIR}$) vs log (PAH EW 17$\mu$m). c) Log (F$_{[OIII,obs]}$/F$_{MIR}$) vs $\alpha_{20-30\mu m}$. d) Log (F$_{[OIII,obs]}$/F$_{MIR}$) vs. D. No strong trends are apparent. The color coding is the same as in Figure \ref{ledd}.}
\end{figure}

\clearpage


\begin{table}[h]
\caption{\label{o3_sample}[OIII] Sample}
\begin{tabular}{lcccrc}
\hline
\hline\noalign{\smallskip}
Galaxy & RA & Dec & \textit{z} & Distance & M$_{BH}$$^1$ \\
       & (J2000) & (J2000) & & Mpc & M$_{\sun}$ \\
\hline\noalign{\smallskip}

NGC 0291                  & 00 53 29.9 &  08 46 04 & 0.019 & 82.6  & 7.10 \\
Mrk 0609                  & 03 25 25.4 &  06 08 37 & 0.034 & 151.3 & 7.80 \\
IC 0486                   & 08 00 21.0 &  26 36 49 & 0.027 & 116.8 & 7.73 \\
2MASX J08035923+2345201   & 08 03 59.2 &  23 45 20 & 0.029 & 128.8 & 7.25 \\
2MASX J08244333+2959238   & 08 24 43.3 &  29 59 24 & 0.025 & 111.0 & 6.99 \\
CGCG 064-017              & 09 59 14.8 &  12 59 16 & 0.034 & 150.9 & 6.77 \\
2MASX J10181928+3722419   & 10 18 19.3 &  37 22 42 & 0.049 & 219.8 & 7.14 \\
2MASX J11110693+0228477   & 11 11 06.9 &  02 28 48 & 0.035 & 154.9 & 7.33 \\
CGCG 242-028              & 11 23 01.3 &  47 03 09 & 0.025 & 110.1 & 7.29 \\
SBS 1133+572              & 11 35 49.1 &  56 57 08 & 0.051 & 229.0 & 7.99 \\
Mrk 1457                  & 11 47 21.6 &  52 26 58 & 0.049 & 216.6 & 7.85 \\
2MASX J11570483+5249036   & 11 57 04.8 &  52 49 04 & 0.036 & 156.7 & 7.17 \\
2MASX J12183945+4706275   & 12 18 39.4 &  47 06 28 & 0.094 & 431.4 & 8.17 \\
2MASX J12384342+0927362   & 12 38 43.4 &  09 27 37 & 0.083 & 377.9 & 8.40 \\
CGCG 218-007              & 13 23 48.5 &  43 18 04 & 0.027 & 119.4 & 7.65 \\
2MASX J13463217+6423247   & 13 46 32.1 &  64 23 24 & 0.024 & 105.6 & 6.81 \\
NGC 5695                  & 14 37 22.1 &  36 34 04 & 0.014 & 61.1  & 7.44 \\
2MASX J15522564+2753435   & 15 52 25.7 &  27 53 43 & 0.075 & 337.6 & 8.05 \\
SBS 1609+490              & 16 10 51.8 &  48 54 39 & 0.045 & 200.0 & 6.87 \\
2MASX J16164729+3716209   & 16 16 47.3 &  37 16 21 & 0.152 & 725.0 & 8.34 \\

\hline
\hline
\multicolumn{6}{l}{Distances based on redshift, assuming a cosmology of}\\
\multicolumn{6}{l}{$H_0 = 70$ km s$^{-1}$, $\Omega_M = 0.27$ and $\Omega_{vac}$ = 0.73.}\\
\multicolumn{6}{l}{$^{1}$Masses based on velocity dispersion:}\\
\multicolumn{6}{l}{M$_{BH}$ = 10$^{8.13}$($\sigma$/200 km s$^{-1}$)$^{4.02}$ M$_{\sun}$ \citep{m-sigma}.}
 \end{tabular}
\end{table}

\begin{table}[h]
\small
\caption{\label{12m_sample}12$\mu$m Sample}
\begin{tabular}{lcrcrcc}
\hline
\hline\noalign{\smallskip}
Galaxy & RA & \multicolumn{1}{c}{Dec} & \textit{z} & Distance  & M$_{BH}$$^1$ & Ref$^{2}$ \\
       & (J2000) & \multicolumn{1}{c}{(J2000)} & & Mpc & M$_{\sun}$ \\
\hline\noalign{\smallskip}

IRAS 00198-7926 & 00 21 53.6 & -79 10 07 & 0.073 & 325.8 & 8.44 & B07 \\
NGC 0424        & 01 11 27.6 & -38 05 00 & 0.012 & 51.2  & 7.78 & B07 \\
NGC 1068        & 02 42 40.7 & -00 00 48 & 0.004 & 16.9  & 7.35 & B07 \\
NGC 1143/4      & 02 55 09.7 & -00 10 40 & 0.028 & 120.8 & 8.29 & B07 \\
NGC 1320        & 03 24 48.7 & -03 02 32 & 0.009 & 38.3  & 7.18 & B07 \\
NGC 1386        & 03 36 46.2 & -35 59 57 & 0.003 & 12.7  & 7.24 & B07 \\
F04385-0828     & 04 40 54.9 & -08 22 22 & 0.015 & 64.1  & 8.77 & W07 \\
NGC 1667        & 04 48 37.1 & -06 19 12 & 0.015 & 64.1  & 7.88 & B07 \\
F05189-2524     & 05 21 01.5 & -25 21 45 & 0.043 & 187.7 & 7.86 & W07 \\
F08572+3915     & 09 00 25.4 &  39 03 54 & 0.058 & 256.0 & 7.95 & V09 \\
NGC 3982        & 11 56 28.1 &  55 07 31 & 0.004 & 16.9  & 6.09 & B07 \\
NGC 4388        & 12 25 46.7 &  12 39 44 & 0.008 & 34.0  & 7.22 & B07 \\
NGC 4501        & 12 31 59.2 &  14 25 14 & 0.008 & 34.0  & 7.86 & B07 \\
TOLOLO 1238-364 & 12 40 52.8 & -36 45 21 & 0.011 & 46.9  & 6.84 & W07 \\
NGC 4968        & 13 07 06.0 & -23 40 37 & 0.010 & 42.6  & 7.27 & P06 \\
MCG-3-34-64     & 13 22 24.4 & -16 43 43 & 0.017 & 72.7  & 7.70 & P06 \\
NGC 5135        & 13 25 44.0 & -29 50 01 & 0.014 & 59.8  & 7.35 & B07 \\
NGC 5194        & 13 29 52.7 &  47 11 43 & 0.002 & 8.5   & 6.95 & B07 \\
NGC 5347        & 13 53 17.8 &  33 29 27 & 0.008 & 34.0  & 6.79 & B07 \\
Mrk 463         & 13 56 02.9 &  18 22 19 & 0.050 & 219.4 & 7.88 & B07 \\
NGC 5506        & 14 13 14.9 & -03 12 27 & 0.006 & 25.5  & 6.65 & B07 \\
NGC 5929        & 15 26 06.1 &  41 40 14 & 0.008 & 34.0  & 7.25 & B07 \\
NGC 5953        & 15 34 32.4 &  15 11 38 & 0.007 & 29.7  & 9.80 & P06 \\
Arp 220         & 15 34 57.1 &  23 30 11 & 0.018 & 77.1  & 7.78 & V09 \\
NGC 6890        & 20 18 18.1 & -44 48 25 & 0.008 & 34.0  & 6.48 & B07 \\
IC 5063         & 20 52 02.3 & -57 04 08 & 0.011 & 46.9  & 7.74 & B07 \\
NCG 7130        & 21 48 19.5 & -34 57 05 & 0.016 & 68.4  & 7.53 & B07 \\
NGC 7172        & 22 02 01.9 & -31 52 11 & 0.009 & 38.3  & 7.67 & B07 \\
NGC 7582        & 23 18 23.5 & -42 22 14 & 0.005 & 21.2  & 7.25 & B07 \\
NGC 7590        & 23 18 54.8 & -42 14 21 & 0.005 & 21.2  & 6.79 & B07 \\
NGC 7674        & 23 27 56.7 &  08 46 45 & 0.029 & 125.3 & 7.56 & B07 \\

\hline
\hline
\multicolumn{7}{l}{Distances based on redshift, assuming a cosmology of}\\
\multicolumn{7}{l}{$H_0 = 70$ km s$^{-1}$, $\Omega_M = 0.27$ and $\Omega_{vac}$ = 0.73.}\\
\multicolumn{7}{l}{Coordinates and redshift values from NED.}\\
\multicolumn{7}{l}{$^{1}$Masses based on velocity dispersion: M$_{BH}$ = 10$^{8.13}$($\sigma$/200 km s$^{-1}$)$^{4.02}$ M$_{\sun}$ \citep{m-sigma},}\\
\multicolumn{7}{l}{except for F08572+3915 which is based on photometry of the host galaxy (see V09 for details).}\\
\multicolumn{7}{l}{The FWHM of the [OIII] line was used as a proxy for $\sigma$ for F04385-0828,}\\
\multicolumn{7}{l}{F05189-2524 and TOLOLO 1238-364 (Wang \& Zhang 2007, Greene \& Ho 2005).}\\
\multicolumn{7}{l}{$^{2}$References: B07 - Bian \& Gu 2007; P06 - Peng et al. 2006; }\\
\multicolumn{7}{l}{V09 - Veilleux et al. 2009; W07 - Wang \& Zhang 2007.}
 \end{tabular}
\end{table}

\clearpage

\begin{table}[h]
\small
\caption{\label{o3_opt}Optical Emission Line Fluxes(10$^{-14}$ erg s$^{-1}$ cm$^{-2}$) and Ratios for [OIII] Sample}
\begin{tabular}{lccccc}
\hline
\hline\noalign{\smallskip}
Galaxy & [OIII]$_{obs}$ & [OIII]$_{corr}$ &  H$\alpha$/H$\beta$ & [NII]/H$\alpha$ & [OIII]/H$\beta$ \\
\hline
NCG 0291                    & 6.47 & 30.4   & 5.06 & 0.70 & 7.94 \\ 
Mrk 0609                    & 7.54 & 31.0   & 4.85 & 0.93 & 5.71 \\ 
IC 0486                     & 4.59 & 17.1   & 4.70 & 1.17 & 9.34 \\ 
2MASX J08035923+2345201     & 5.22 & 6.45   & 3.32 & 1.02 & 10.1 \\ 
2MASX J08244333+2959238     & 8.76 & 19.31  & 3.98 & 0.37 & 9.30 \\ 
CGCG 064-017                & 8.46 & 12.33  & 3.49 & 0.24 & 6.63 \\ 
2MASX J10181928+3722419     & 4.05 & 6.01   & 3.51 & 0.84 & 11.2 \\ 
2MASX J11110693+0228477     & 4.04 & 5.45   & 3.41 & 0.91 & 9.31 \\ 
CGCG 242-028                & 4.50 & 7.92   & 3.71 & 0.75 & 7.66 \\ 
SBS 1133+572                & 4.71 & 27.3   & 5.41 & 0.90 & 6.28 \\ 
Mrk 1457                    & 4.27 & 16.2   & 4.72 & 0.78 & 4.63 \\ 
2MASX J11570483+5249036     & 4.95 & 7.17   & 3.49 & 0.54 & 11.1 \\ 
2MASX J12183945+4706275     & 4.94 & 10.4   & 3.93 & 0.21 & 10.6 \\ 
2MASX J12384342+0927362     & 5.48 & 10.2   & 3.77 & 0.50 & 11.3 \\ 
CGCG 218-007                & 4.12 & 21.3   & 5.22 & 0.96 & 6.24 \\ 
2MASX J13463217+6423247     & 4.15 & 9.60   & 4.04 & 1.24 & 10.1 \\ 
NGC 5695                    & 5.93 & 9.30   & 3.57 & 1.36 & 9.56 \\ 
2MASX J15522564+2753435     & 5.87 & 7.69   & 3.38 & 0.57 & 12.4 \\ 
SBS 1609+490                & 4.68 & 8.22   & 3.71 & 0.86 & 10.6 \\ 
2MASX J16164729+3716209     & 5.03 & 7.21   & 3.47 & 0.35 & 10.3 \\ 

\hline
\hline
\end{tabular}
\end{table}

\clearpage
\begin{landscape}
\begin{deluxetable}{lcccccccc}
\small
\tablewidth{0pt}
\tablecaption{\label{12m_opt}Optical Emission Line Fluxes (10$^{-14}$ erg s$^{-1}$ cm$^{-2}$) and Ratios for 12$\mu$m Sample}
\tablehead{
\colhead{Object} &  \colhead{[OIII]$_{obs}$} & \colhead{[OIII]$_{corr}$} &
 \colhead{H$\alpha$/H$\beta$} & \colhead{Ref\tablenotemark{1}} & 
\colhead{[NII]/H$\alpha$} & \colhead{Ref} & \colhead{[OIII]/H$\beta$}  & \colhead{Ref} }

\startdata

IRAS 00198-7926  & 3.88  & 38.5  & 6.41 & D92 & 0.63 & V93 & 11.5 & V93 \\

NGC 0424         & 70.0  & 121   & 3.69 & V97 & 0.30 & V97 & 4.55 & V97 \\

NGC 1068         & 1367 & 21280  & 7.39 & S95 & 1.80 & S95 & 13.0 & S95 \\

NGC 1144         & 4.30 & 28.7   & 5.65 & D88 & 1.73 & M06 & 7.10 & D88 \\

NGC 1320         & 14.0 & 58.0   & 4.86 & T03 & 0.60 & D92 & 8.98 & D92 \\

NGC 1386         & 79.2 & 1116   & 7.16 & S89 & 1.28 & S89 & 16.1 & S89 \\

F04385-0828      & 0.35 & 10.9   & 9.20 & T03 & $>$0.65 & D92 & $>$2.15 & D92  \\

NGC 1667         & 6.38 & 238    & 9.74 & S95 & 0.69 & S95 & 3.48 & S95 \\

F05189-2524      & 15.3 & 348    & 8.33 & V99 & 0.46 & V99 & 44.0 & V99 \\

F08572+3915      & 0.12 & 1.41   & 6.71 & SDSS & 0.40 & SDSS & 0.97 & SDSS \\

NGC 3982         & 20.0 & 64.9   & 4.50 &  T03 & ... & ...  & 17.4 & T03 \\

NGC 4388         & 80.0 & 557    & 5.73 & D92  & 2.47 & D92 & 11.8 & D92 \\

NGC 4501         & 3.70 & 5.48   & 3.51 & T03 &  1.86 & S82 & 5.82 & S82 \\

TOLOLO 1238-364  & 45.7 & 289    & 5.56 & S95 &  0.69 & S95 & 5.31 & S95\\

NGC 4968         & 17.7 & 1255   & 11.9 & D92 &  1.01 & D92 & 21.5 & D92 \\

M-3-34-64        & 151  & 3190   & 8.14 & D92 & 0.99 & D92 & 19.4 & D92 \\

NGC 5135         & 37.5 & 738    & 7.96 & S95 & 0.87 & S95 & 4.30 & S95 \\

NGC 5194         & 5.14 &  45.0  & 6.16 & M06 &  2.40 & M06 & 6.43 & M06 \\

NGC 5347         & 4.43 & 16.2   & 4.67 & SDSS & 0.66 & SDSS & 7.79 & SDSS \\

Mrk 463          & 56.3 & 150.3  & 4.23 & D92 &  0.40 & D92 & 9.01 & D92 \\

NGC 5506         & 127  & 959    & 5.88 & W10 &  0.70 & W10 & 7.29 & W10 \\

NGC 5929         & 6.92 & 21.6   & 4.44 & SDSS & 0.59 & SDSS & 3.42 & SDSS \\

NGC 5953         & 4.73 & 80.1   & 7.59 & V95  & 0.78 & V95  & 2.00 & V95 \\

Arp 220          & 0.18 & 33.7   & 16.2 & SDSS & 1.41 & SDSS & 2.33 & SDSS \\

NGC 6890         & 23.3 & 86.1   & 4.69 & S90 & 0.76 & S90  & 10.2 & S90 \\

I5063            & 175  & 224    & 3.35 & S90 &  0.54 & S90  & 9.71 & S90 \\

NGC 7130         & 34.1 & 668    & 7.95 & S95 & 0.75 & S95  & 4.84 & S95 \\

NGC 7172         & 3.70 & 214    & 11.2 & V97 & 0.30 & V97  & 10.0 & V97 \\

NGC 7582         & 51.3 & 385    & 5.87 & S95 &  0.78 & S95 & 3.21 & S95 \\

NGC 7590         & 2.91 & 17.3   & 5.45 & S95 & 0.65 & S95 & 1.07 & S95 \\

NGC 7674         & 71.8 & 366    & 5.19 & D92 & 0.64 & D92 & 11.3 & D92 \\

\enddata

\tablenotetext{1}{Optical References: \\
D88: Dahari \& De Robertis, 1988; D92: de Grijp et al. 1992; M06: Moustakas \& Kennicutt 2006;  
S89: Storchi-Bergmann et al. 1989; S90: Storchi-Bergman et al. 1990; S95: Storchi-Bergmann et al. 1995; 
 T03: Tran 2003; V97: Vaceli et al 1997, V95: Veilleux et al 1995; V99: Veilleux et al. 1999;
W10: Winter et al. 2010 }

\end{deluxetable}
\end{landscape}
\clearpage

\clearpage
\begin{deluxetable}{lcccc}
\small
\tablewidth{0pt}
\tablecaption{\label{Spitzer_info}{\it Spitzer} Observation Summary\tablenotemark{1}}
\tablehead{
\colhead{Galaxy} & \colhead{Program ID} & \colhead{Mode} & \multicolumn{2}{c}{Extraction Area\tablenotemark{2}} \\
\cline{4-5}
\colhead{} & \colhead{} & \colhead{} & \colhead{''} & \colhead{kpc}}

\startdata

IRAS 00198-7926  & 30291\tablenotemark{b} & Staring & ... & ... \\

NGC 0424         & 30291\tablenotemark{b} & Staring & ... & ... \\
                 & 03269 & Mapping & 15.3$\times$20.4 & 3.5$\times$4.6 \\

NGC 1068         & 00014 & Staring & ... & ... \\

NGC 1144         & 00021\tablenotemark{b} & Staring & ... & ... \\
                 & 00021 & Mapping & 20.4$\times$20.4 & 11.3$\times$11.3\\

NGC 1320         & 30291 & Staring & ... & ... \\
                 & 03269 & Mapping & 15.3$\times$20.4 & 2.6$\times$3.5 \\

NGC 1386         & 00086 & Staring & ... & ... \\
                 & 30572\tablenotemark{b} & Staring & ... & ...  \\
                 
F04385-0828      & 30291\tablenotemark{b} & Staring & ... & ... \\
                 & 03269 & Mapping & 15.3$\times$20.4 & 4.7$\times$6.2 \\

NGC 1667         & 30291\tablenotemark{b} & Staring & ... & ... \\
                 & 03269 & Mapping & 15.3$\times$20.4 & 4.7$\times$6.3 \\

F05189-2524      & 00105 & Staring & ... & .. \\

F08572+3915      & 00105 & Staring & ... & ... \\

NGC 3982         & 30291\tablenotemark{b} & Staring & ... & ... \\
                 & 03269 & Mapping & 15.3$\times$20.4 & 1.3$\times$1.8 \\

NGC 4388         & 00086 & Staring & ... & ... \\
                 & 30572\tablenotemark{b} & Staring & ... & ... \\

NGC 4501         & 30291\tablenotemark{b} & Staring & ... & ... \\
                 & 03269 & Mapping & 15.3$\times$20.4 & 2.7$\times$3.7 \\

TOLOLO 1238-364  & 30291\tablenotemark{b} & Staring & ... & ...  \\
                 & 03269 & Mapping & 15.3$\times$20.4 & 3.9$\times$5.2 \\

NGC 4968         & 30291\tablenotemark{b} & Staring & ... & ... \\
                 & 03269 & Mapping & 15.3$\times$20.4 & 3.4$\times$4.6 \\

M-3-34-64        & 30323 & Staring & ... & ... \\

NGC 5135         & 00086 & Staring & ... & ... \\
                 & 30572\tablenotemark{b} & Staring & ... & ... \\

NGC 5194         & 00159 & Mapping & 20.4$\times$35.7 & 0.90$\times$1.6 \\

NGC 5347         & 00086 & Staring & ... & ... \\
                 & 30572\tablenotemark{b} & Staring & ... & ... \\

Mrk 463          & 00105 & Staring & ... & ... \\

NGC 5506         & 00086 & Staring & ... & ... \\
                 & 30572\tablenotemark{b} & Staring & ... & ... \\

NGC 5929         & 20140\tablenotemark{b} & Staring & ... & ... \\
                 & 03269 & Mapping & 15.3$\times$20.4 & 2.7$\times$3.6 \\

NGC 5953         & 00059 & Staring & ... & ... \\
                 & 03269 & Mapping & 15.3$\times$20.4 & 2.2$\times$2.9 \\

Arp 220          & 00105 & Staring & ... & ... \\

NGC 6890         & 30291\tablenotemark{b} & Staring & ... & ... \\
                 & 03269 & Mapping & 15.3$\times$20.4 & 2.4$\times$3.2 \\

IC 5063          & 00086 & Staring & ... & ... \\
                 & 30572\tablenotemark{b} & Staring & ... & ... \\

NGC 7130         & 00086 & Staring & ... & ... \\
                 & 30572\tablenotemark{b} & Staring & ... & ... \\

NGC 7172         & 00086 & Staring & ... & ... \\
                 & 30572\tablenotemark{b} & Staring & ... & ... \\

NGC 7582         & 00024 & Staring & ... & ... \\
                 & 40936\tablenotemark{b} & Staring & ... & ... \\

NGC 7590         & 30291\tablenotemark{b} & Staring & ... & ... \\
                 & 03269 & Mapping & 15.3$\times$20.4 & 1.4$\times$1.9 \\

NGC 7674         & 03605 & Staring & ... & ... \\
                 & 03269 & Mapping & 15.3$\times$20.4 & 8.5$\times$11.4 \\

\enddata

\tablenotetext{1}{The Program ID for the Sy2s in the [OIII] sample is 30773.}
\tablenotetext{2}{For low resolution spectral mapping observations.}
\tablenotetext{b}{Galaxies that had dedicated off-source high-resolution observations and were therefore background subtracted. See Section 3.2 for details.}
\end{deluxetable}

\begin{table}[h]
\footnotesize
\caption{\label{oiii_ir} MIR Flux and PAH EW values for [OIII] Sample}
\begin{tabular}{lcccccc}
\hline

\hline\noalign{\smallskip}

Galaxy &[NeII]12.81$\mu$m & [OIV]25.89$\mu$m &  F$_{MIR}$ & \multicolumn{2}{c}{EW} & $\alpha_{20-30\mu m}$\\
\cline{5-6} & & & & 11.3$\mu$m & 17$\mu$m \\
  & \multicolumn{2}{c}{10$^{-14}$ erg s$^{-1}$ cm$^{-2}$} & 10$^{-11}$ erg s$^{-1}$ cm$^{-2}$ 
  & $\mu$m & $\mu$m   \\

\hline\noalign{\smallskip}

NCG 0291                  & 20.6$\pm$1.2  & 26.9$\pm$0.7  & 4.87 &  0.823 & 0.471 & 2.76   \\ 
Mrk 0609                  & 19.9$\pm$3.5  & 7.82$\pm$1.20 & 5.64 &  1.021 & 0.831 & 2.22   \\ 
IC 0486                   & 4.20$\pm$0.83 & 11.6$\pm$0.7  & 4.59 &  0.139 & 0.147 & 1.51   \\ 
2MASX J08035923+2345201   & 0.96$\pm$0.71 & 3.18$\pm$0.51 & 0.63 &  0.232 & 0.552 & 0.44   \\ 
2MASX J08244333+2959238   & 3.07$\pm$1.08 & 9.87$\pm$1.08 & 8.36 &  0.044 & 0.094 & 0.74   \\ 
CGCG 064-017              & 2.52$\pm$0.27 & 4.44$\pm$2.59 & 0.73 &  0.458 & 0.373 & 1.56   \\ 
2MASX J10181928+3722419   & 0.71$\pm$0.69 & 2.82$\pm$1.4  & 1.10 &  0.163 & 0.107 & 1.22   \\ 
2MASX J11110693+0228477   & 1.15$\pm$0.68 & 2.83$\pm$1.11 & 1.56 &  0.100 & 0.272 & 0.14   \\ 
CGCG 242-028              & 1.59$\pm$0.77 & 2.14$\pm$0.37 & 0.39 &  0.110 & 0.675 & 0.26   \\ 
SBS 1133+572              & 13.0$\pm$0.9  & 16.1$\pm$3.8  & 9.63 &  0.135 & 0.131 & 1.57   \\ 
Mrk 1457                  & 9.87$\pm$0.89 & 3.22$\pm$0.78 & 2.93 &  0.631 & 0.421 & 1.92   \\ 
2MASX J11570483+5249036   & 0.91$\pm$0.67 & 5.53$\pm$0.93 & 2.33 &  0.054 & 0.115 & 1.52   \\ 
2MASX J12183945+4706275   & $<$1.23       & 4.31$\pm$0.18 & 1.32 &  0.048 & 0.018 & 0.76   \\ 
2MASX J12384342+0927362   & $<$2.25       & 6.46$\pm$1.13 & 3.02 &  0.007 & 0.090 & 0.34   \\ 
CGCG 218-007              & 10.7$\pm$0.7  & 17.0$\pm$0.8  & 2.95 &  0.887 & 0.435 & 2.66   \\ 
2MASX J13463217+6423247   & 2.10$\pm$1.19 & 2.02$\pm$0.47 & 0.86 &  0.266 & 0.183 & 1.50   \\ 
NGC 5695                  & 2.94$\pm$0.58 & 7.64$\pm$0.20 & 0.87 &  0.459 & 0.666 & 1.83   \\ 
2MASX J15522564+2753435   & -             & 8.04$\pm$1.21 & 1.44 &  0.019 & 0.059 & 0.86   \\ 
SBS 1609+490              & $<$1.53       & 8.45$\pm$0.84 & 2.35 &  0.037 & 0.098 & 0.60   \\ 
2MASX J16164729+3716209   & 1.78$\pm$0.79 & 5.11$\pm$0.54 & 1.73 &  0.051 & 0.066 & 1.38   \\ 

\hline
\hline
 \end{tabular}
\end{table}

\clearpage

\begin{table}[h]
\caption{\label{12m_ir} MIR Flux and PAH EW values for 12$\mu$m Sample}
\begin{tabular}{lcccccc}
\hline

\hline\noalign{\smallskip}

Galaxy & [NeII]12.81$\mu$m & [OIV]25.89$\mu$m &  F$_{MIR}$ & \multicolumn{2}{c}{EW}  & $\alpha_{20-30\mu m}$ \\
\cline{5-6} & & & & 11.3$\mu$m &17$\mu$m   \\
  & \multicolumn{2}{c}{10$^{-14}$ erg s$^{-1}$ cm$^{-2}$} & 10$^{-11}$ erg s$^{-1}$ cm$^{-2}$ 
  & $\mu$m & $\mu$m   \\

\hline\noalign{\smallskip}

IRAS 00198-7926 & 2.57$^{+0.19}_{-0.18}$ & 25.7$^{+2.7}_{-2.4}$   & -     & -     & -     & -     \\
NGC 0424        & 9.23$^{+0.70}_{-0.66}$ & 25.2$^{+3.3}_{-3.0}$   & 18.2  & 0.058 & 0.130 & -0.08 \\
NGC 1068        & 149$^{+19}_{-17}$      & 1763$^{+221}_{-170}$   & -     & -     & -     & -     \\
NGC 1144        & 6.00$^{+0.61}_{-0.59}$ & 4.91$^{+0.98}_{-0.70}$ & 4.28  & 0     & 0.347 & 1.92  \\
NGC 1320        & 2.51$^{+0.26}_{-0.20}$ & 21.6$^{+1.0}_{-0.9}$   & 8.09  & 0.135 & 0.117 & 0.51  \\
NGC 1386$^{1}$  & 4.74$^{+0.66}_{-0.59}$ & 82.0$^{+7.0}_{-6.6}$   & 6.59  & 0.131 & 0.239 & 1.57  \\
F04385-0828     & 4.28$^{+0.42}_{-0.40}$ & 8.24$^{+1.95}_{-1.50}$ & 10.3  & 0     & 0.131 & 1.00  \\
NGC 1667        & 3.30$^{+0.30}_{-0.30}$ & 2.87$^{+0.36}_{-0.35}$ & 2.76  & 0.578 & 1.35  & 1.45  \\
F05189-2524     & 7.89$^{+0.65}_{-0.60}$ & $<$69                  & 7.98  & 0     & 0.129 & 2.74  \\
F08572+3915     & 2.45$^{+0.44}_{-0.37}$ & $<$33                  & 4.19  & 0     & 1.85  & 4.13  \\
NGC 3982        & 3.10$^{+0.17}_{-0.17}$ & 4.02$^{+0.52}_{-0.53}$ & 2.39  & 1.63  & 0.933 & 1.53  \\
NGC 4388$^{1}$  & 24.6$^{+1.4}_{-1.4}$   &  262$^{+6}_{-6}$       & 6.44  & 0.083 & 0.187 & 1.79  \\
NGC 4501        & 1.63$^{+0.13}_{-0.12}$ & 2.71$^{+0.23}_{-0.21}$ & 1.41  & 0.947 & 1.86  & 1.35  \\
TOLOLO 1238-364 & 13.9$^{+0.4}_{-0.4}$   & 14.9$^{+4.2}_{-2.7}$   & 11.7  & 0.253 & 0.065 & 0.82  \\
NGC 4968        & 8.08$^{+0.46}_{-0.46}$ & 26.3$^{+2.0}_{-1.9}$   & 7.78  & 0.174 & 0.105 & 0.12  \\
M-3-34-64       & 16.9$^{+0.80}_{-0.75}$ & 91.1$^{+11.9}_{-9.9}$  & 11.7  & 0.010 & 0.035 & 0.85  \\
NGC 5135$^{2}$  & 32.6$^{+3.7}_{-3.6}$   & 56.7$^{+10.4}_{-11.0}$ & 4.16  & 0.387 & 0.322 & 2.46  \\
NGC 5194        & 22.9$^{+0.8}_{-0.8}$   & 27.2$^{+2.4}_{-1.9}$   & 8.07  & 1.71  & 1.63  & 2.06  \\
NGC 5347$^{3}$  & 1.60$^{+0.24}_{-0.17}$ & 6.39$^{+1.53}_{-1.08}$ & 4.04  & 0.086 & 0.060 & 0.54  \\
Mrk 463         & 2.99$^{+0.24}_{-0.23}$ & 54.9$^{+3.3}_{-3.1}$   & 6.72  & 0     & 0.016 & 0.58  \\
NGC 5506$^{1}$  & 21.0$^{+1.4}_{-1.3}$   & 216$^{+16}_{-16}$      & 17.1  & 0.048 & 0.257 & 1.37  \\
NGC 5929        & 3.08$^{+0.09}_{-0.12}$ & 4.38$^{+0.25}_{-0.24}$ & 0.58  & 1.15  & 0.792 & 1.61  \\
NGC 5953        & 17.8$^{+0.8}_{-0.9}$   & 15.8$^{+1.2}_{-1.1}$   & 4.17  & 2.28  & 1.82  & 2.55  \\
Arp 220         & 21.8$^{+1.0}_{-1.0}$   & $<$88                  & 7.45  & 0     & 5.512 & 6.27  \\
NGC 6890        & 3.48$^{+0.20}_{-0.20}$ & 8.07$^{+0.59}_{-0.56}$ & 4.05  & 0.380 & 0.357 & 0.91  \\
IC 5063         & 7.81$^{+0.84}_{-0.80}$ & 97.0$^{+8.1}_{-7.2}$   & 14.2  & 0.009 & 0.063 & 0.97  \\
NGC 7130$^{1}$  & 22.3$^{+1.2}_{-1.3}$   & 15.8$^{+2.0}_{-1.8}$   & 4.46  & 0.138 & 0.113 & 2.51  \\
NGC 7172        & 11.2$^{+1.6}_{-1.3}$   & 34.9$^{+1.7}_{-1.6}$   & 3.13  & 0.170 & 2.31  & 2.53  \\
NGC 7582        & 77.9$^{+4.6}_{-4.7}$   & 87.1$^{+4.5}_{-4.4}$   & 5.06  & 1.51  & 0.275 & 2.70  \\
NGC 7590        & 2.42$^{+0.14}_{-0.16}$ & 3.61$^{+0.43}_{-0.43}$ & 1.79  & 1.95  & 1.95  & 1.80  \\
NGC 7674        & 6.94$^{+0.49}_{-0.45}$ & 37.2$^{+4.0}_{-3.5}$   & 11.1  & 0.012 & 0.103 & 0.64  \\

\hline
\hline 
\multicolumn{6}{l}{$^1$Program ID 30572 only for F$_{MIR}$.}\\ 
\multicolumn{6}{l}{$^2$Program ID 00086 only for F$_{MIR}$. }\\ 
\multicolumn{6}{l}{$^3$Program IDs 30572 and 00086 averaged together for F$_{MIR}$.}
\end{tabular}
\end{table}

\clearpage

\begin{table}[h]
\small
\caption{\label{diag_results1} Diagnostic Ratios: Optical \& Infrared}
\begin{tabular}{llcc}
\hline
\hline\noalign{\smallskip}
Diagnostic Ratio & \multicolumn{1}{c}{Sample} & Mean & $\sigma$ \\
                 &        & dex  & dex      \\
\hline\noalign{\smallskip}
F$_{[OIII],corr}$/F$_{[OIII],obs}$ & Both (51 Sy2s)          & 0.73 & 0.52 \\
                                   & [OIII]-only (20 Sy2s)   & 0.34 & 0.22 \\
                                   & 12$\mu$m-only (31 Sy2s) & 0.97 & 0.51 \\
                                   & Sy2s (Winter et al. 2010, 24) & 0.43 & 0.49 \\
                                   & Sy1s (Winter et al. 2010, 29) & 0.28 & 0.48 \\

F$_{[OIV]}$/F$_{[OIII],corr}$      & Both (48 Sy2s)          & -0.59 & 0.48 \\
                                   & [OIII]-only (20 Sy2s)   & -0.28 & 0.22 \\
                                   & 12$\mu$m-only (28 Sy2s) & -0.82 & 0.49 \\
                                   & Sy2s$^1$ (8)            & -0.38 & 0.56 \\
                                   & Sy1s$^{1}$ (10)         & -0.90 & 0.58 \\

F$_{[OIV]}$/F$_{[OIII],obs}$       & Both (48 Sy2s)          & 0.08 & 0.41 \\
                                   & [OIII]-only (20 Sy2s)   & 0.06 & 0.29 \\
                                   & 12$\mu$m-only (28 Sy2s) & 0.09 & 0.48 \\
                                   & Sy2s (Mel\'endez et al. 2008, 12)     & 0.60 & 0.74 \\
                                   & Sy2s (Diamond-Stanic et al. 2009, 56) & 0.57 & 0.67 \\
                                   & Sy1s (Mel\'endez et al. 2008, 18)     & -0.21 & 0.42 \\
                                   & Sy1s (Diamond-Stanic et al. 2009, 16) & -0.07 & 0.53 \\

F$_{MIR}$/F$_{[OIII],corr}$        & Both (49 Sy2s)          & 1.89 & 0.63 \\
                                   & [OIII]-only (20 Sy2s)   & 2.24 & 0.26 \\
                                   & 12$\mu$m-only (29 Sy2s) & 1.66 & 0.70 \\

F$_{MIR}$/F$_{[OIII],obs}$         & Both (49 Sy2s)          & 2.61 & 0.63 \\
                                   & [OIII]-only (20 Sy2s)   & 2.58 & 0.37 \\
                                   & 12$\mu$m-only (29 Sy2s) & 2.62 & 0.77 \\
                                   & Sy1s$^2$ (12)           & 2.56 & 0.50 \\
                         
F$_{MIR}$/F$_{[OIV]}$              & Both (46 Sy2s)          & 2.46 & 0.38 \\
                                   & [OIII]-only (20 Sy2s)   & 2.52 & 0.26 \\
                                   & 12$\mu$m-only (26 Sy2s) & 2.41 & 0.44 \\
                                   & Sy1s$^3$ (24)           & 2.59 & 0.38 \\

\hline
\hline
\multicolumn{4}{l}{\normalsize\textbf{Ratios in log space}}\\
\multicolumn{4}{l}{$^1$Sy1 and Sy2 [OIII] values from Winter et al. (2010) and [OIV] values from Weaver et al. (2010)}\\
\multicolumn{4}{l}{$^2$Sy1 MIR values from Deo et al. (2009) and [OIII] values from Mel\'endez et al. (2008)}\\
\multicolumn{4}{l}{$^3$Sy1 MIR values from Deo et al. (2009) and [OIV] values from Tommasin et al. (2010) \& }\\
\multicolumn{4}{l}{Mel\'endez et al. (2008)}

 \end{tabular}
\end{table}

\begin{table}[h]
\small
\caption{\label{diag_results2} Diagnostic Ratios: Optical \& Infrared vs Radio$^1$ \& X-ray$^{2}$}
\begin{tabular}{llcc}
\hline
\hline\noalign{\smallskip}
Diagnostic Ratio & \multicolumn{1}{c}{Sample} & Mean & $\sigma$ \\
                 &        & dex  & dex      \\
\hline\noalign{\smallskip}

F$_{[OIII],corr}$/F$_{8.4GHz}$     & 12$\mu$m-only (26 Sy2s) & 3.56 & 0.87 \\
F$_{[OIII],obs}$/F$_{8.4GHz}$      & 12$\mu$m-only (26 Sy2s) & 2.53 & 0.99 \\
                                   & Sy1s (12)               & 2.94 & 0.93 \\

F$_{[OIV]}$/F$_{8.4GHz}$           & 12$\mu$m-only (23 Sy2s) & 2.83 & 0.64 \\
                                   & Sy1s (25)               & 2.84 & 0.90 \\

F$_{MIR}$/F$_{8.4GHz}$             & 12$\mu$m-only (25 Sy2s) & 5.15 & 0.66 \\
                                   & Sy1s (22)               & 5.41 & 0.78 \\

F$_{14-195keV}$/F$_{[OIII],corr}$  & Both          (51 Sy2s) & 0.50  & ... \\        
                                   & [OIII]-only   (20 Sy2s) & 1.79  & ... \\
                                   & 12$\mu$m-only (31 Sy2s) & 0.48  & ... \\
                                   & Sy2s (Winter et al. 2010, 24) & 2.30 & 0.68 \\
                                   & Sy1s (Winter et al. 2010, 29) & 1.74 & 0.54 \\

F$_{14-195keV}$/F$_{[OIII],obs}$   & Both          (51 Sy2s) & 1.48  & ... \\
                                   & [OIII]-only   (20 Sy2s) & 2.37  & ... \\
                                   & 12$\mu$m-only (31 Sy2s) & 1.53  & ... \\ 
                                   & Sy2s (Winter et al. 2010, 24)  & 2.73  & 0.57 \\
                                   & Sy1s (Winter et al. 2010, 29)  & 2.02  & 0.54 \\

F$_{14-195keV}$/F$_{[OIV]}$        & Both          (48 Sy2s) & 1.53  & ... \\
                                   & [OIII]-only   (20 Sy2s) & 2.14  & ... \\
                                   & 12$\mu$m-only (28 Sy2s) & 1.54  & ... \\
                                   & Sy2s (Weaver et al. 2010, 33) & 2.30 & 0.47 \\
                                   & Sy1s (Weaver et al. 2010, 37) & 2.60 & 0.47 \\
                                   & Sy1s (Rigby et al. 2009, 17) & 2.68 & 0.39 \\

F$_{14-195keV}$/F$_{MIR}$          & Both          (49 Sy2s) & -0.78 & ... \\
                                   & [OIII]-only   (20 Sy2s) & -0.47 & ... \\
                                   & 12$\mu$m-only (29 Sy2s) & -0.72 & ... \\
                                   & Sy1s$^3$ (21)           & -0.28 & 0.26 \\

F$_{14-195keV}$/F$_{8.4GHz}$       & 12$\mu$m-only (26 Sy2s) & 4.27  & ... \\
                                   & Sy1s (21)               & 5.30  & 0.61 \\
\hline
\hline
\multicolumn{4}{l}{\normalsize\textbf{Ratios in log space}}\\
\multicolumn{4}{l}{$^1$Sy1 radio data from Thean et al. (2000) with [OIII] and [OIV] data from}\\
\multicolumn{4}{l}{Mel\'endez et al. (2008) and Diamond-Stanic et al. (2009), MIR values from}\\
\multicolumn{4}{l}{Deo et al. (2009) and 14-195 keV data from Tueller et al. (2009) and}\\
\multicolumn{4}{l}{Mel\'endez et al. (2008).}\\
\multicolumn{4}{l}{$^{2}$Mean of Sy2 X-ray flux ratios calculated using survival}\\
\multicolumn{4}{l}{analysis (Kaplan-Meier estimator) to include upper limits.}\\
\multicolumn{4}{l}{The KM-estimator requires at least 2 detections, so one of the}\\
\multicolumn{4}{l}{[OIII]-sample upper limits was converted to a detection which biases}\\
\multicolumn{4}{l}{the mean for this sub-sample.}\\
\multicolumn{4}{l}{$^3$Sy1 MIR data from Deo et al (2009) with 14-195 keV values from}\\
\multicolumn{4}{l}{Mel\'endez et al. (2008) and Tueller et al. (2009).}
 \end{tabular}
\end{table}

\begin{table}[h]
\caption{\label{Kuiper} Results of Two Sample Tests between Sy1s and Sy2s}
\begin{tabular}{lrcccc}
\hline
\hline\noalign{\smallskip}
& & \multicolumn{2}{c}\textbf{Kuiper Test} & \multicolumn{2}{c}\textbf{KS Test}  \\
Diagnostic Ratio & $N_{eff}$ &  D  & Probability & D & Probability \\
\hline\noalign{\smallskip}
F$_{[OIV]}$/F$_{[OIII],obs}$    & 19.9  & 0.310 & 0.223 & 0.301 & 0.042 \\
F$_{MIR}$/F$_{[OIII],obs}$      & 8.49  & 0.274 & 0.912 & 0.160 & 0.949 \\
F$_{MIR}$/F$_{[OIV]}$           & 15.8  & 0.264 & 0.674 & 0.203 & 0.487 \\
F$_{[OIII],obs}$/F$_{8.4GHz}$   & 8.21  & 0.385 & 0.544 & 0.340 & 0.242 \\
F$_{[OIV]}$/F$_{8.4GHz}$        & 12.0  & 0.318 & 0.576 & 0.200 & 0.684 \\
F$_{MIR}$/F$_{8.4GHz}$          & 11.7  & 0.344 & 0.457 & 0.253 & 0.389 \\
\hline
\hline
\multicolumn{4}{l}{$N_{eff} = n_1 \times n_2/(n_1 + n_2)$ where $n_1$ and $n_2$}\\
\multicolumn{4}{l}{are the number of data points in each sample.}
 \end{tabular}
\end{table}

\begin{table}[h]
\caption{\label{lum_dep} Correlation of Diagnostic Ratios with Eddington Parameters}
\begin{tabular}{lcc}
\hline
\hline\noalign{\smallskip}
Diagnostic Ratio & $\rho$ & P$_{uncorr}$ \\
\hline\noalign{\smallskip}
Optical \& Infrared \\
\cline{1-3}\\
F$_{[OIII],corr}$/F$_{[OIII],obs}$ & -0.224 & 0.115 \\

F$_{[OIV]}$/F$_{[OIII],corr}$      & -0.131 & 0.375 \\
F$_{[OIV]}$/F$_{[OIII],obs}$       & -0.351 & 0.015 \\

F$_{MIR}$/F$_{[OIII],corr}$        & -0.164 & 0.259 \\
F$_{MIR}$/F$_{[OIII],obs}$         & -0.376 & 0.008 \\
F$_{MIR}$/F$_{[OIV]}$              & -0.074 & 0.625 \\

\hline
\multicolumn{3}{l}{Radio and X-ray}\\
\cline{1-3}\\
F$_{[OIII],corr}$/F$_{8.4GHz}$     &  0.008 & 0.967 \\
F$_{[OIII],obs}$/F$_{8.4GHz}$      &  0.142 & 0.491 \\
F$_{[OIV]}$/F$_{8.4GHz}$           & -0.220 & 0.314 \\
F$_{MIR}$/F$_{8.4GHz}$             & -0.415 & 0.039 \\

F$_{14-195keV}$/F$_{[OIII],corr}$  &  ...   & 0.329 \\      
F$_{14-195keV}$/F$_{[OIII],obs}$   &  ...   & 0.194 \\
F$_{14-195keV}$/F$_{[OIV]}$        &  ...   & 0.914 \\
F$_{14-195keV}$/F$_{MIR}$          &  ...   & 0.449 \\
F$_{14-195keV}$/F$_{8.4GHz}$       &  ...   & 0.283 \\
\hline
\hline
\multicolumn{3}{l}{X-ray correlations computed using Cox Hazard Model}\\
\multicolumn{3}{l}{in ASURV software package.}\\
 \end{tabular}
\end{table}

\clearpage

\appendix
\section{Aperture Bias}
As the optical data for the 12$\mu$m sample were culled from the literature, we examined the data to see if an aperture bias was evident. The aperture sizes used for the optical data ($\theta_{OIII}$) ranged from 3 to 14'' for the 12$\mu$m sources and was 3'' for all the SDSS Sy2s. For several of our sources, we found the size of the NLR ($\theta_{NLR}$) from Schmitt et al (2003a) and estimated the NLR for the remainder using log $R_{maj}$ = (0.31 $\pm$ 0.04) $\times$ log $L{[OIII]}$ - 10.08 $\pm$ 1.80 (Schmitt et al 2003b). As shown in Figure \ref{ap_nlr} (same color coding as Figure 1 in the main text), the aperture was large enough to encompass the full NLR for all sources. Hence, we are not ``missing'' any of the NLR optical flux.

But are we observing too much [OIII] flux, perhaps from starburst contamination in the host galaxy which would affect [OIII] emission more than [OIV]? If this is the case, we would expect the F$_{[OIV]}$/F$_{[OIII],obs}$ ratio to decrease as the projected size of the aperture increases and the ``PAH-strong'' sources to lie at systematically higher aperture sizes. However, neither of these trends are apparent (Figure \ref{o3_o4_ap}), reaffirming the results in the main text where we find no evidence for starburst bias on [OIII] emission. We also note that the opposite effect is also absent, namely increase of F$_{[OIV]}$/F$_{[OIII],obs}$ with aperture size. This indicates that though the [OIV] flux is collected from the Long-High module which has a larger aperture than the [OIII] data, the [OIV] flux is not produced in regions outside of the NLR. Hence the dispersion present in the flux ratios is not due to sampling the galaxies at different scales between the separate IR and optical observations.

\section{Starburst Contribution to the MIR}
As discussed in the main text, in many cases the Spitzer IRS data show evidence for the presence of both an AGN and active star formation. This implies that the MIR continuum will include emission from warm dust heated by both the AGN and massive stars. Since we are interested in using the MIR luminosity as an indicator of the intrinsic luminosity of the AGN, we would need a way to subtract off the contribution from the dust heated by stars. 

We attempted to make this correction by using a simple dilution model based on the EW of the PAH features bracketing the continuum around 15.5$\mu$m (the PAH EW complexes at 11.3$\mu$m and 17$\mu$m, c.f. Genzel et al. 1998, Wu et al. 2009). This approach assumes that for starbursts there is a simple linear relationship between the PAH luminosity and that of the MIR continuum, and that the only effect of the AGN is to produce an additional source of MIR continuum emission, while not affecting the PAH luminosity.  

As we have shown, the PAH EWs are indeed correlated with other indicators of star formation activity. We therefore compared the average PAH EWs for starburst galaxies ($<EW_{11.3\mu m,SB}>$=2.87 and $<EW_{17\mu m,SB}>$=1.74, O'Dowd et al. 2009; Treyer private communication) to the measured EWs for our Sy2s and used this to derive the fraction of the MIR emission that is attributed to the AGN heating of the torus (i.e. $f_{AGN} = 1 - EW_{Sy2}/<EW_{SB}>$). In most cases, this correction was negligible. However, in the composite (“strong PAH”) AGN, this procedure seemed to subtract too much MIR flux, leading to poorer agreement with the other proxies and strong systematic trends with AGN luminosity (e.g. Figure \ref{raw_mir} as compared with Figure \ref{mir2_oiv}).

Such results indicate that accurately isolating the amount of MIR flux due to the AGN in strongly composite systems requires more detailed modeling that takes account of the possible influence of the AGN on the PAHs themselves. In the tight linear correlation we see between the uncorrected MIR and [OIV] luminosities (Figure 3) there is no evidence that the strong-PAH sources have systematically higher MIR luminosities. Taken at face value this would imply that in these sources the starburst does not dominate the MIR continuum. This would in turn imply that the EW of these long-wavelength PAHs with respect to the {\it pure starburst} MIR continuum is unusually high. This can not be understood as resulting from the destruction of PAHs by the AGN (e.g. Treyer et al. 2010), as the effect is in the opposite direction.


\setcounter{figure}{0}
\renewcommand{\thefigure}{A.\arabic{figure}}

\begin{figure}
\epsscale{0.80}
\plotone{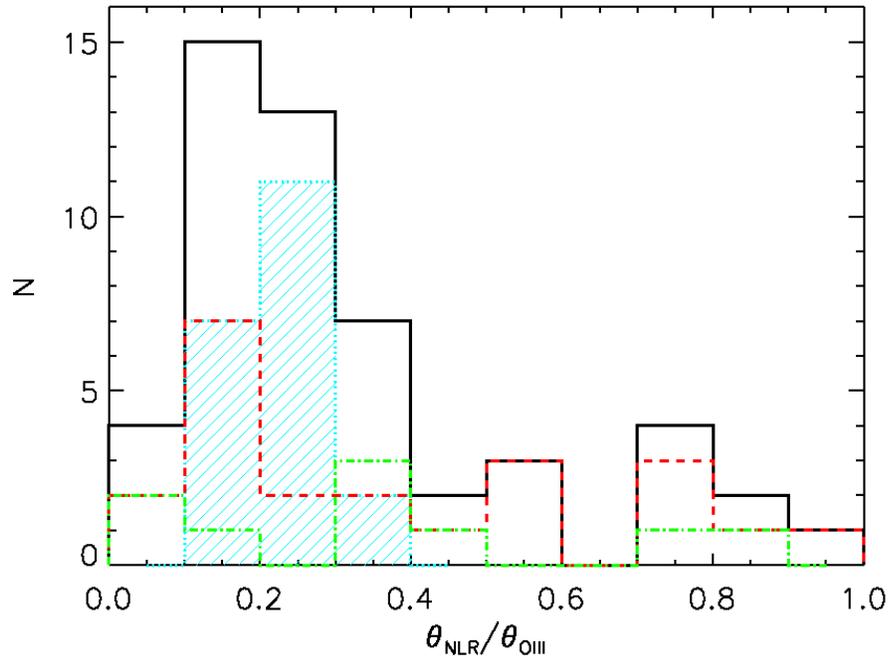}
\caption{\label{ap_nlr} Ratio of the NLR size to the optical aperture size used for the [OIII] flux and H$\alpha$ and H$\beta$ values. In all cases, the optical aperture encompasses the full NLR.}
\end{figure}

\begin{figure}
\epsscale{1.3}
\plottwo{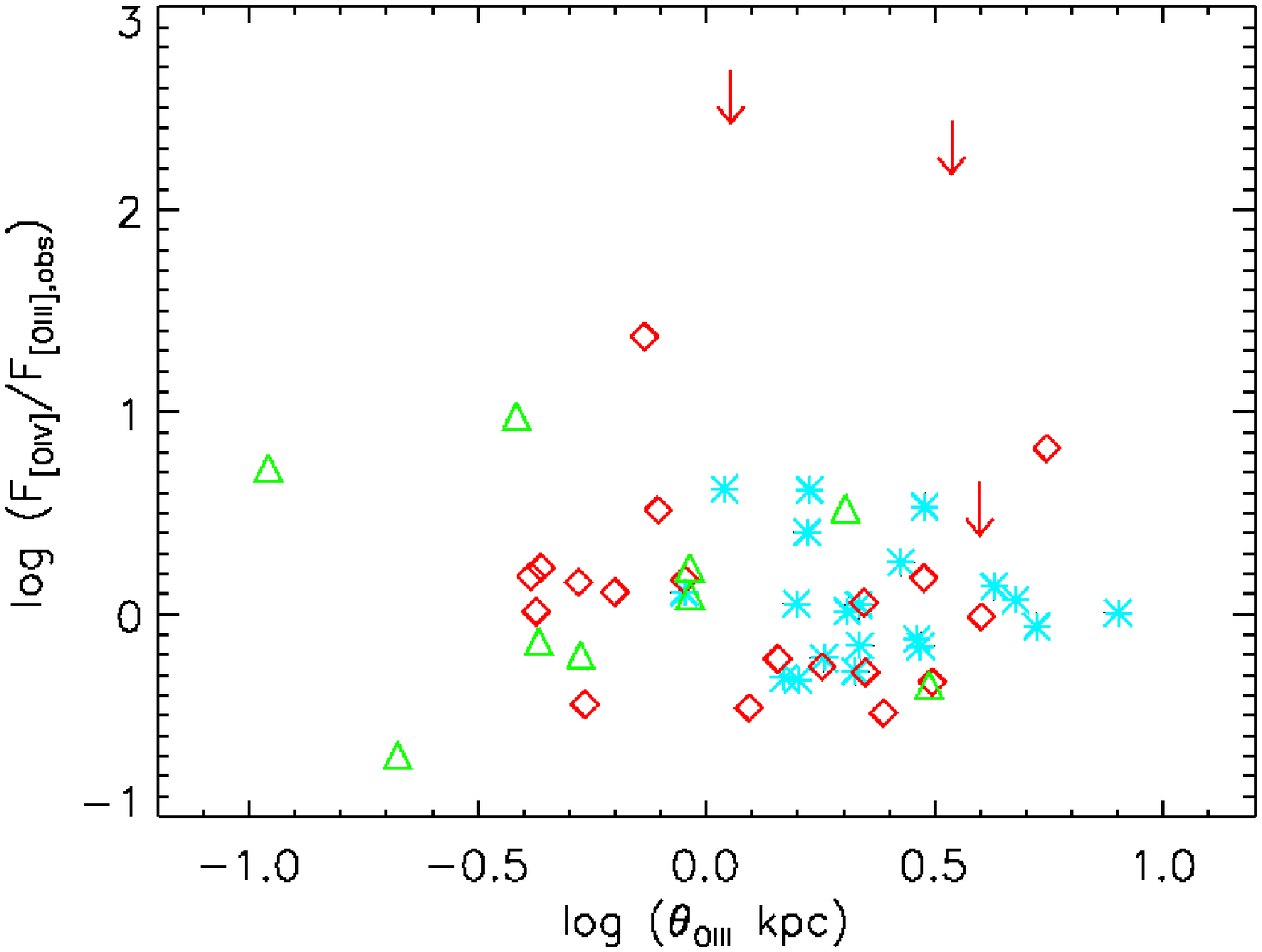}{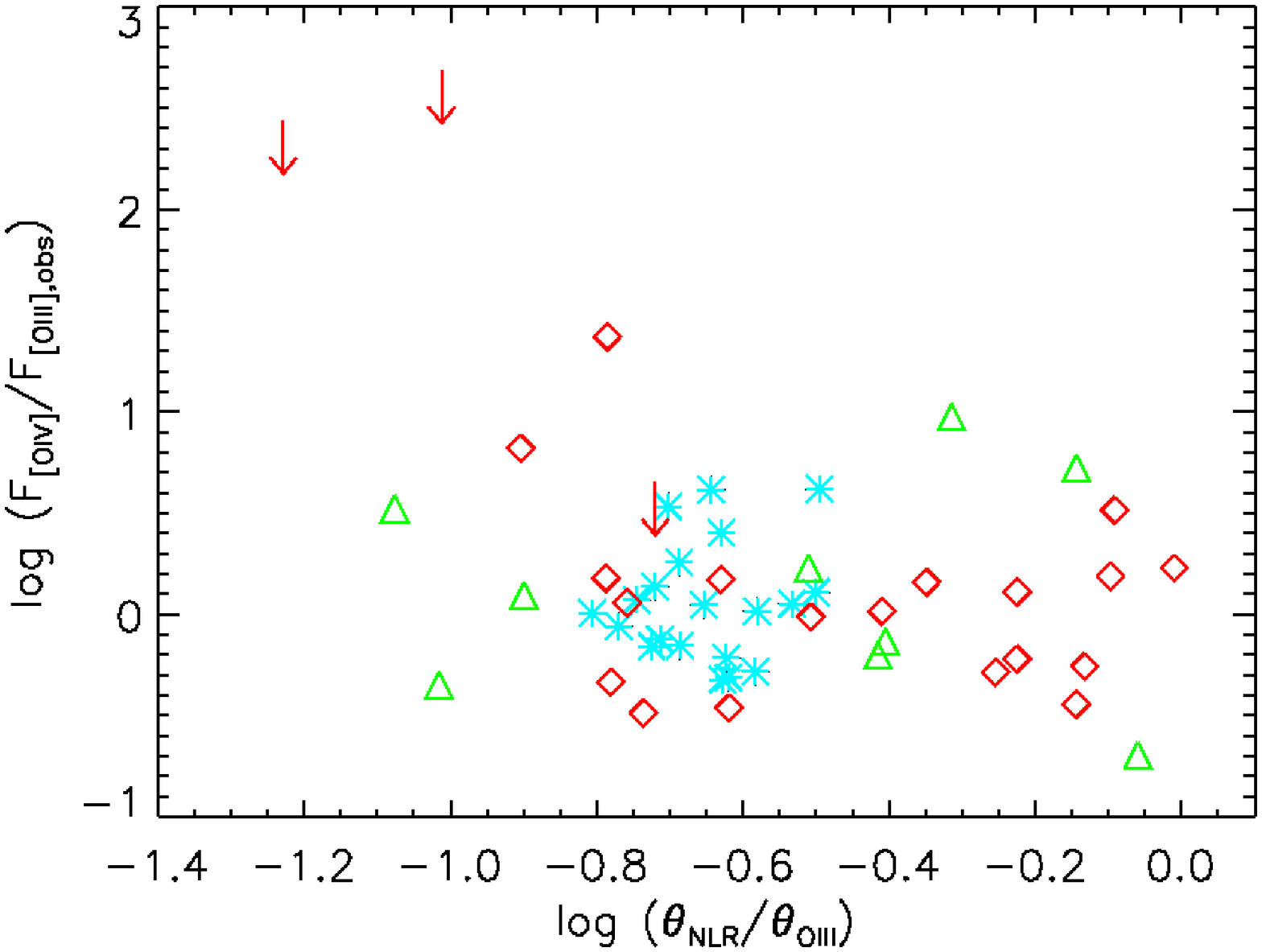}
\caption{\label{o3_o4_ap}Left: Log (F$_{[OIV]}$/F$_{[OIII],obs}$) vs log ($\theta_{OIII}$). Right: Log (F$_{[OIV]}$/F$_{[OIII],obs}$) vs log ($\theta_{NLR}$/$\theta_{OIII}$). The lack of significant trends suggest that the [OIV] flux originates in the NLR, consistent with the [OIII] flux, and that star formation processes in the host galaxy are not enhancing [OIII] emission.}
\end{figure}

\begin{figure}
\epsscale{1.3}
\plottwo{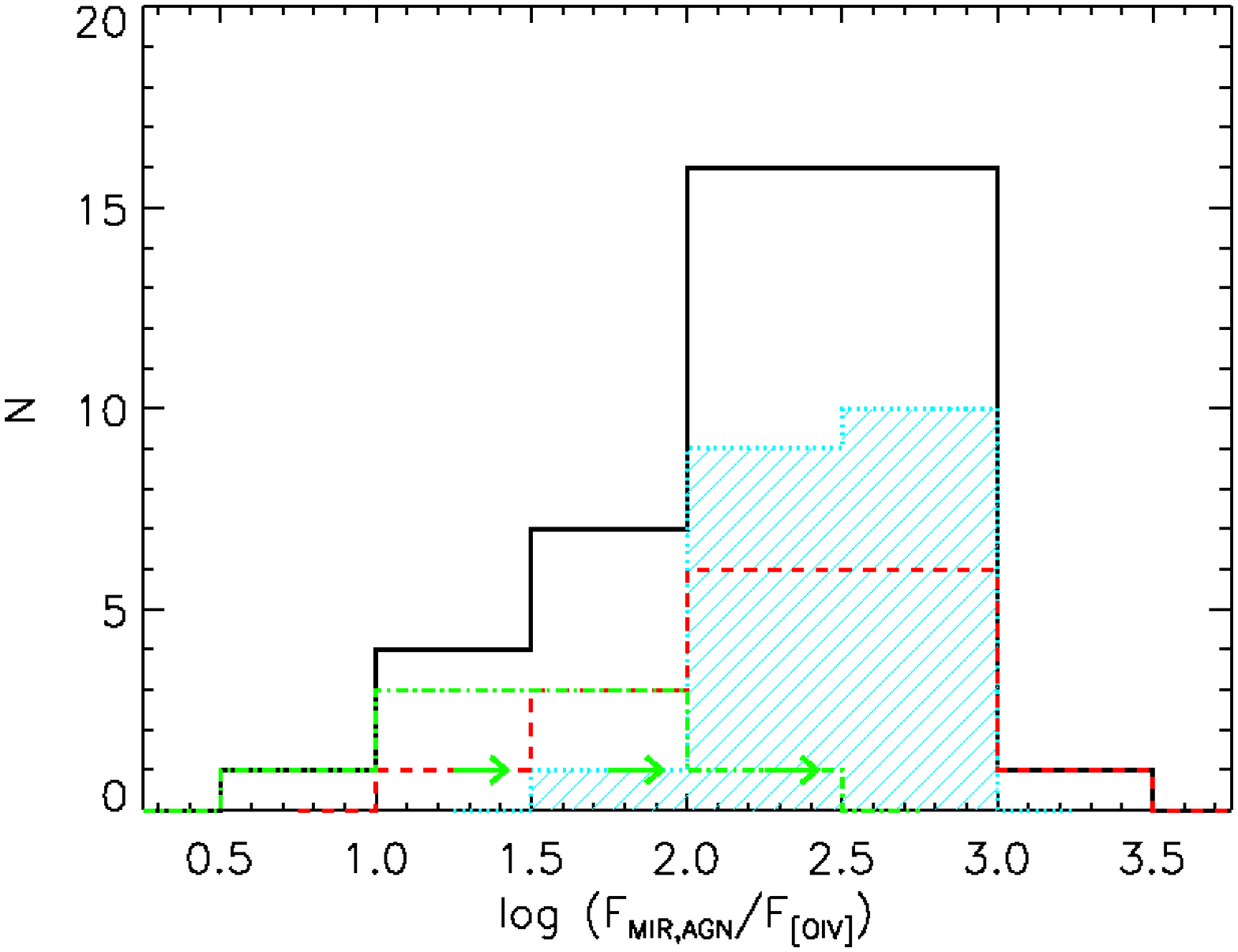}{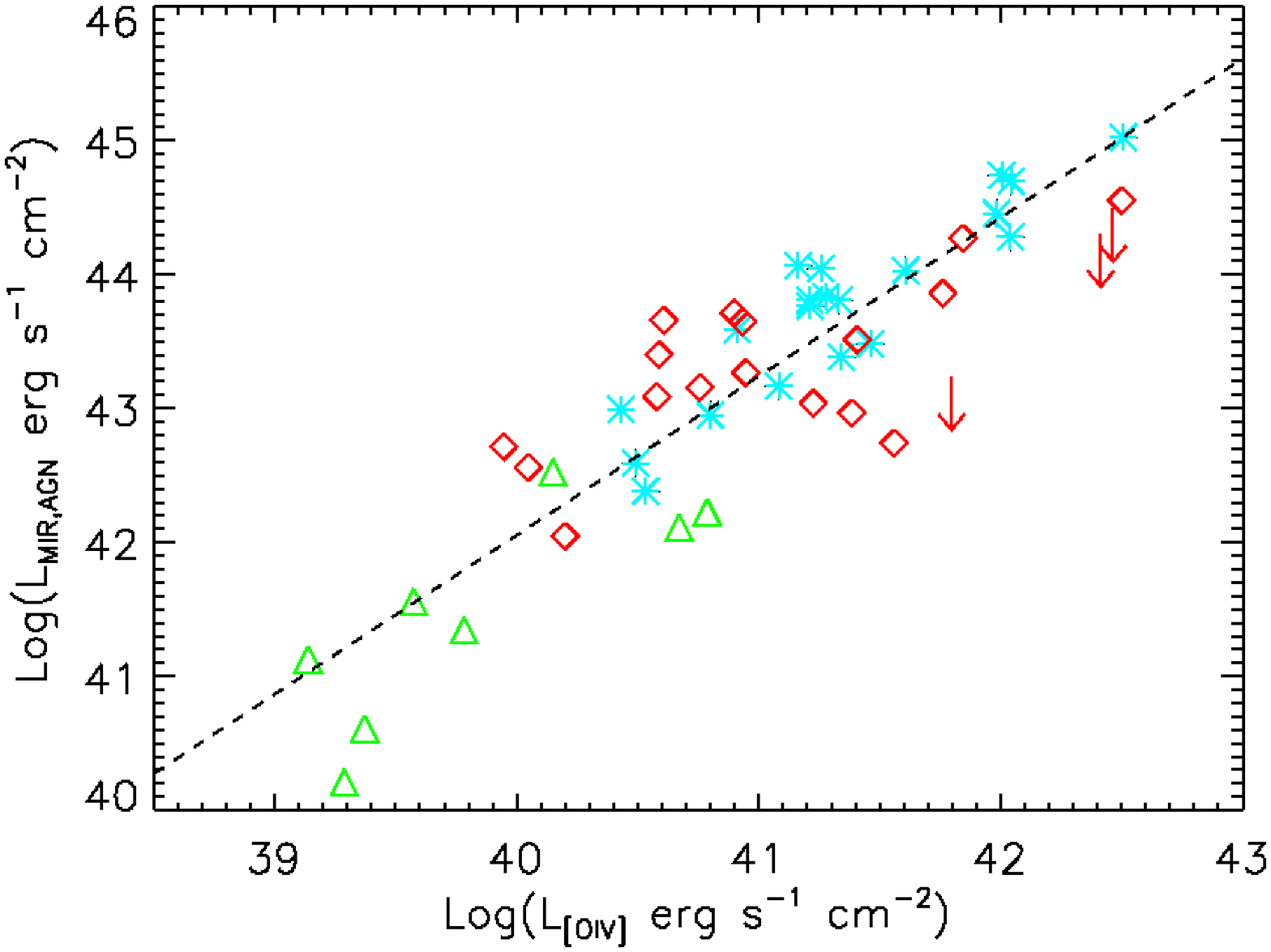}
\caption{\label{raw_mir}Left: Distribution of F$_{MIR,AGN}$/F$_{[OIV]}$, where F$_{MIR,AGN}$ subtractions out the starburst contribution using the simple dilution model in the text. Right: Log (F$_{MIR,AGN}$) vs log (F$_{[OIV]}$). Compared with the raw MIR flux, this corrected flux leads to a poorer agreement with the [OIV] flux and seems to over-substract the AGN contribution to the MIR flux for the ``PAH-strong'' sources.}
\end{figure}

\end{document}